\documentclass{aastex63}

\usepackage{natbib}
\usepackage{rotating}
\pdfoutput=1

\received{4 August 2020}
\revised{28 August 2020}
\accepted{31 August 2020}

\shorttitle{ORBARC}
\shortauthors{Vrijmoet et al.}

\begin{document}

\title{The Solar Neighborhood. XLVI. Revealing New M Dwarf Binaries and their Orbital Architectures}

\correspondingauthor{Eliot Halley Vrijmoet}
\email{vrijmoet@astro.gsu.edu}

\author[0000-0002-1864-6120]{Eliot Halley Vrijmoet}
\altaffiliation{Visiting Astronomer, Cerro Tololo Inter-American Observatory. CTIO is operated by AURA, Inc., under contract to the National Science Foundation.}
\affiliation{Department of Physics and Astronomy,
Georgia State University,
Atlanta, GA 30303, USA}
\affiliation{RECONS Institute,
Chambersburg, PA 17201, USA}

\author[0000-0002-9061-2865]{Todd J.\ Henry}
\altaffiliation{Visiting Astronomer, Cerro Tololo Inter-American Observatory. CTIO is operated by AURA, Inc., under contract to the National Science Foundation.}
\affiliation{RECONS Institute,
Chambersburg, PA 17201, USA}

\author[0000-0003-0193-2187]{Wei-Chun Jao}
\altaffiliation{Visiting Astronomer, Cerro Tololo Inter-American Observatory. CTIO is operated by AURA, Inc., under contract to the National Science Foundation.}
\affiliation{Department of Physics and Astronomy,
Georgia State University,
Atlanta, GA 30303, USA}

\author[0000-0001-7210-3912]{Serge B.\ Dieterich}
\altaffiliation{Visiting Astronomer, Cerro Tololo Inter-American Observatory. CTIO is operated by AURA, Inc., under contract to the National Science Foundation.}
\affiliation{Space Telescope Science Institute,
3700 San Martin Dr. Baltimore, MD 21218}
\affiliation{RECONS Institute,
Chambersburg, PA 17201, USA}

\begin{abstract}
We use 20 years of astrometric data from the RECONS program on the
CTIO/SMARTS 0.9-m to provide new insight into multiple star systems in the solar neighborhood.  
We provide new and updated parallaxes for 210 systems and derive nine 
high-quality astrometric orbits with periods of 2.49--16.63 years.
Using a total of 542 systems' parallaxes from RECONS, we compare systems within 25 parsecs to \textit{Gaia} DR2 to define criteria for selecting unresolved astrometric multiples from the DR2 results.
We find that
three out of four unresolved multi-star red dwarf systems within 25 pc in DR2
have \texttt{parallax\_error} $\geq 0.32$ mas,
\texttt{astrometric\_gof\_al} $\geq 56$,
\texttt{astrometric\_excess\_noise\_sig} $\geq 108.0$, \texttt{ruwe}
$\geq 2.0$, and parallaxes more than $\sim$10\% different than the long-term
RECONS results.  These criteria have broad applications to any work
targeting nearby stars, from studies seeking binary systems to efforts
targeting single stars for planet searches.
\end{abstract}


\section{Introduction} 
\label{sec:intro}

The orbit of a binary star pair provides a host of useful information about the system, but many such orbits, when considered together, add another dimension to their utility by providing statistical clues to the formation and evolution of these systems. The size and shape of any orbit we observe today is the product of a suite of processes pertaining to dynamical gravitational interactions, magnetic fields, radiative feedback, and gas dynamics and accretion \citep[for a thorough review see][]{Bat15}. The balance of these processes depends on the initial conditions at formation, including the initial stellar masses, their angular momenta, and ultimately the environment of the molecular cloud that led to these configurations. 

Untangling the roles and dependencies of each of these processes requires assembling a representative set of observed multi-star systems to which we can compare models of multiple star formation mechanisms \citep[e.g.,][]{Bat12}. Such observational results would also provide realistic inputs for multi-star evolution modeling \citep[e.g.,][]{Par14}. These applications have been well covered for solar-type and more massive stars, most notably by \cite{Moe17}. At the low-mass end of the stellar main sequence, the red dwarf systems, also known as M dwarfs, represent a particularly important application because they span a factor of eight in mass \citep[0.08~M$_\odot$ to 0.62~M$_\odot$;][]{Ben16} and, consequently, display a wide range of complex effects on their surfaces and interiors. This is evident in their remarkable scatter in luminosity at optical wavelengths \citep{Cle17}. 
The distribution of separations of stars in red dwarf binaries have been observed to peak at a few tens of AU \citep{Win19}, hinting at the distribution of their semi-major axes and thus orbital periods \citep{Moe19}, but statistics derived from true orbits remain largely as described in the review of \cite{Duc13}.
Hence the goal of our work is to assemble a rich set of orbits that, when taken together, can constrain the formation and evolution models of stellar multiples for these complicated low-mass stars in particular. This paper represents the beginning of that effort.

The challenge of measuring M dwarf orbits has long been these systems' intrinsic faintness. This faintness has prevented the late M dwarfs in particular from being observable with most spectroscopic instruments until recently \citep[e.g.,][]{CAR15,Win20}.
The diminutive masses of these stars also increase the orbital period for a given semi-major axis, so observing orbits larger than a few AU requires a decades-long time commitment for the smallest M dwarfs. For many instruments these targets must also be restricted to within a few dozen parsecs to ensure all are sufficiently bright.

The REsearch Consortium On Nearby Stars\footnote{\url{www.recons.org}} (RECONS) has been observing M dwarfs within 25 pc since 1999 via astrometry and photometry at Cerro Tololo Inter-American Observatory (CTIO). 
Astrometric observations of these systems provide a beautiful complement to spectroscopic observations, as these methods are sensitive to different types of binaries (different mass ratios and separations). 
Furthermore, astrometry can achieve comparable signal-to-noise with less light, so high-quality observations can be obtained on systems as faint as $VRI$ 22 mag using a small aperture telescope such as the CTIO/SMARTS 0.9~m. 
With these benefits of the method and an observational program that has been observing the same set of stars for as long as 20 years, in this work RECONS is creating a catalog of multi-star system orbits that will complement existing work in a way that is critically important for these small stars in particular.

In this paper we present several orbits from our ongoing astrometry program at CTIO, representing the first infusion of M dwarf orbits for a broader project, described in $\S$\ref{sec:theproject}. 
The sample of systems in our observing program is described in $\S$\ref{sec:sample}, and the astrometric observations and reductions are summarized in $\S$\ref{sec:observations}. Because every astrometric orbit is built upon accurate characterization of the system's parallactic and proper motions, 
$\S$\ref{sec:results} presents 210 systems' new or updated trigonometric parallaxes from our long-term program, 
and $\S$\ref{sec:orbits} presents nine astrometric orbits.
In $\S$\ref{sec:gaia} we compare 542 of our ground-based parallax results to those of the space-based \textit{Gaia} mission's Data Release 2 \citep[DR2;][]{GAI18a}
and establish four criteria to 
select likely unresolved multi-star systems from DR2, which could benefit any work that intends to target (or, alternatively, avoid) unresolved multiples among nearby red dwarfs. Future work, including observations of these systems to map orbits for this project, are briefly discussed in $\S$\ref{sec:SOARvalidation}.
Our conclusions are summarized in $\S$\ref{sec:conclusions}.

%
\section{Orbital Architectures Project} \label{sec:theproject}
%

The larger project launched by this work, dubbed the Orbital Architectures project, intends to bring together $\sim$120 M dwarf orbits to establish the distributions of orbital periods and eccentricities for massive, intermediate, and low-mass M dwarfs. 
Because these structures are the end results of billions of years of dynamical evolution compounded on their configurations at formation, this study of orbits provides several quantities that directly constrain key aspects of stellar formation models. The fraction of systems that form multiple stars (i.e., multiplicity) is the end product of the number of stars produced by each stellar core via fragmentation, modulated by the influences of magnetic fields, radiative feedback, and the dynamical environment during evolution \citep{Duc13}.
The deeper statistical properties of these systems, such as their distributions of mass ratios, separations, periods, and eccentricities, constrain dominant processes in their formation. 
Observations are often used to inform inputs to star cluster formation models, with the outputs compared to additional observations to evaluate the models' validity.
Also notable is that the more unusual multi-star configurations observed act as crucial tests for those formation models, as these \textbf{outliers} must not be ruled out as physically impossible.

Previous investigations into these distributions have primarily focused on solar-type stars \citep[e.g., ][]{Hal05,Duq91}. Attention turned to the M dwarfs mainly though efforts to determine their mass-luminosity relation \citep{Hen93,Hen99,Del00}, and any investigations of their orbital distributions are usually presented as side notes. \cite{Udr00} present early results of a volume-limited all-sky search for M dwarf multiples with the CORAVEL spectrometers, forming a preliminary period versus eccentricity distribution using 13 of these systems supplemented by 17 from the literature. They note evidence of a circularization timescale that matches that of G- and K-type stars, as well as hints of a paucity of circular orbits with long periods up to 20 years.  
Although updates to this work have not been published, it remains a good starting point to which we can compare our study of stellar companions.

The M dwarfs' smaller masses, radii, and luminosities have also made them attractive targets for exoplanet searches across all major detection regimes: radial velocities with CARMENES \citep{Qui16}, transits with the \textit{Transiting Exoplanet Survey Satellite}  \citep[\textit{TESS};][]{Ric14,Sta18}, and direct imaging with the anticipated \textit{James Webb Space Telescope} \citep[\textit{JWST};][]{Gar06} and the Nancy Grace Roman Space Telescope\footnote{Formerly known as the \textit{Wide Field Infrared Survey Telescope} or \textit{WFIRST}.}. The M dwarfs' lower luminosities place their habitable zones closer to the stellar surface than to those of larger stars, increasing the radial velocity semiamplitudes and probability of transits for those habitable planets. Their lower luminosities also benefit direct imaging observations by decreasing the star-planet contrast ratios. 
Reliable characterization of any detected exoplanets, however, depends critically upon precise information of their host stellar systems, including accurate stellar properties and identification of stars in the aperture \citep[see, e.g.,][]{Cia15,Fre13,Fur17}.  
The significance of these effects (and the observing follow-up required) has inspired most surveys to carefully omit multi-star systems from their samples of potential exoplanet hosts via extensive literature searches and sometimes even pre-survey observations of potential targets \citep[e.g.,][]{Cor17}. This step demonstrates the value of surveys that detect and characterize stellar multiples with a wide range of potential orbital diversity, and in particular those surveys that combine different observational techniques to break free of any single method's limitations.

It is with these requirements in mind that we have begun this study of M dwarf systems' orbits. By assembling nearby systems from our own astrometry, and later adding radial velocity studies and a new speckle interferometry program at SOAR (see $\S$\ref{sec:SOARvalidation}), we will form a complete picture of M dwarf orbits out as far as 10 AU, while simultaneously representing all members of this expansive spectral type (from 0.08 M$_\odot$ to 0.62 M$_\odot$). This wide survey across stellar type as well as orbit size is the key element that will make this work useful for constraining formation and evolution models as well as providing insight for upcoming exoplanet work.

%
\section{Sample} \label{sec:sample}
%


The sample presented here is composed of red dwarfs within 25 parsecs visible from the southern sky. These limits are enforced as trigonometric parallax $\pi \geq 40$ mas, $V$ band absolute magnitude $9 \lesssim M_V \lesssim 24$, and declination $\delta \lesssim +30^\circ$. 
Table~\ref{tab:astrRECONS} includes 210 systems (of 220 objects) for which we present new or updated parallaxes in this work.  
Those within 25~pc are also included in Table~\ref{tab:RECONS-GDR2} along with many previously published parallaxes from RECONS, forming the sample of 582 systems that we compare with \textit{Gaia} DR2.
The absolute magnitude limits correspond to M dwarf mass limits of $0.08$~M$_\odot$ and $0.62$~M$_\odot$ using the mass-luminosity relation (MLR) of \cite{Ben16} for the $V$ band. For systems that have no reliable $V$ band photometry available, we instead require $5.3 \lesssim M_K \lesssim 12.0$ in 2MASS $K_S$ band. 

The growth of the sample has followed the growth of the RECONS observing program at the CTIO/SMARTS 0.9~m, which began in 1999 under the auspices of the NOAO surveys program. With the goal of identifying ``missing'' members of the Solar neighborhood, initial RECONS astrometry targets were red and brown dwarfs that were deemed likely to be nearby but were missing precise trigonometric parallaxes. These targets were selected from proper motion surveys and photometric distance measurements. After the RECONS program graduated from the NOAO surveys program, it continued fulfilling the spirit of that effort, providing a database of time-series astrometric and photometric observations that have been fundamental to investigations of several aspects of these nearby M stars beyond their distances. These studies, collected in \textit{The Solar Neighborhood} series of papers, include work on M dwarf 
populations~\citep{Hen06,Win17,Hen18},
ages~\citep{Rie10,Rie14,Rie18}, 
metallicities~\citep{Jao05,Jao11,Jao17}, 
surface activity and long-term photometric variability~\citep{Hos15,Cle17}, and 
multiplicity~\citep{Win19}, 
as well as 
white dwarfs~\citep{Sub09,Sub17}, 
the stellar-substellar boundary~\citep{Die14}, and
exoplanet searches~\citep{Lur14}.

%
\section{Astrometry Observations and Reductions} \label{sec:observations}
%

The parallax results and orbits presented here come from the astrometric monitoring program at the CTIO/SMARTS 0.9~m telescope. In this section we focus on the details and capabilities of those observations.

\subsection{Observing Red Dwarfs at the CTIO/SMARTS 0.9~m}
All RECONS astrometry is currently carried out at the CTIO/SMARTS 0.9~m telescope, with the same camera and CCD setup used for that program since its inception in 1999. The CCD is a Tektronics 2048 $\times$ 2048 with $401$ mas/pixel, with only the central quarter (6$\farcm$8 $\times$ 6$\farcm$8) used for astrometry observations in order to minimize coma and other distortions. Observations for each target are taken in either the $V$, $R$, or $I$ filter\footnote{$V$, $R$, and $I$ here and thoughout this paper refer specifically to the Johnson $V$ and Kron-Cousins $R$ and $I$ filters, respectively.}; the only discontinuity in the use of these filters is the period from March 2005 to August 2009 when the Tek \#2 filter, which had become cracked, was replaced by the effectively identical Tek \#1 filter. The Tek\#1 filter that matches the Tek \#2 filter photometrically to 1\%, but resulted in systematic offsets in the astrometry (see \cite{Sub09} and \cite{Rie10} for details), so we returned the Tek \#2 filter to service in August 2009. 
Those offsets are now avoided in each astrometry reduction by choosing reference stars located near to the target star on the CCD where possible, or by omitting the Tek \#1 frames for systems that have a sufficient number of Tek \#2 frames.

The specifics of the observations are given briefly here, but also described in more detail in \cite{Jao05} and \cite{Hen06}. Each target is placed on the CCD such that the number of useful reference stars is maximized (most fields have 5--10), and is observed in a single filter ($V$, $R$, or $I$) chosen to maximize the number of counts in that target star and reference stars. 
Each target is visited at least twice per year, with 3--5 frames taken at each visit, each within 120 minutes of the target's transit of the meridian to minimize the correction needed for differential color refraction (DCR). 
Exposure times vary from 30 to 300 seconds, with some exceptional systems requiring up to 900 seconds, and exposures are adjusted on-the-fly by the observer to accommodate minute-by-minute variations in seeing and targets of different brightness in different filters.

The full observing list consists of $\sim$700 red, brown, and white dwarfs,
observed in 4--6 runs per year of 10--16 nights each run. 
Proper motions and parallaxes are considered reliable when the data span at least two years and 60 frames and about 12 visits. 
Many targets remain on the observing list after this point for long-term astrometric and photometric studies.

\subsection{Astrometry Reductions: Characterizing Proper Motion, Parallax, and Orbital Motion}
\label{sec:reductions}

Astrometry reductions are conducted as described in \cite{Jao05}, so only the basic steps are summarized here. All frames are first bias-subtracted and flat-fielded in IRAF using the bias frames and dome flat frames taken nightly prior to observations. Astrometric reductions then proceed for each system using all frames accumulated for it during the program as follows:
\begin{enumerate}
\item Reference stars and the target star are tagged and centroided in each frame using SExtractor \citep{Ber96}.

\item A representative, high-quality ``trail plate'' is chosen, and that field is matched to the 2MASS catalog \citep{Cut03,Skr06} to determine rotation and scaling for that frame.

\item Target and reference star positions in all frames are measured relative to that trail plate and corrected for differential color refraction (DCR) using the empirical relation determined for our specific program \citep[described in][]{Jao05}.

\item Using GaussFit \citep{Jef87}, a least-squares optimization is performed to determine the plate constants for each frame and relative proper motions of the reference stars and science target star, under the assumption that the reference stars' proper motions sum to zero.

\item The above GaussFit optimization also determines the relative parallax of the science star. This value is then corrected to absolute parallax using the photometric distances of the reference stars.

\end{enumerate}
The result of this process is proper motion, parallax, tangential velocity, and the time-series residuals of the proper motion and parallax fit for each tagged star in the field. For single stars, these residuals are flat, with no long- or short-term trends. 
Two dozen single stars, spread evenly across all hour of R.A., are monitored to evaluate trends in residuals and thus confirm the astrometric stability of the telescope and instruments.
The median deviation in the nightly mean points for these ``flatline'' systems is 2.36~mas and 2.55~mas in Right Ascension and Declination, respectively, after proper motion and parallax fits. The median parallax error for systems in the 25~pc sample is 1.40~mas.

 Unresolved multiple systems are detectable in our data by periodic motions of the system's photocenter
superimposed on the parallactic and proper motion, corresponding to the photocenter's orbit around the system's center of mass. In these cases our usual method is to fit the proper motion and parallax using the pipeline described above, then fit the orbital motion left in the residuals using the algorithm of \cite{Har89}. 
This preliminary orbital motion is then subtracted from the residuals, and the proper motion and parallax are fit again to secure a more precise solution. 
 The orbit fit in most of these cases 
does not represent the final photocentric orbit 
unless more than one cycle has been observed. 
More robust orbital results are possible using an updated algorithm from \cite{Die18} that fits the parallax, proper motion, and photocentric orbital motion simultaneously, hence that is the procedure we have employed for the orbits presented in $\S$\ref{sec:orbits}.

%
\section{RECONS Parallax Results} \label{sec:results}
%

    The decades-long baseline of this astrometry program has allowed it to fill a unique niche in stellar astrophysics through both the astrometry and the photometry available in these data. In its first decade, the program focused on filling the paucity of nearby red dwarf parallaxes, improving the total number of stellar systems known to be within 10 pc by 15\%. As the \textit{Gaia} mission filled gaps in the 25 pc sample and promises to continue adding to and validating these parallaxes in future data releases, the RECONS astrometry program has shifted toward harnessing the strength of up to 20 years of observations of these targets. 
These time-series observations have already opened doors for more comprehensive multiplicity surveys \citep{Win19}, analyses of which systems do \textit{not} have low-mass companions~\citep{Lur14}, and studies of multi-year photometric variability cycles on these typically active stars \citep{Hos15}.

It is with this focus on system characterization that we present the parallaxes for 210 systems in Table~\ref{tab:astrRECONS}, which includes 146 new values and 64 updates to the RECONS catalog since the last publication in this series \citep{Hen18}. The final column of Table~\ref{tab:astrRECONS} notes if a preliminary orbit has been fit to this astrometry data to improve the results (``orbit''), or if the time-series astrometric residuals have a perturbation to which we have not fit an orbit (``PB''). In most of these cases we have no reason to suspect that the perturbation is not astrophysical, but the signal shape is not yet well defined enough to permit an orbit fit. That column also notes if this result is an update of previously published parallax in this series (``update''), defined as a change in absolute parallax of more than 2.0 mas, parallax error improvement by a factor of 2 or more, or parallax error that fell from above 3.0 mas to below 2.0 mas. For each system, Table~\ref{tab:astrRECONS} gives the name (column 1), Right Ascension and Declination (columns 2 and 3), filter of our observations (4), number of seasons (5) and frames (6) over which it has been observed, dates of time coverage (7) and duration of time coverage (8). Also listed is the number of reference stars used in the final astrometry reduction (column 9), relative parallax (10), correction to parallax based on reference star photometric distances (11), and final absolute parallax (12). The proper motion (column 13), position angle of proper motion (14), and tangential velocity (15) are also results of our parallax solutions.

%
\section{RECONS Orbit Results} \label{sec:orbits}
%

\begin{figure}
\gridline{\rightfig{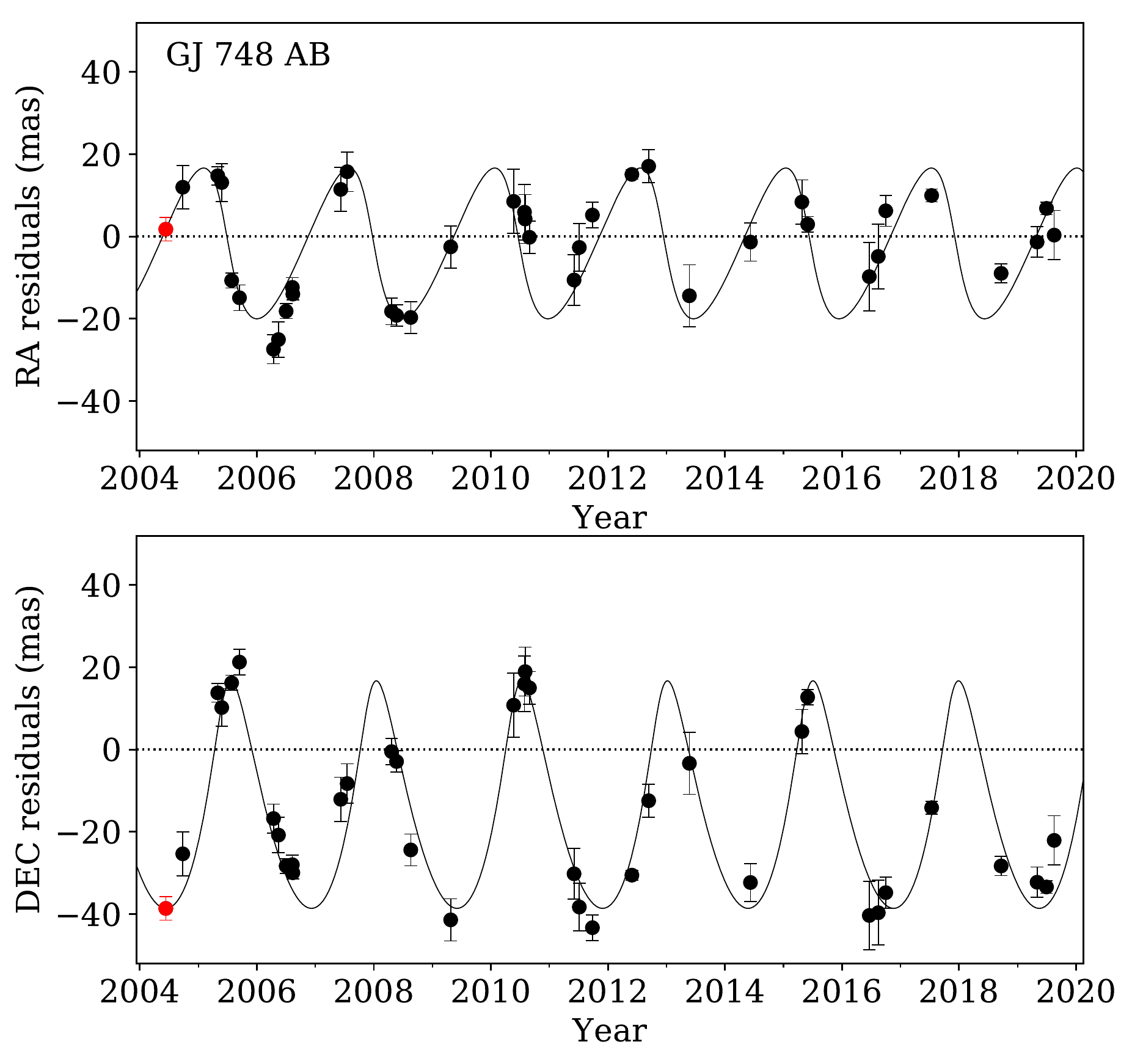}{0.4\textwidth}{}
\leftfig{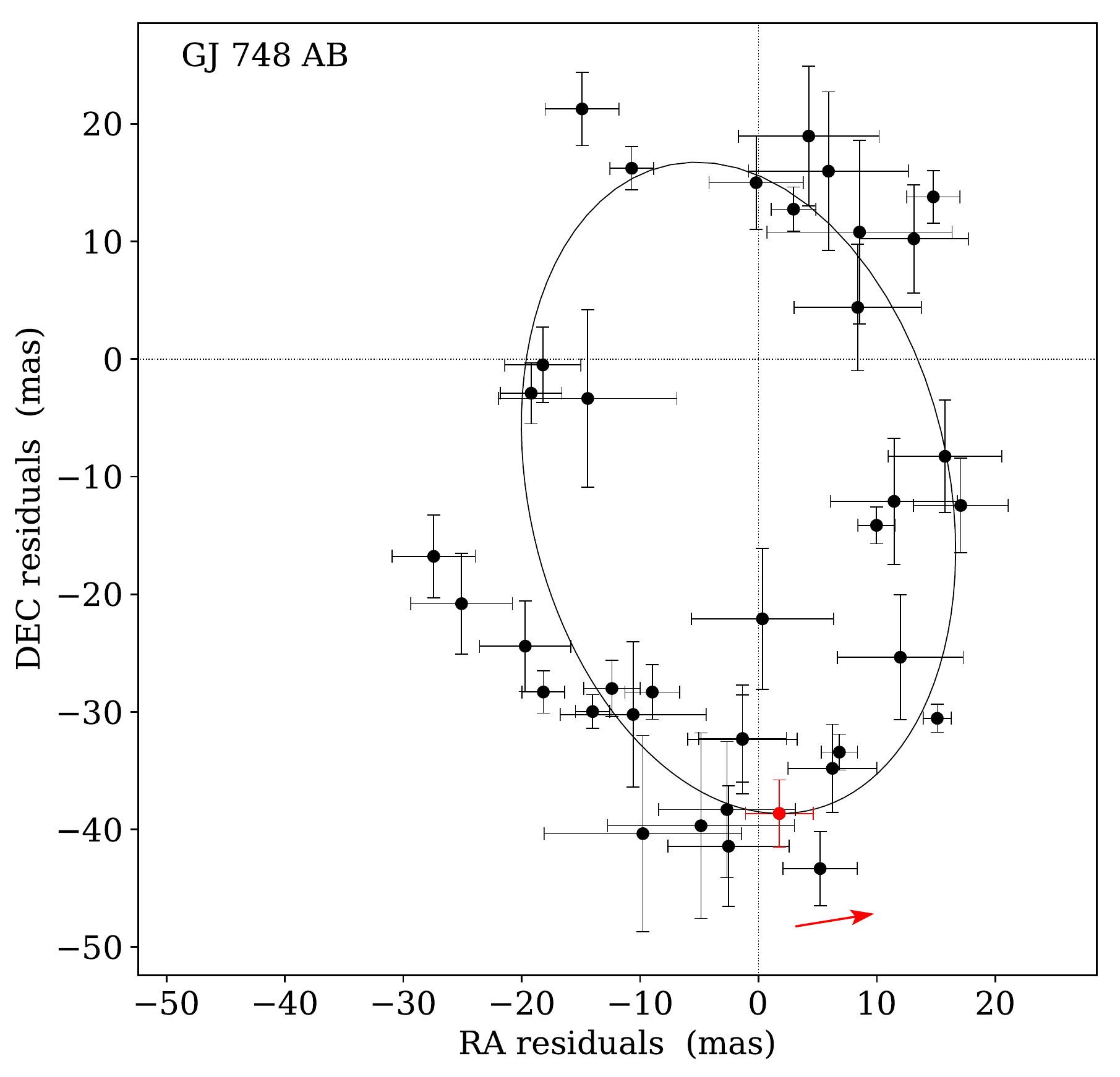}{0.38\textwidth}{}}
\gridline{\rightfig{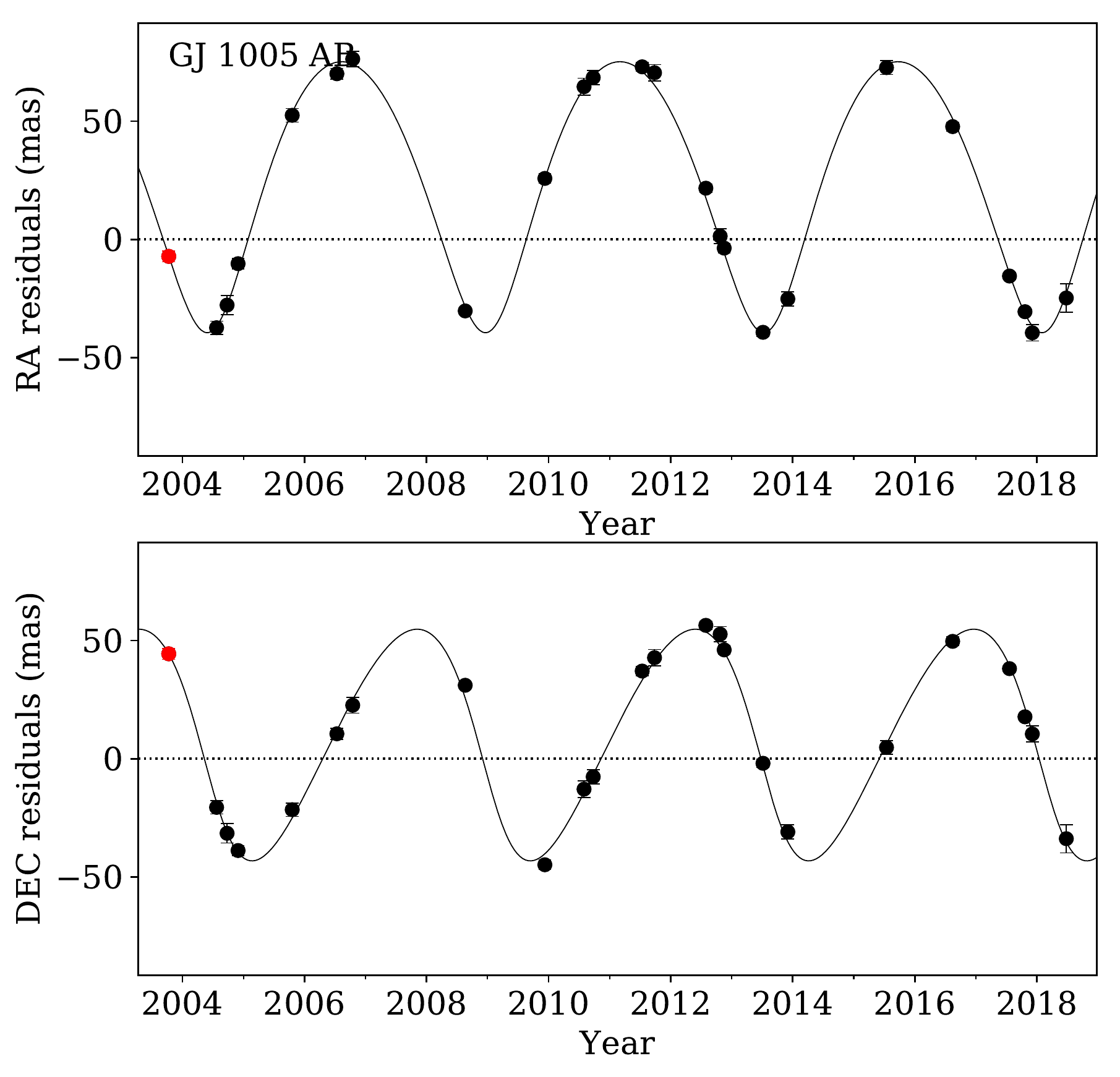}{0.4\textwidth}{}
\leftfig{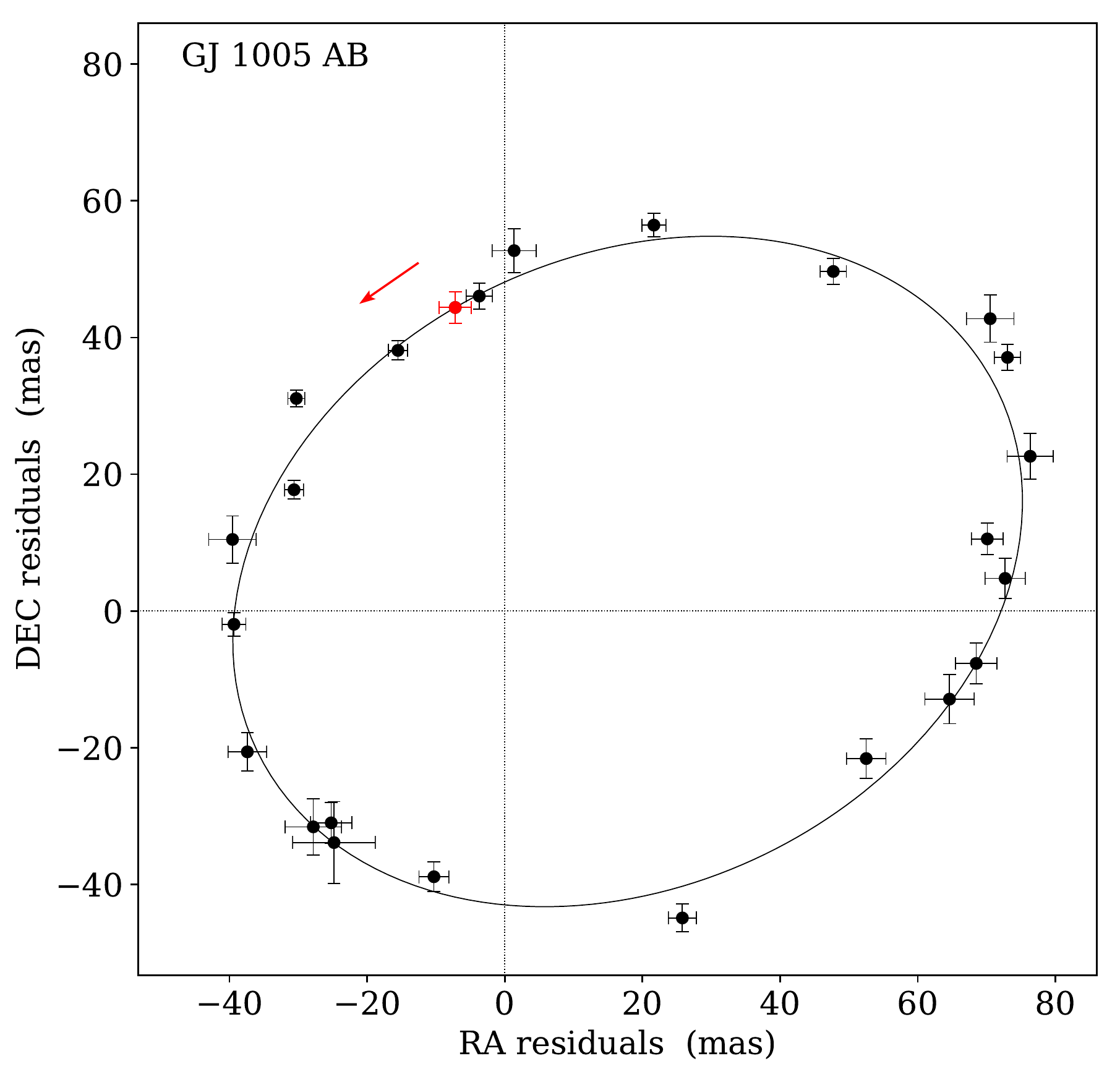}{0.38\textwidth}{}}
\gridline{\rightfig{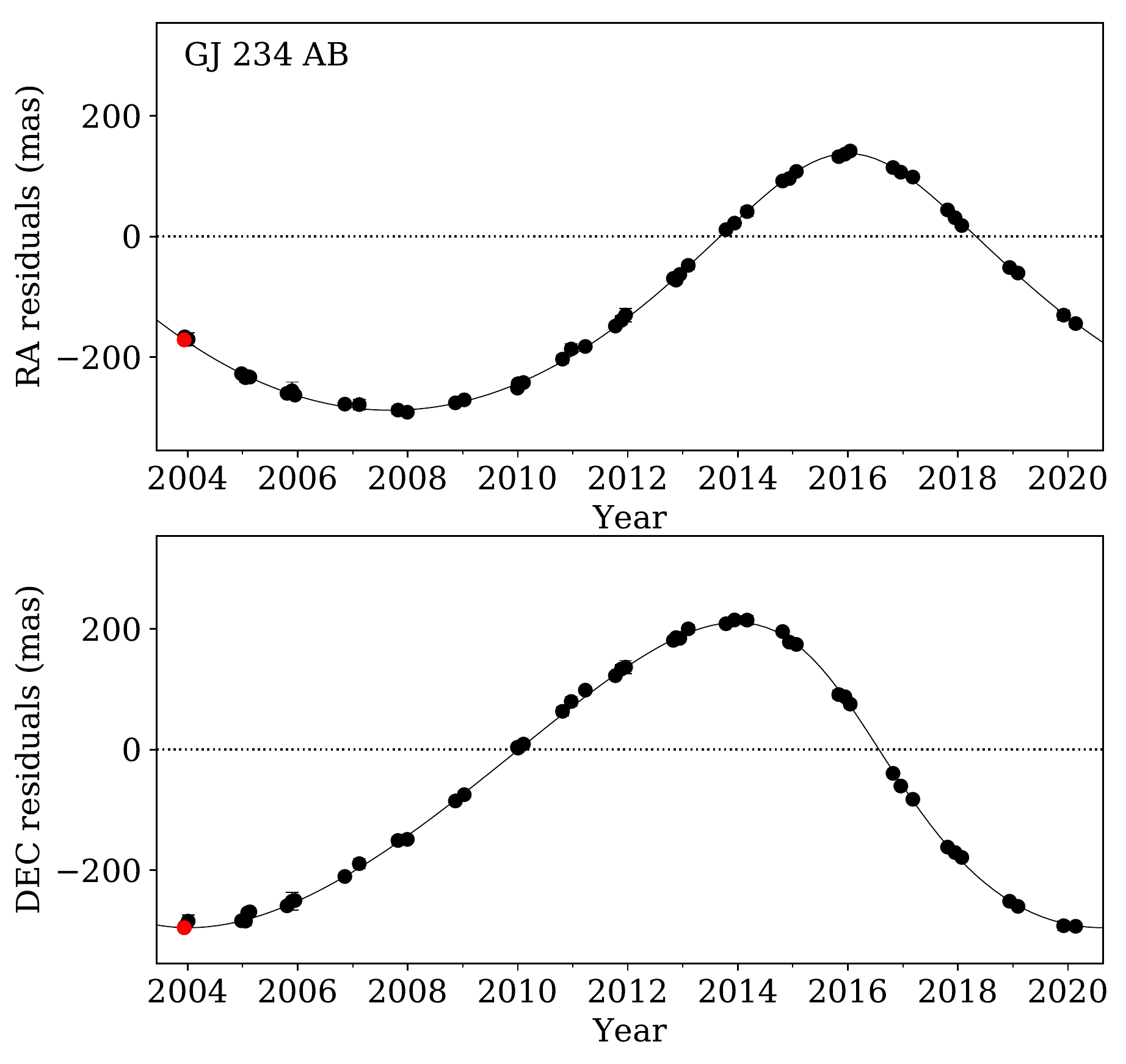}{0.4\textwidth}{}
\leftfig{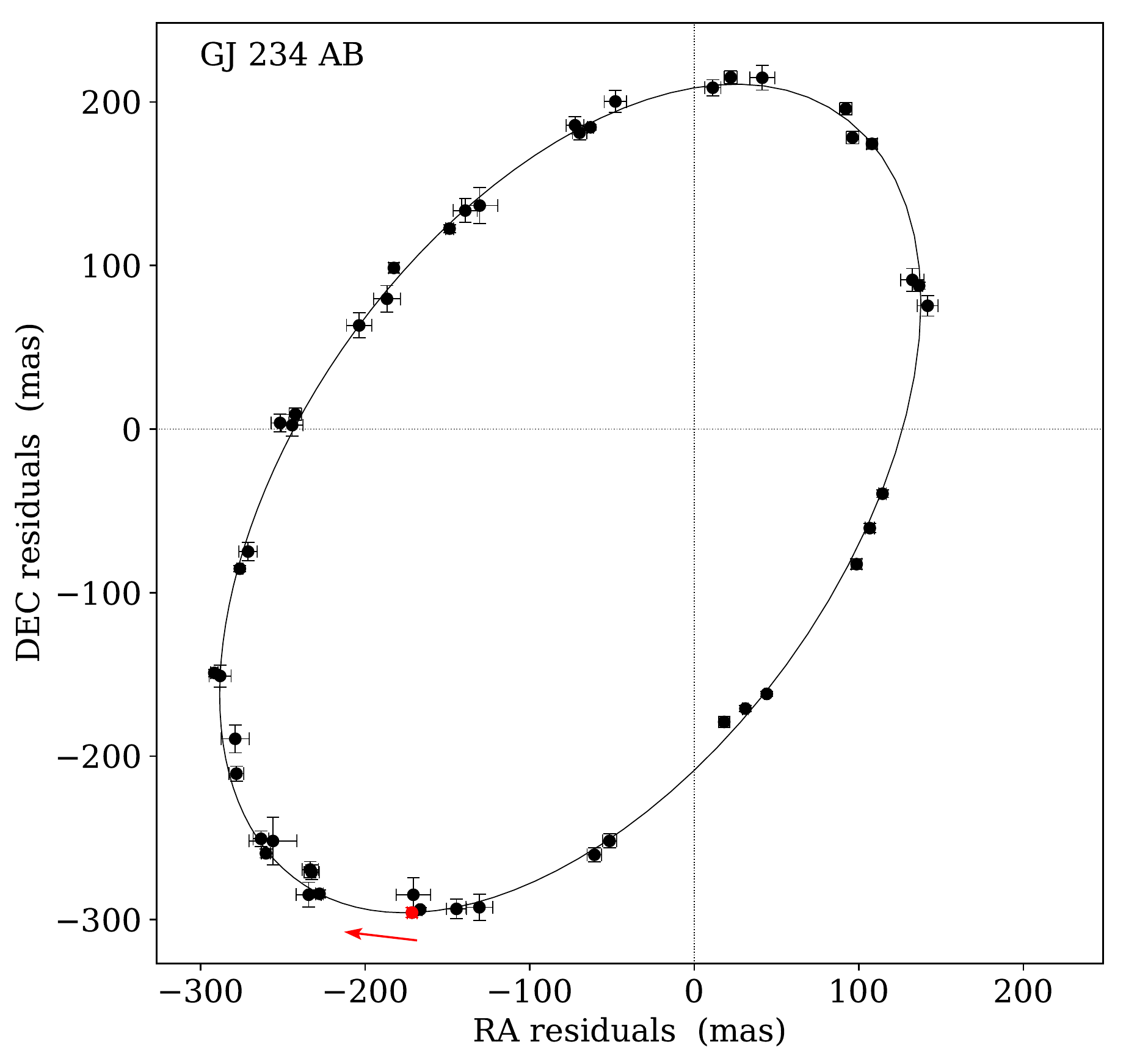}{0.38\textwidth}{}}
\caption{\label{fig:orbits_cal} \scriptsize 
Astrometric residuals, after proper motion and parallax have been removed, for three nearby red dwarf systems showing perturbations indicative of orbiting companions. In each panel, the solid line represents the orbit fit to that system's photocentric motion, for which the best-fit elements are given in Table~\ref{tab:orbelements}. 
The first epoch is marked with a red point, and the red arrow indicates the direction of motion. In the right column plots, north is up and east is to the right.
These systems have well-known solutions in the literature, making them calibration systems for our observations and fitting routine. 
\textit{Top to bottom:} GJ~748~AB ($P_\mathrm{orb} = 2.49$ years), GJ~1005~AB (4.56 years), and GJ~234~AB (11.16 years).
}
\end{figure}

\begin{figure}
\gridline{\rightfig{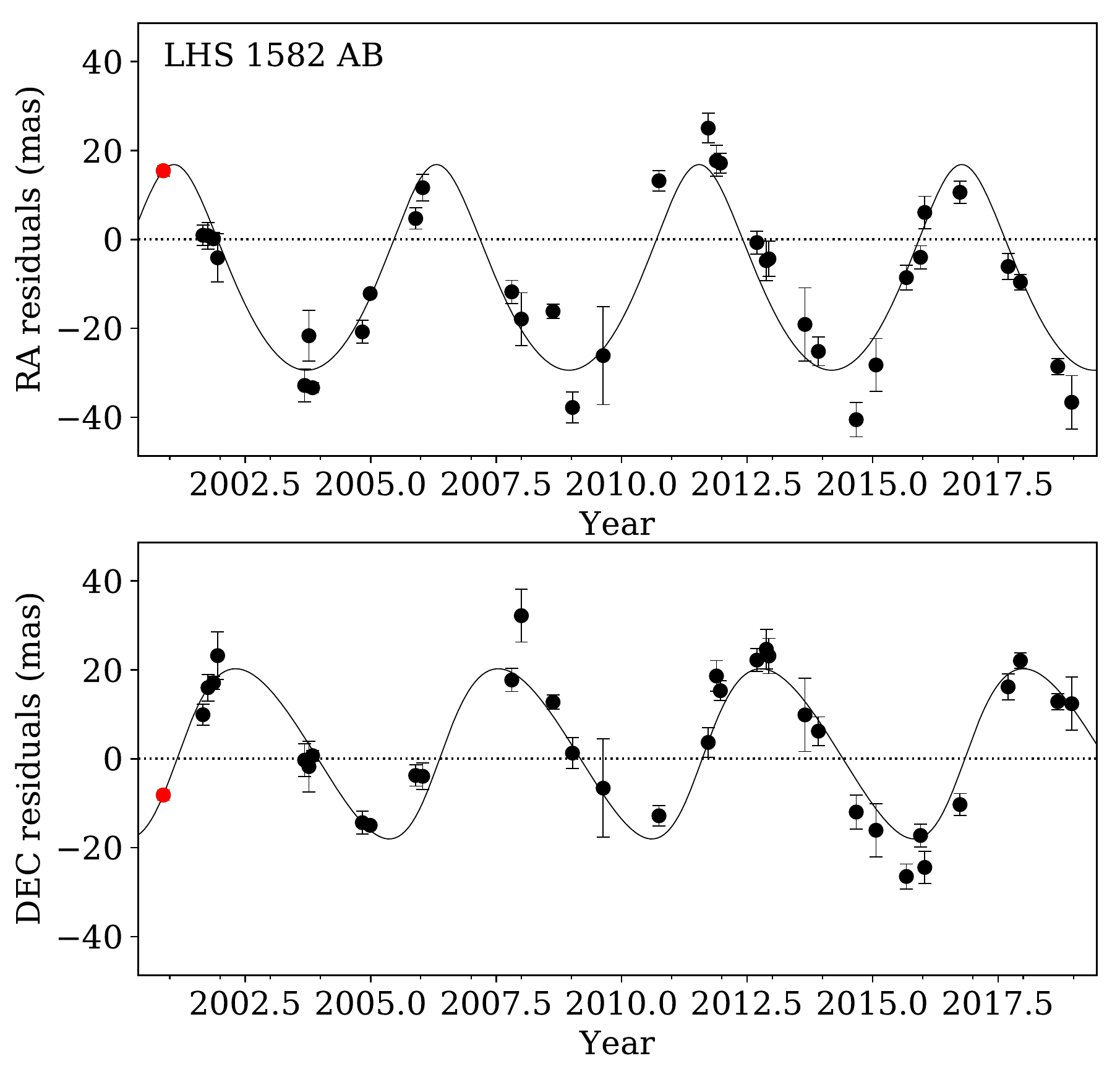}{0.4\textwidth}{}
\leftfig{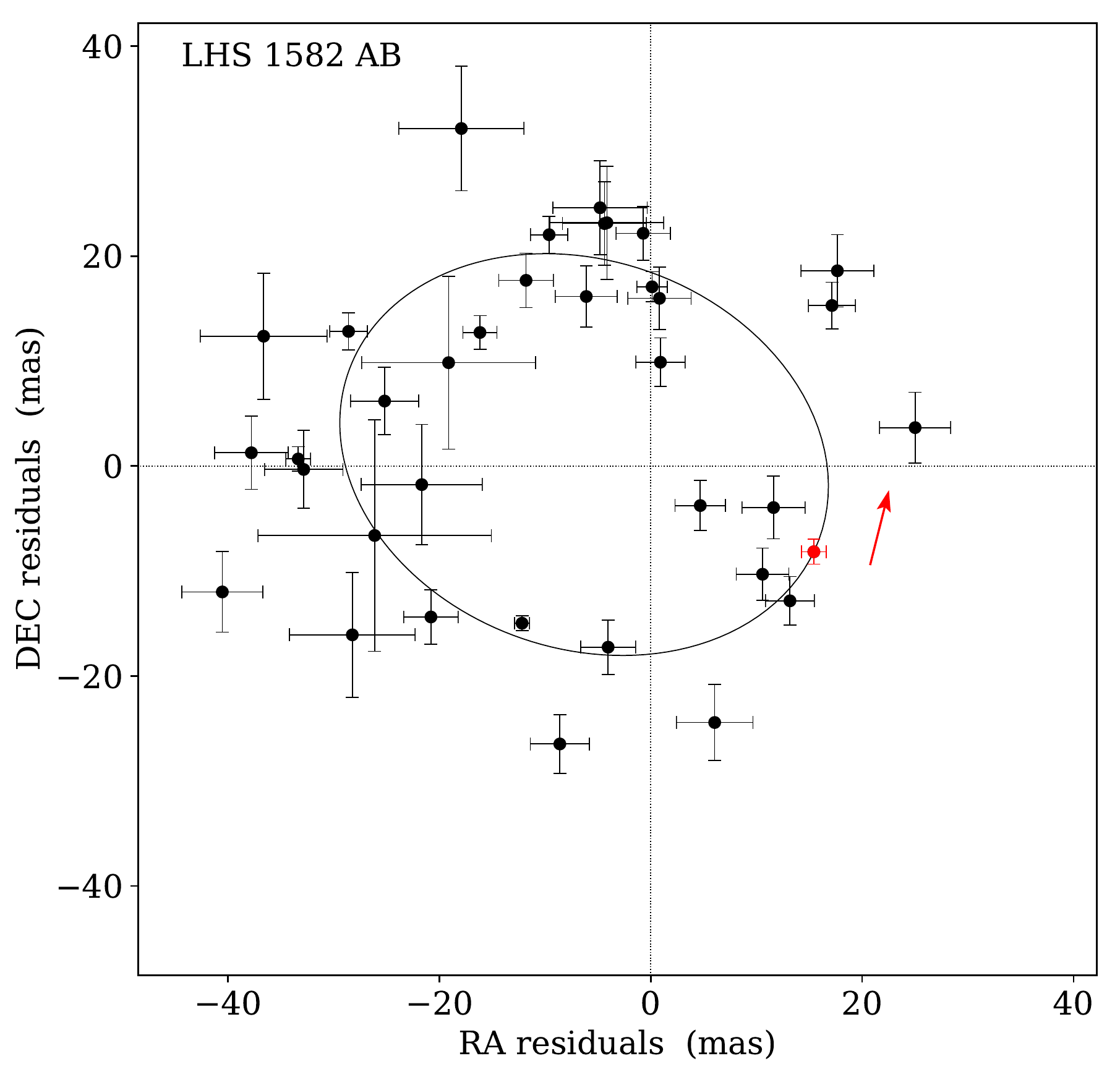}{0.38\textwidth}{}}
\gridline{\rightfig{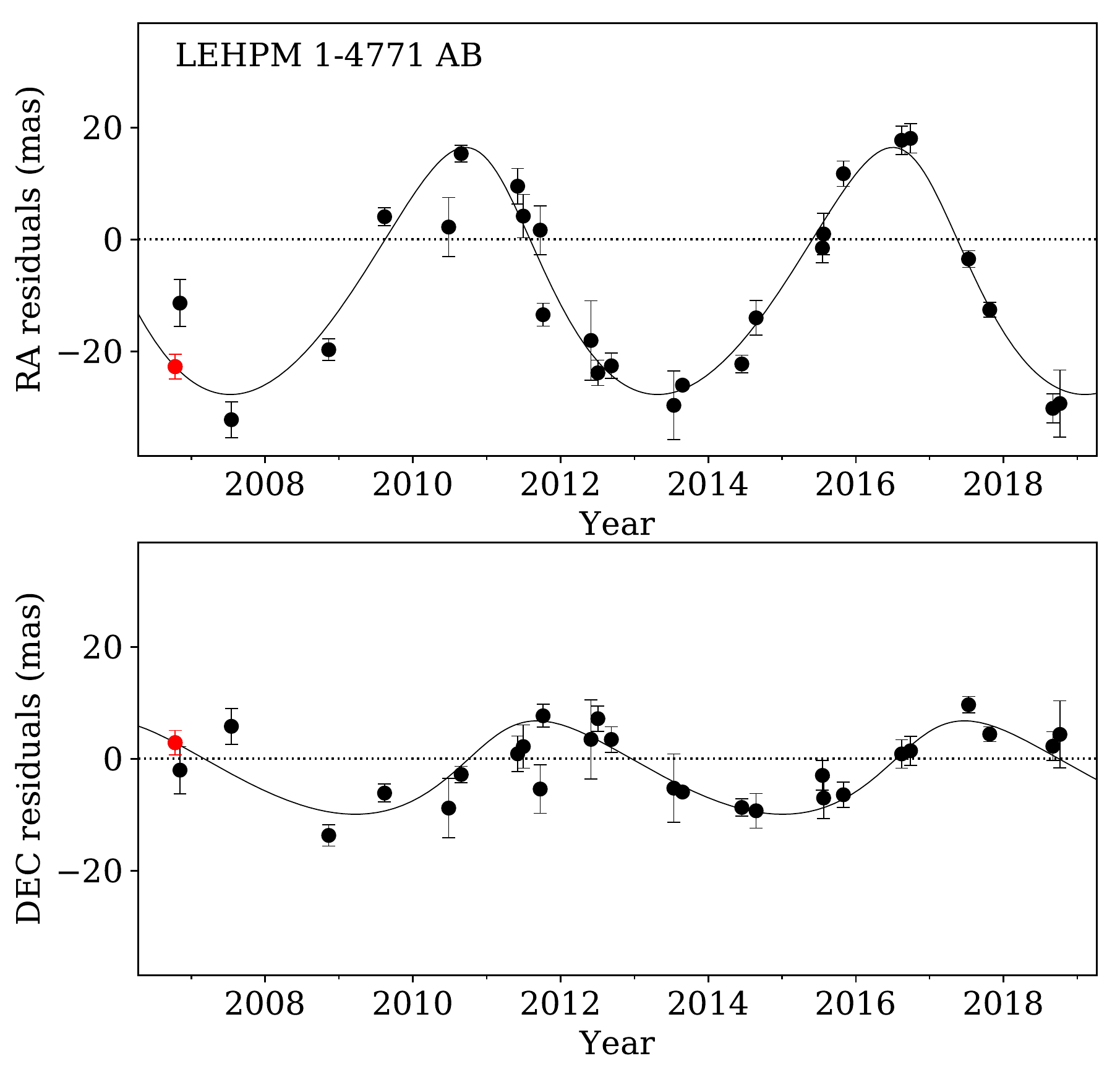}{0.4\textwidth}{}
\leftfig{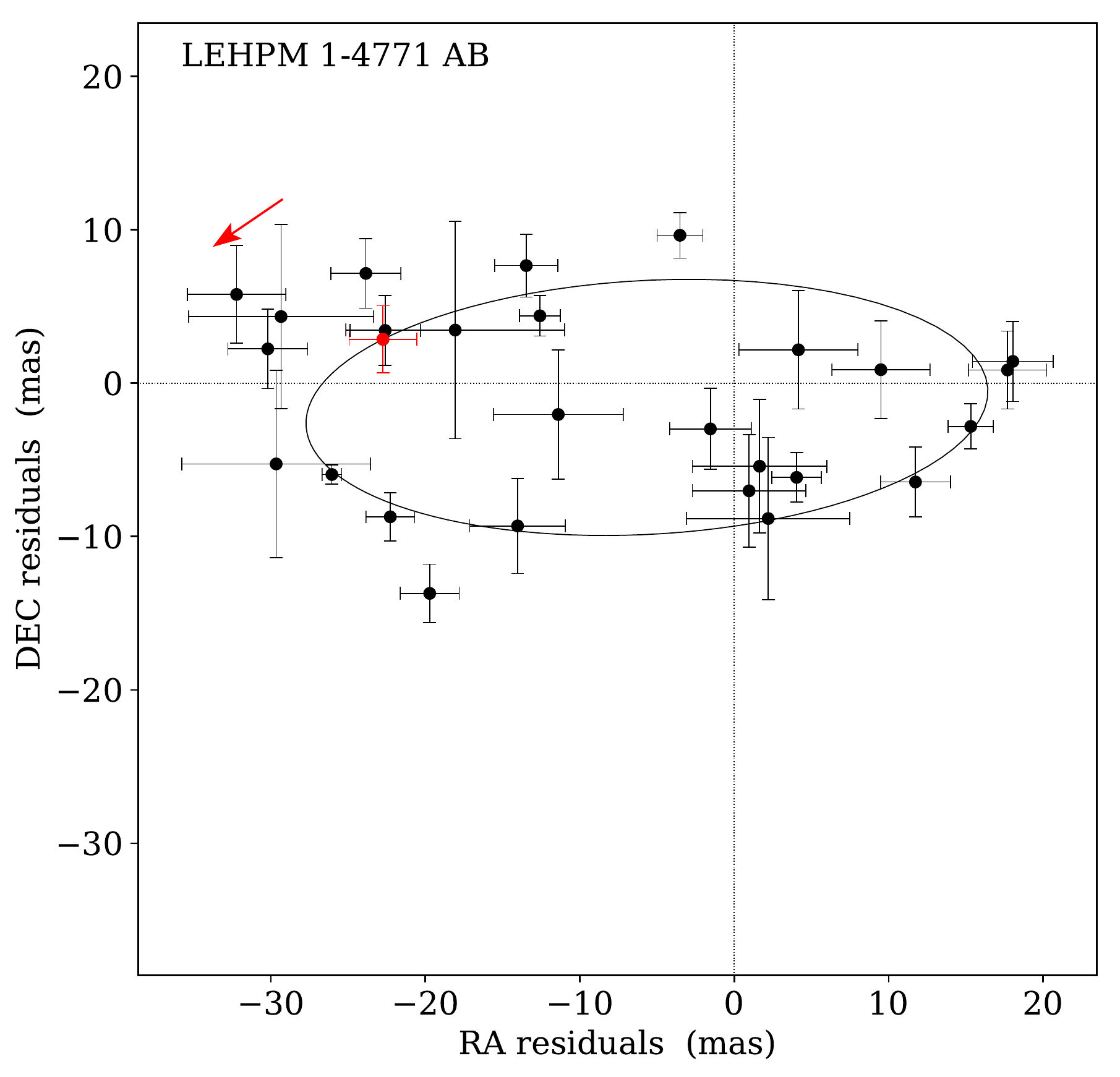}{0.38\textwidth}{}}
\gridline{\rightfig{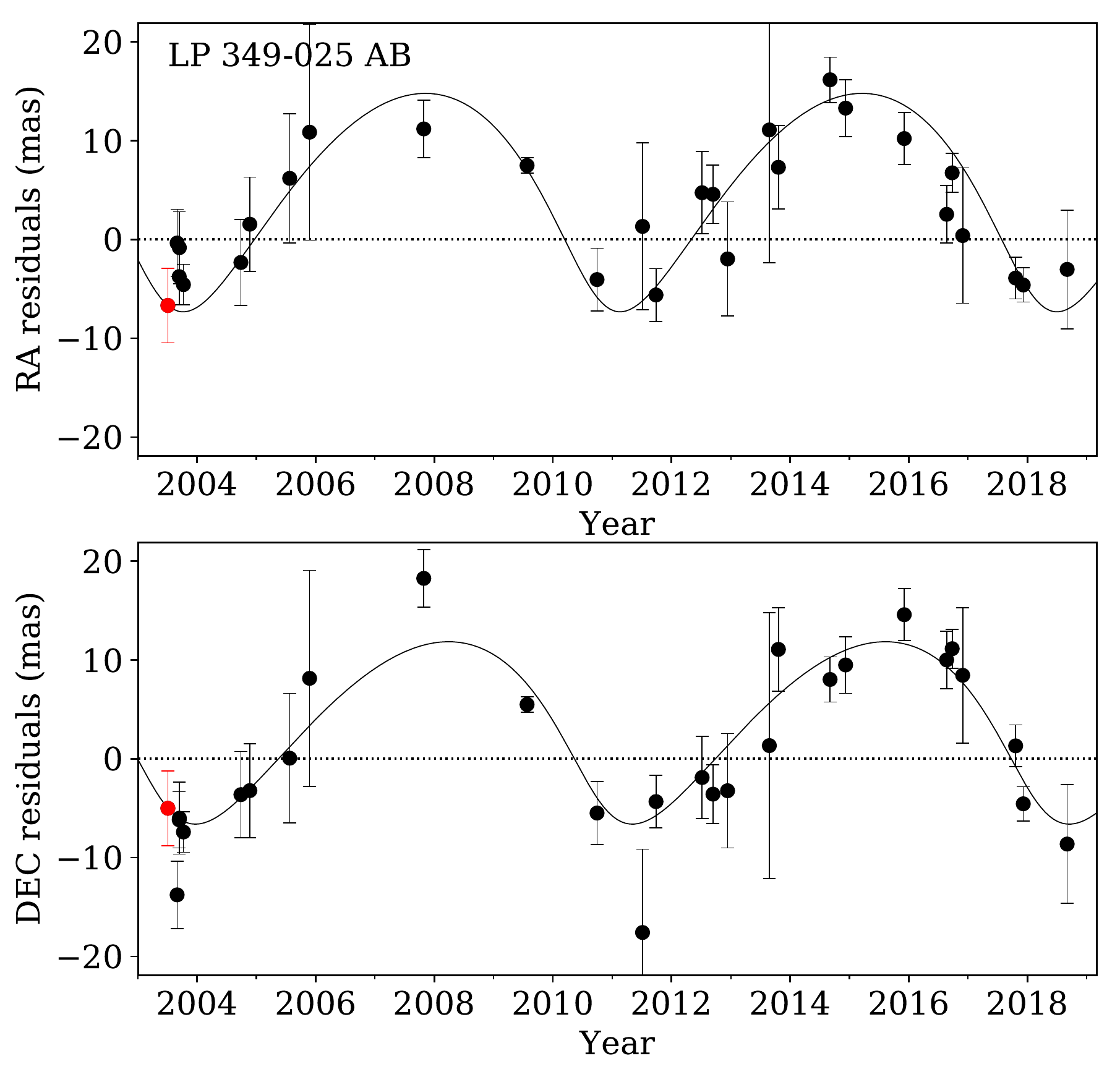}{0.4\textwidth}{}
\leftfig{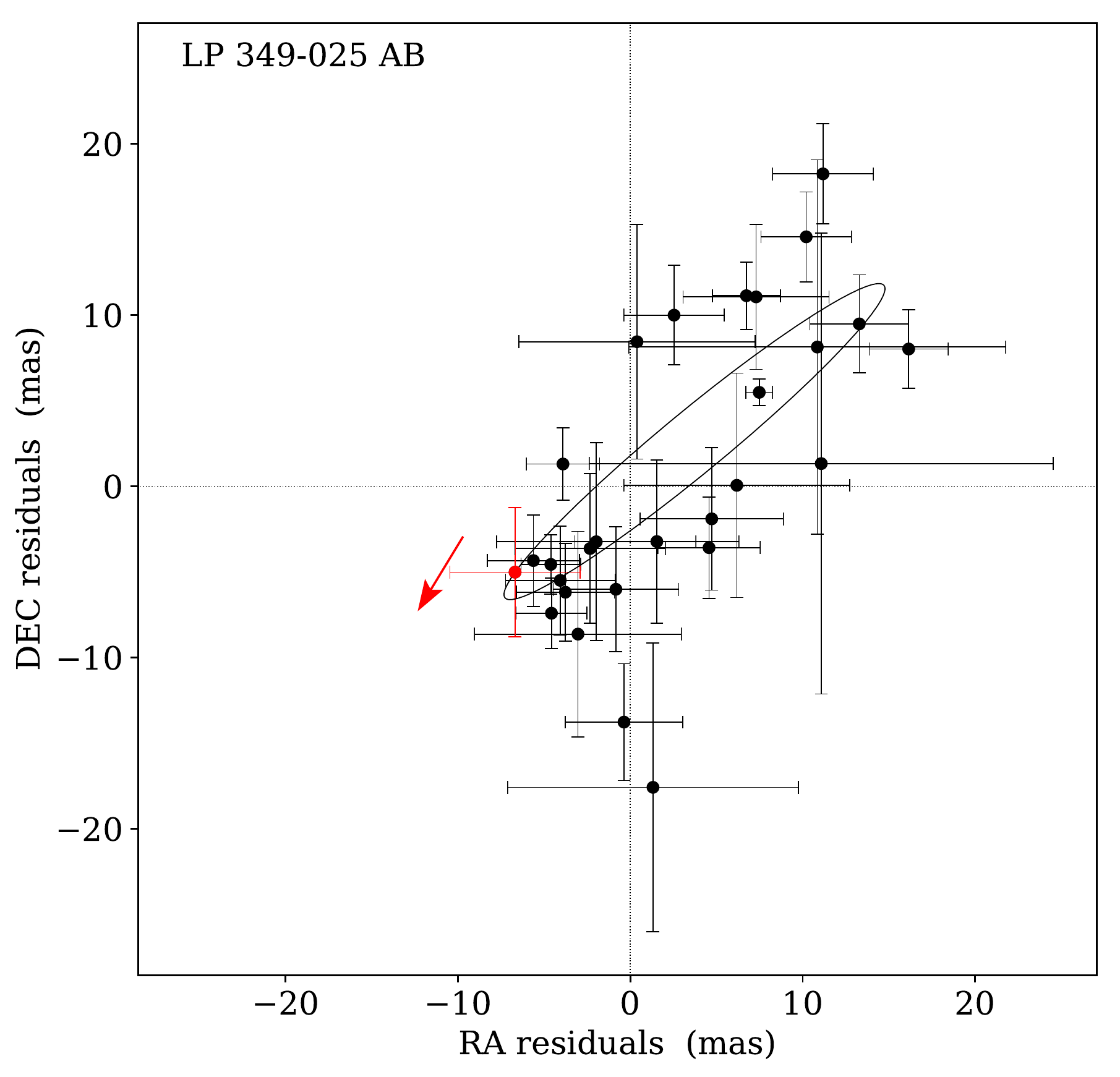}{0.38\textwidth}{}}
\caption{\label{fig:orbits1} \scriptsize 
Astrometric residuals, after proper motion and parallax have been removed, for three nearby red dwarf systems showing perturbations indicative of orbiting companions. In each panel, the solid line represents the orbit fit to that system's photocentric motion, for which the best-fit elements are given in Table~\ref{tab:orbelements}. 
The first epoch is marked with a red point, and the red arrow indicates the direction of motion. In the right column plots, north is up and east is to the right.
\textit{Top to bottom:} LHS~1582~AB ($P_\mathrm{orb} = 5.23$ years), LEHPM~1-4771~AB (5.79 years), and LP~349-025~AB (7.37 years).
}
\end{figure}

\begin{figure}
\gridline{\rightfig{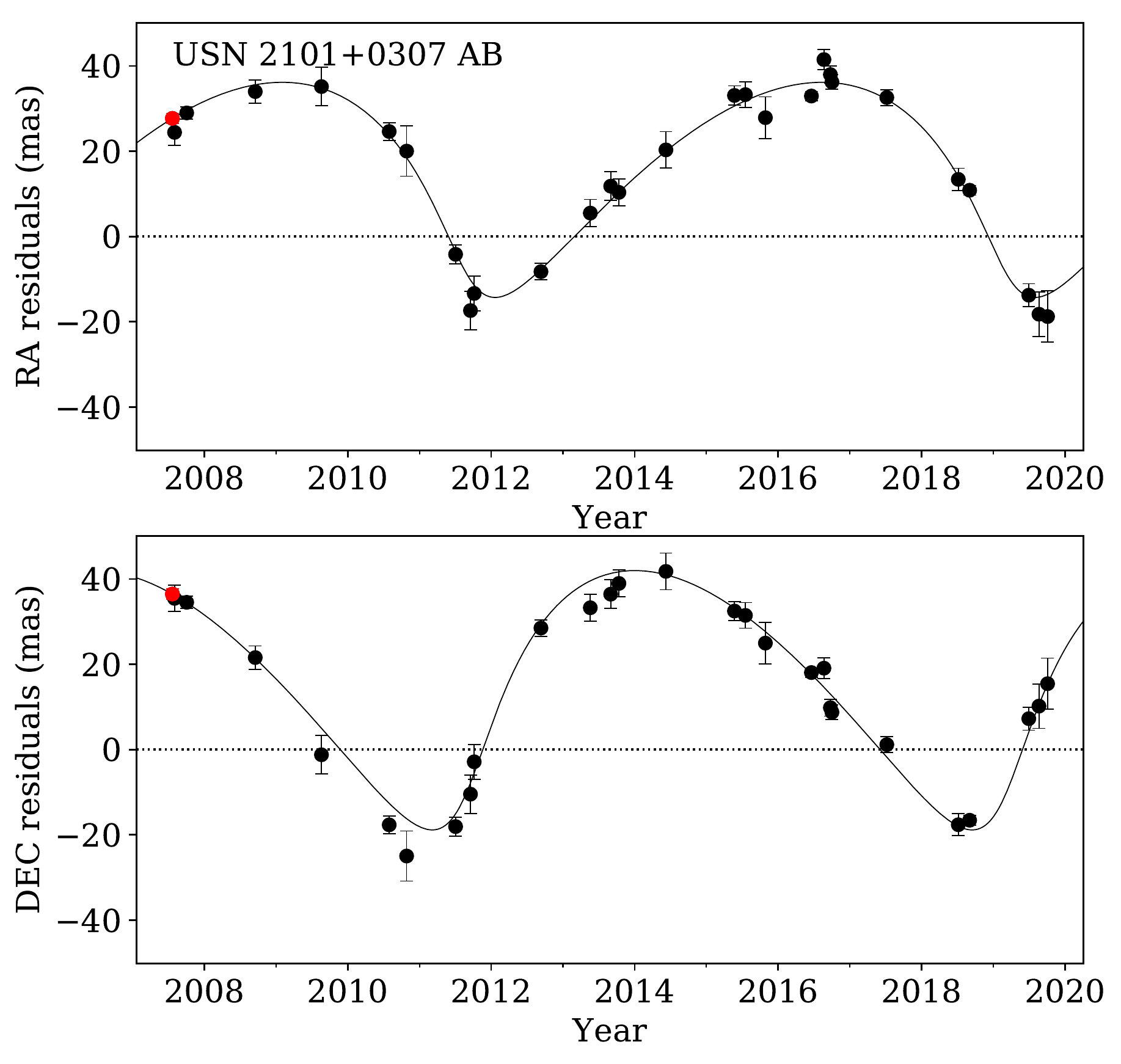}{0.4\textwidth}{}
\leftfig{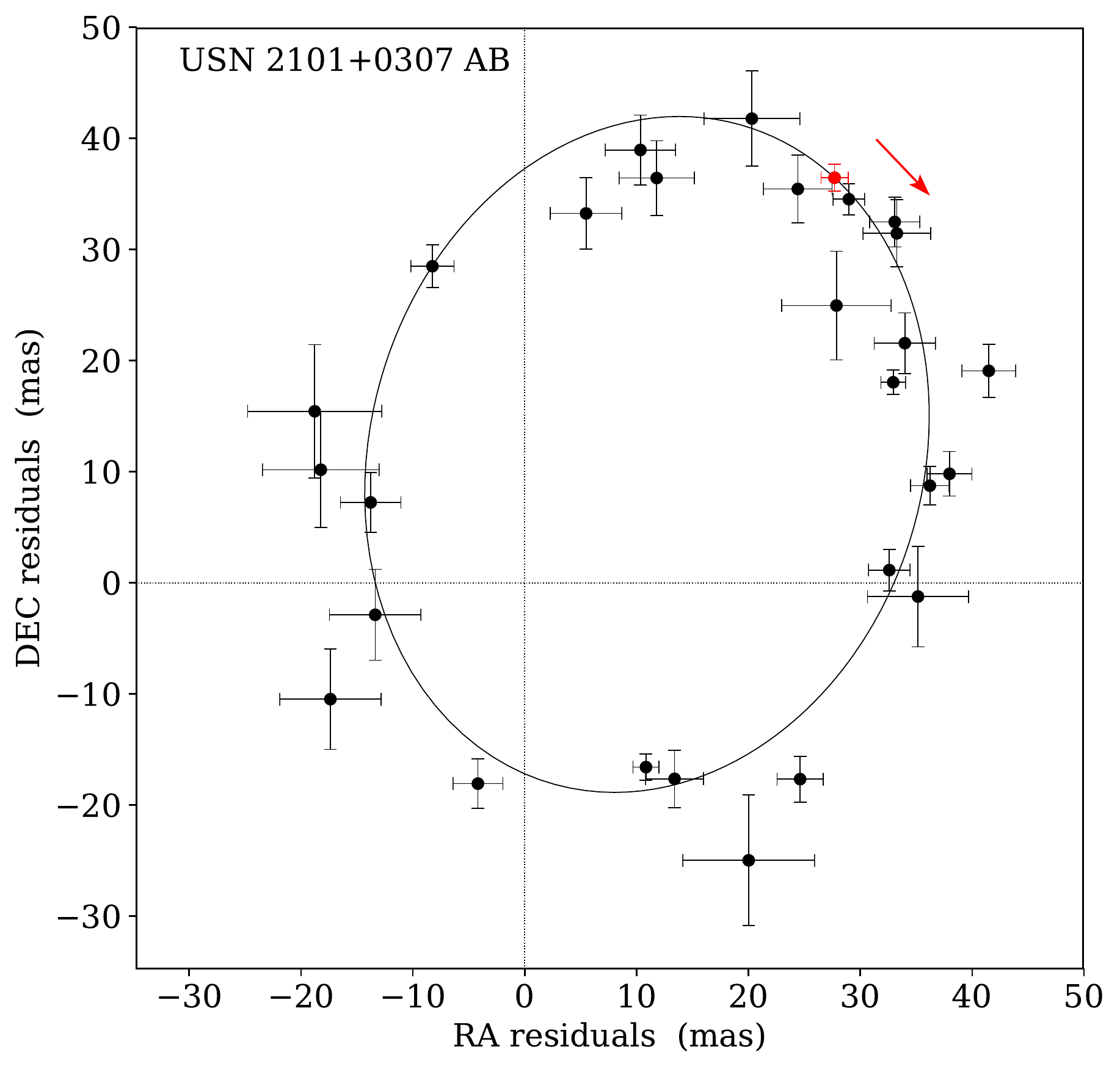}{0.38\textwidth}{}}
\gridline{\rightfig{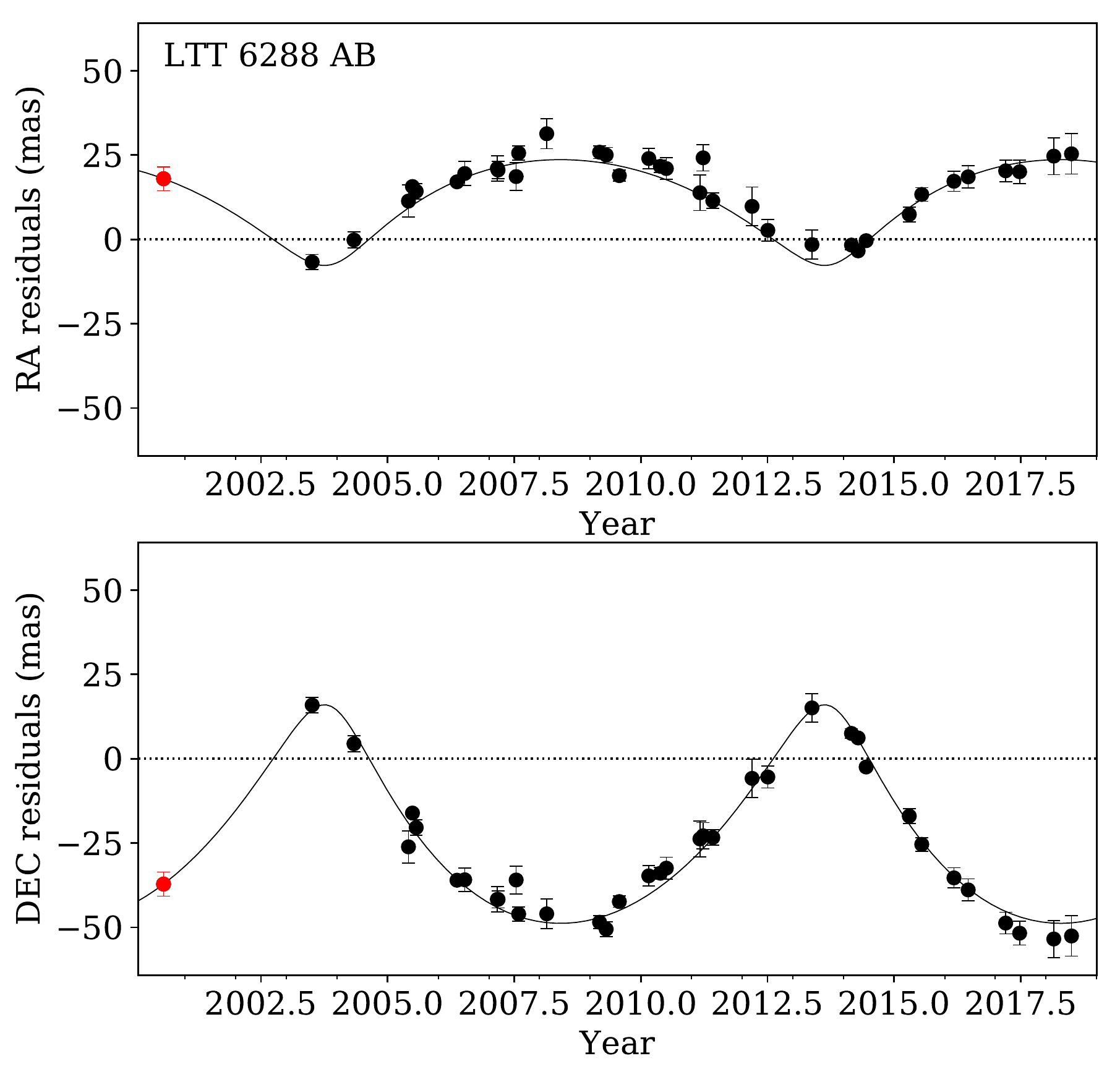}{0.4\textwidth}{}
\leftfig{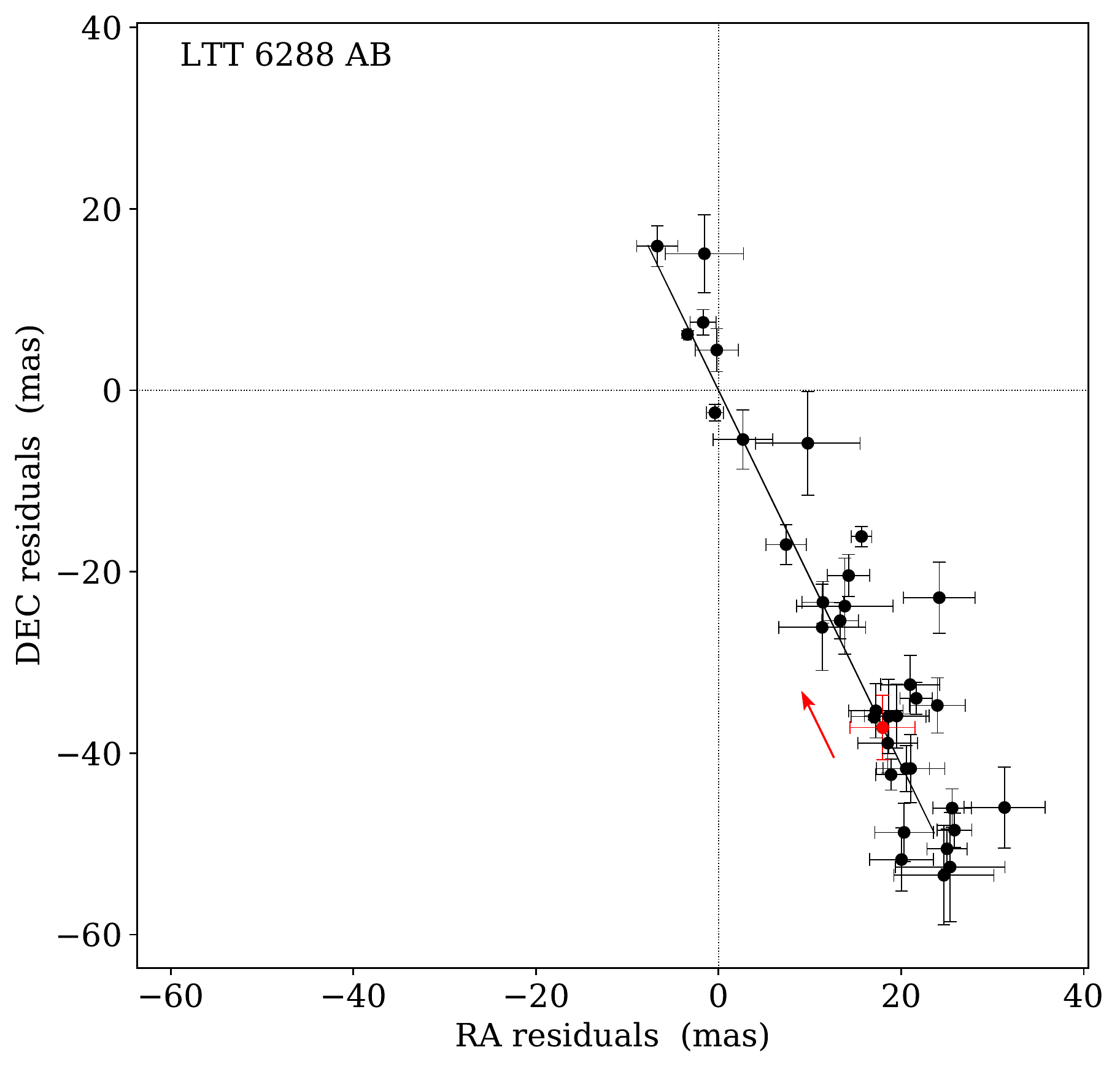}{0.38\textwidth}{}}
\gridline{\rightfig{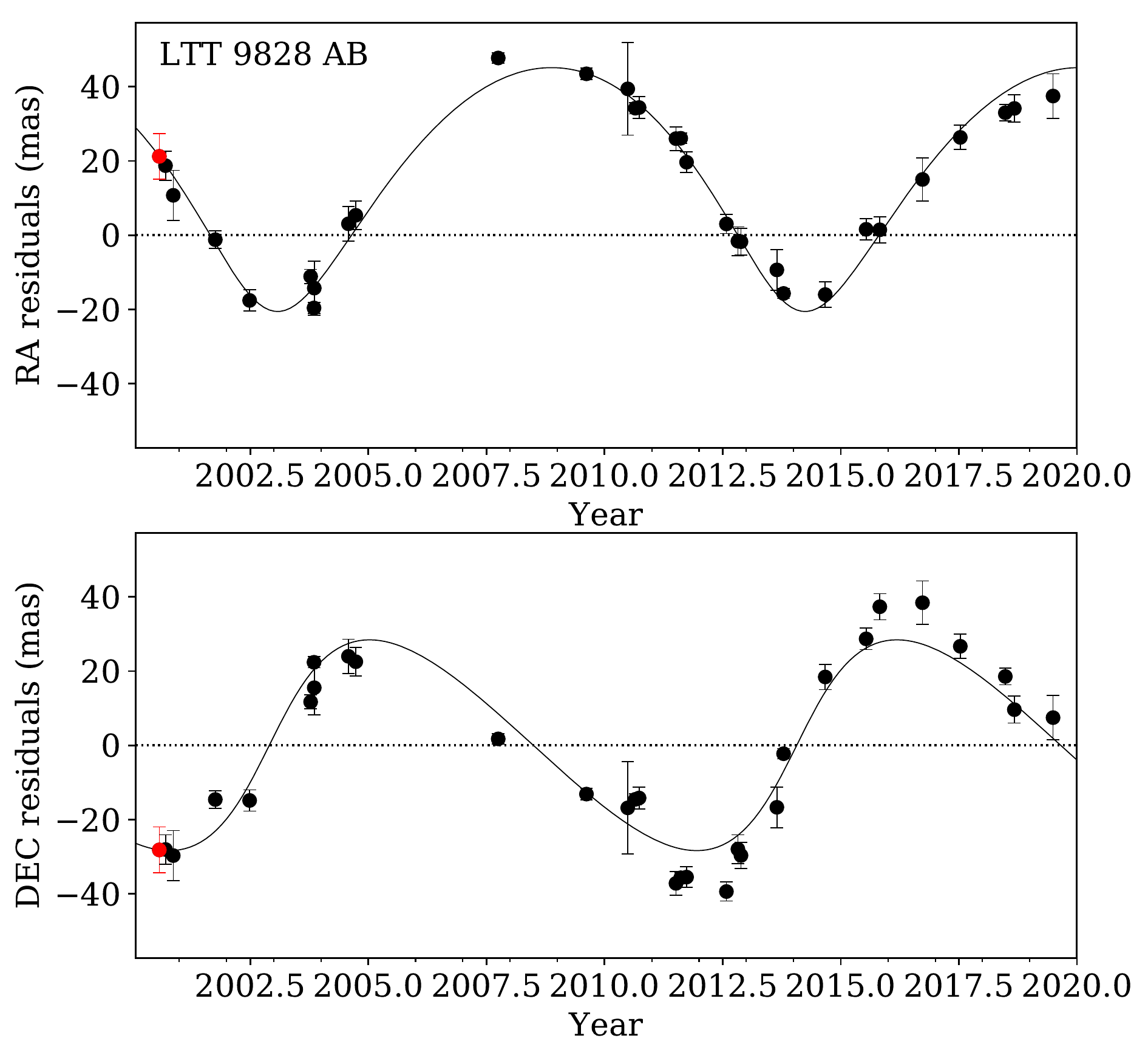}{0.4\textwidth}{}
\leftfig{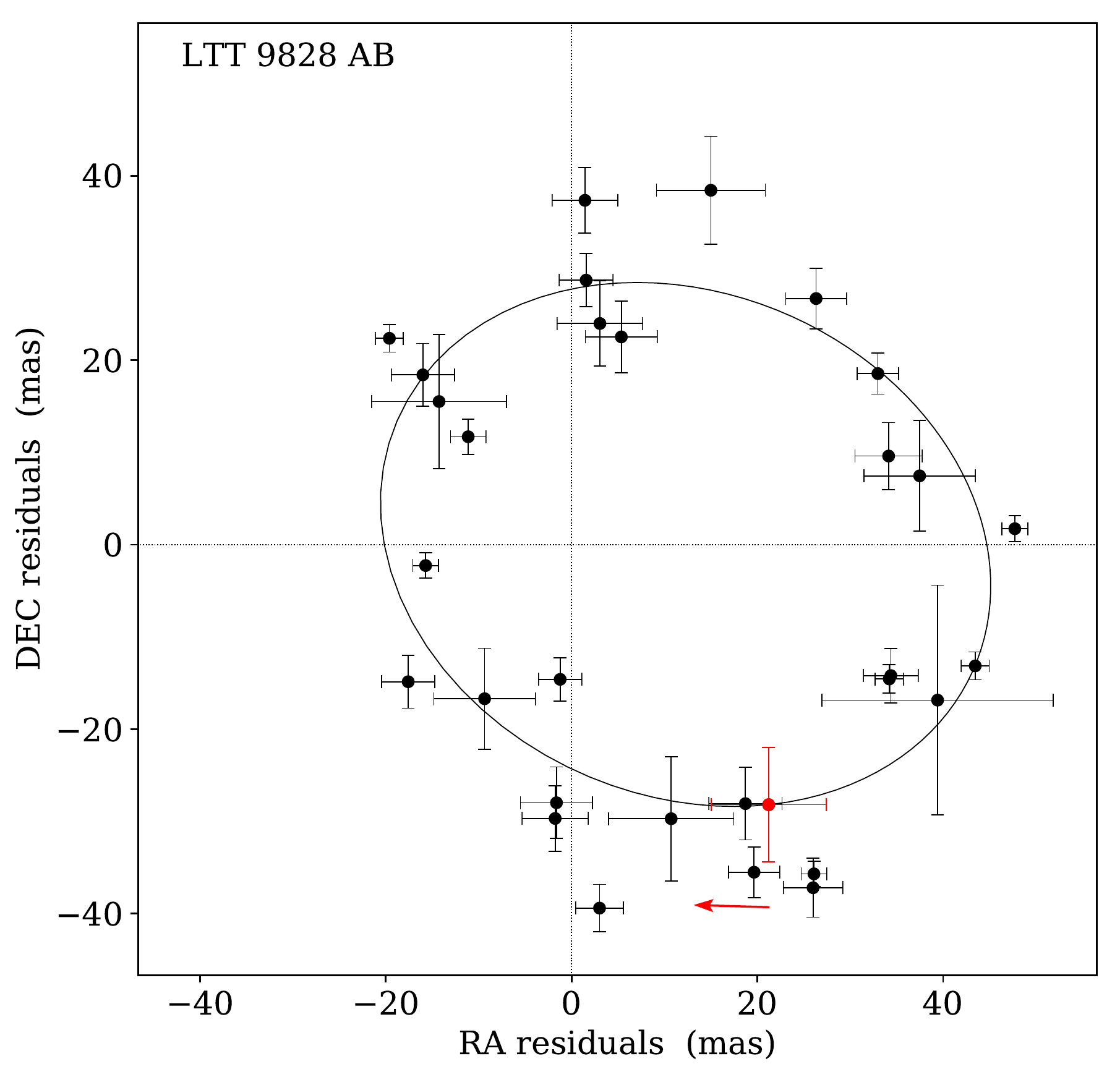}{0.38\textwidth}{}}
\caption{\label{fig:orbits2} \scriptsize 
Astrometric residuals, after proper motion and parallax have been removed, for three nearby red dwarf systems showing perturbations indicative of orbiting companions. In each panel, the solid line represents the orbit fit to that system's photocentric motion, for which the best-fit elements are given in Table~\ref{tab:orbelements}. 
The first epoch is marked with a red point, and the red arrow indicates the direction of motion. In the right column plots, north is up and east is to the right.
\textit{Top to bottom:} USN~2101+0307~AB ($P_\mathrm{orb} = 7.53$ years), LTT~6288~AB (10.10 years), and LTT~9828~AB (11.58 years).
}
\end{figure}

Orbital motion and fits of nine systems selected from RECONS
astrometry are shown in Figures~\ref{fig:orbits_cal},
\ref{fig:orbits1}, and \ref{fig:orbits2}.  For each system, the left
panel shows Right Ascension and Declination residuals plotted against
time, after proper motions and parallaxes have been determined and
causative shifts removed; deviations from a flat line indicate orbital
motion.  In the right column the orbits are shown on the plane of the
sky.  In both views the points represent mean positions from typically
five observations on a night and the best-fit orbit is the solid
curve.  Note that each orbit represents motion of the photocenter,
i.e., the center of light.  The semi-major axis of this orbit, as a
fraction of the relative orbit of star B around star A, is directly
proportional to the mass of star B relative to the total system mass,
and inversely proportional to the flux of B relative to
A\footnote{Appendix B of \cite{Die18} illustrates the mechanics of
  photocentric orbits.}, following the prescription by \cite{vanK67}.


The orbits are the result of images processed with the usual RECONS
pipeline, as described in $\S$\ref{sec:reductions}, through the
step where the target positions are measured relative to the sidereal
frame and corrected for DCR (Step 3 in $\S$\ref{sec:reductions}).  The orbit fits were
then derived using a different method than those used in previous
publications in \textit{The Solar Neighborhood} series.  Here we use
the Markov chain Monte Carlo (MCMC) algorithm of \cite{Die18}, where a
thorough description of the technique can be found.  Briefly, a fit is
made for proper motion, parallax, and the seven orbital elements
simultaneously, resulting in astrometry that reliably attributes the
three different motions of the photocenter.  The relative positions,
together with their observation epochs and parallax factors, are the
input for the MCMC fitting code, which is typically run with 51 chains
of 200{,}000 steps each to identify the most likely values for the ten
parameters (proper motion in R.A.\ and Decl., parallax, plus the seven
orbital motion parameters).  The code varies the parameters over given
ranges uniformly, with step sizes set such that no one parameter's
convergence dominates the others.  For these systems, the input
parameter ranges were set initially to broad uniform priors for all
but the parallax, which was informed by the preliminary RECONS values.
After this initial run, we fit each system again using narrower
parameter ranges based on the results of the preliminary run.
Convergence was judged by plotting the probability density functions
based on the last $10{,}000$ chains; Gaussian distributions indicate
good convergence.

The orbital elements for each fit are given in
Table~\ref{tab:orbelements}.  Figure~\ref{fig:orbits_cal} illustrates
three systems used for calibration that have well-known orbits of
short, medium, and long duration, demonstrating the capabilities of
our fitting procedure over each of these timescales: GJ 748 AB ($P =
2.49$ years), GJ 1005 AB ($4.56$ years), and GJ 234 AB ($16.63$
years).  These three systems were observed using (primarily)
interferometric measurements from a long-term \textit{Hubble Space
  Telescope} Fine Guidance Sensors program, augmented with radial
velocities from McDonald Observatory, as described in \cite{Ben16}.
Comparisons of the relative orbits in \cite{Ben16} to our orbits, both
included in Table~\ref{tab:orbelements}, indicate that most elements
match to within the error bars for all expected, except the argument
of periastron ($\omega$) and longitude of the ascending node
($\Omega$) for GJ 748 AB, which differ by 65.2$^\circ$ and
17.9$^\circ$, respectively.  Note that for each calibration system our
semi-major axis ($a$) is not expected to match that of \cite{Ben16}
because our data are for photocentric orbits rather than relative
orbits, and $\Omega$ and $\omega$ will differ in quandrant by
180$^\circ$.

Figures \ref{fig:orbits1} and \ref{fig:orbits2} illustrate six new
orbits for red dwarf binaries within 25 pc.  Several are updates
from previous orbits in this series of papers; these new orbits are
more reliable given the new technique of fitting for proper motion,
parallax, and orbital motion simultaneously.  We consider all six
orbits to be quite robust, with orbital periods of 5.23--11.17 years
and errors of only 0.02--0.19 years.  Observations of these systems at
the CTIO/SMARTS 0.9~m will continue in order to improve the orbital
elements further.

%
\section{Comparison of \textit{Gaia} DR2 and RECONS Astrometry Results} \label{sec:gaia}
%

The Orbital Architectures project ($\S$\ref{sec:theproject}) requires nearby multiples with months-long orbital periods as well as years-long periods from the RECONS astrometry program (as in $\S$\ref{sec:orbits}). 
With this in mind, in our comparisons with DR2 we search for evidence of how unresolved multiples' photocentric orbital motion affects their DR2 solutions, with the intention of discerning criteria to identify new potential unresolved multiples. 
A new era of space-based astrometry has been initiated by ESA's \textit{Gaia} mission \citep{GAI16}, and with the April 2018 release of 1.7 billion parallaxes from \textit{Gaia}'s first 22 months of observations \citep[Data Release 2;][]{GAI18a}, the RECONS team now has an accurate but short-term yardstick against which to compare our own results.
Both RECONS and \textit{Gaia} provide proper motions and parallaxes, but RECONS also provides orbital and multiplicity information not yet presented in the \textit{Gaia} results. 
Each DR2 solution is computed by fitting a five-parameter single-star astrometric model for that source \citep{Lin18}, and each entry in DR2 includes additional parameters describing the quality of the observations and of the subsequent astrometric fit.
Eventually, all multiples with separations less than $\sim$100 mas are expected to be unresolved to the point where only the photocenter (not individual components) is detected \citep{Lin18}. In DR2, multiples with separations greater than $0\farcs5$ tend to be well resolved \citep{Are18}. 

In the sections that follow, we compare the RECONS astrometry results to those of \textit{Gaia} DR2 for M dwarfs within 25 pc, using the DR2 astrometric fit quality parameters to form criteria for blindly selecting potential unresolved multiples. $\S$\ref{sec:picomparisonsample} introduces the sample of 542 nearby red dwarf systems and our procedure for matching them to DR2 sources. $\S$\ref{sec:picomparison} compares the RECONS and DR2 parallaxes, and $\S$\ref{sec:paramcomparison} compares the astrometric fit quality of unresolved multiples to resolved and single sources, where we define ``cutoff'' values for systems likely to be multiple. 
$\S$\ref{sec:missingsystems} briefly discusses the systems missing from \textit{Gaia} DR2. Finally, other works selecting unresolved multiples from DR2 are discussed in $\S$\ref{sec:otherwork}.

\subsection{Preparing the Comparison Sample and Matching to DR2} \label{sec:picomparisonsample}

The systems we compare to \textit{Gaia} DR2 are M dwarfs within 25 parsecs, as determined by one or both of RECONS and DR2 parallax $\pi \geq 40$ mas. These 542 systems, as listed in Table~\ref{tab:RECONS-GDR2}, include those listed in Table~\ref{tab:astrRECONS} with distances within 25 pc, as well as several hundred additional 25 pc members previously published in \textit{The Solar Neighborhood} series. 
Column~4 of Table~\ref{tab:RECONS-GDR2} gives the RECONS parallax, noted in Column~5 as either a new value (asterisk) or previously published value (reference given); this parallax we compare to the DR2 parallax of Column~6. 
Columns~7 through 14 reproduce the astrometric fit parameters and $G$ magnitude from DR2 that we investigate in more detail in $\S$\ref{sec:paramcomparison}. Column~15 notes the system classification if there is evidence that it is not single: individual component of a resolved multiple (``res''), unresolved multiple (``unr''), or system with a perturbation in its RECONS astrometric residuals (``PB''). Column~16 marks those meeting all of our criteria for potential unresolved multiplicity (described later in $\S$\ref{sec:paramcomparison}).

Comparing RECONS to \textit{Gaia} DR2 parallaxes first requires carefully matching targets between these two catalogs. Because a key aspect of this analysis involves comparing systems with differing degrees of poor astrometric results in DR2 and multi-star systems that may or may not be resolved by either catalog, care was taken to ensure that even systems without full five-parameter solutions in DR2 were considered in the pool of potential matches. Starting with the full DR2 catalog extracted from the CDS/VizieR \citep{VIZI}, with no cuts for quality, the matching proceeded as follows:
\begin{enumerate}
    \item For each RECONS result's Right Ascension and Declination (J2000), find all DR2 solutions within a 1.0 arcmin radius. The DR2 solutions' coordinates are converted to J2000 (computed automatically by VizieR) and ranked by proximity to the RECONS target.
    
    \item Any DR2 solution within $\sim$2 arcsec of the RECONS solution on the sky and with parallax within 10 mas of the RECONS parallax is automatically considered a match. The vast majority of targets fell into this category.
    
    \item For RECONS systems with no obvious DR2 source as described above, nearby DR2 sources lacking five-parameter solutions were considered carefully. Where possible, their $B_G$ and $R_G$ magnitudes\footnote{The \textit{Gaia} DR2 magnitudes given in that catalog as \texttt{BPmag} and \texttt{RPmag} are here referred to as $B_G$ and $R_G$. In other works they are sometimes referred to as $BP$ and $RP$ or $B_P$ and $R_P$.} were checked for similarity to $V$ and $I$, respectively, and their positions were compared in Aladin to images from 2MASS and DSS2. This process identified several dozen targets with DR2 sources lacking full solutions.

    \item If the system was flagged in the RECONS catalog as having \textbf{a} perturbation in its astrometric residuals and/or an orbit fit, the potential matches within the 1.0 arcsec radius were also screened to identify any match to the secondary component. Some secondary matches were clear by their proper motion and parallax matching the primary star, but many others were missing full five-parameter solutions (i.e., had no proper motion and parallax). When no secondary component was apparent within 1.0 arcsec, an additional search was performed by eye within 3.0--5.0 arcsec using Aladin to visualize the DR2 catalog on background images from 2MASS and DSS2. When no secondary companion was found using these strategies, the system was flagged as an unresolved multiple.

\end{enumerate}

\subsection{Comparison of Parallaxes} \label{sec:picomparison}

\begin{figure} \centering
\plottwo{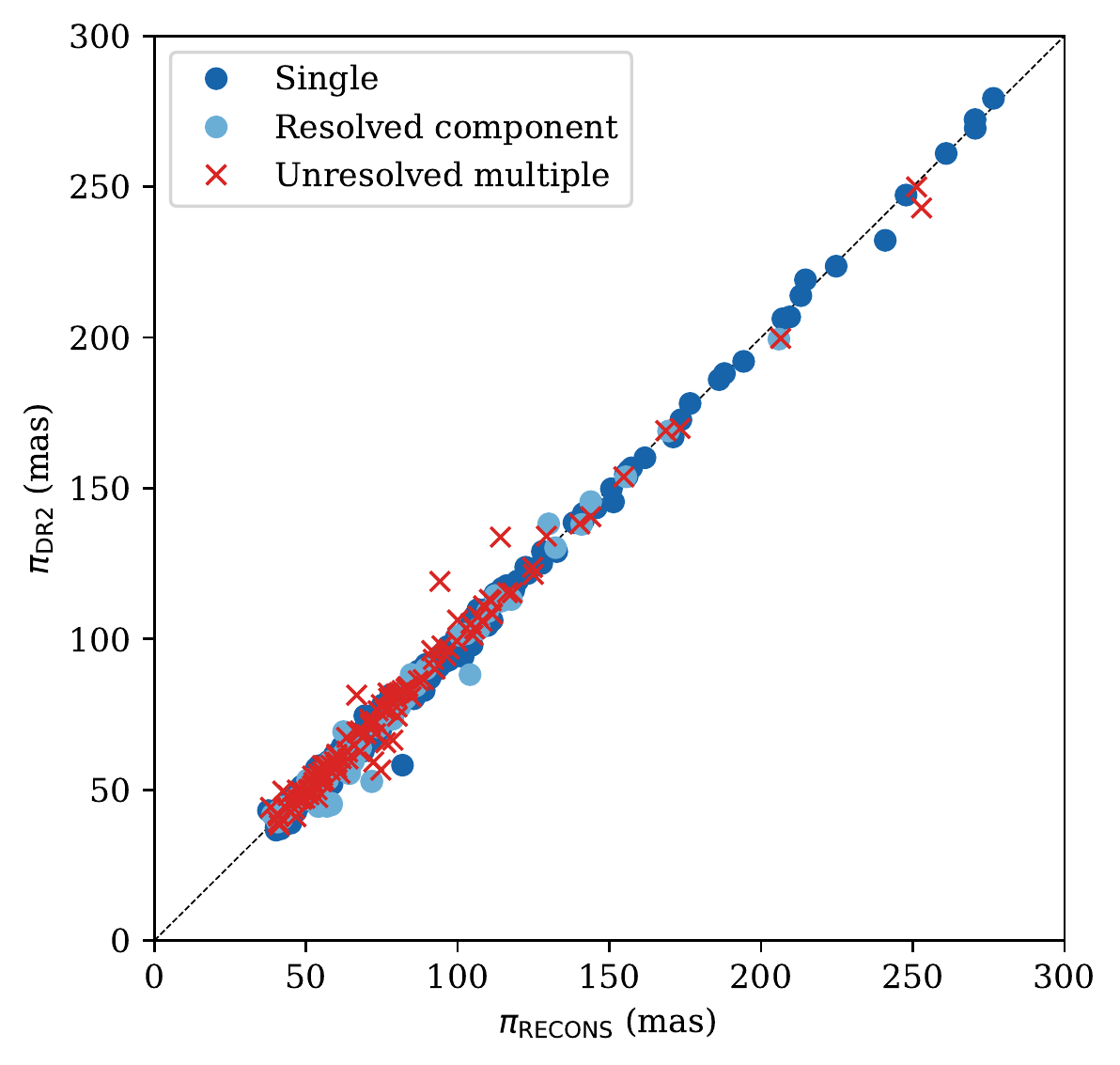}{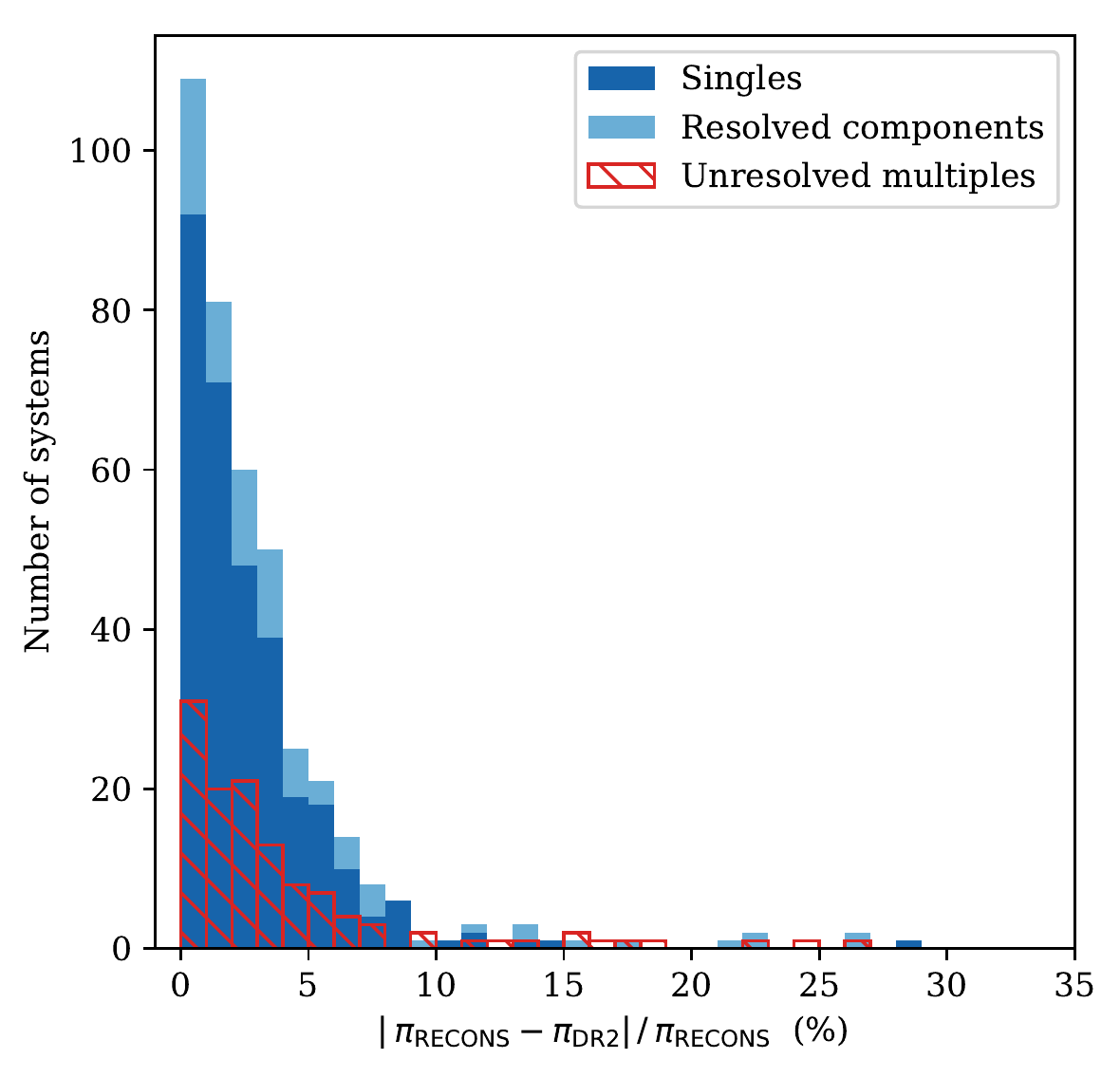}
\caption{\label{fig:pipi} \scriptsize Parallax comparisons of RECONS astrometry to \textit{Gaia} DR2 using M dwarfs within 25 pc, divided by their multiplicity status. Unresolved binaries or multi-star systems are indicated with red symbols (\textit{left panel}) and hatched bars (\textit{right panel}), resolved components of multi-star systems are light blue dots and bars, and presumed single stars are dark blue dots and bars.
\textit{Left panel:} RECONS parallax versus DR2 parallax, with the dashed black line indicating the position of perfect agreement between those quantities. 
\textit{Right panel:} difference (absolute value) between RECONS and DR2 parallax values, scaled by the corresponding RECONS parallaxes and expressed as percentages of those values. 
The singles and resolved components have been combined into a single histogram here, but the bars are colored to indicate the fraction belonging to each subset.}
\end{figure}

We then compare the parallaxes of the targets common to these two catalogs in Figure~\ref{fig:pipi}, with the full table of these values given in Table~\ref{tab:RECONS-GDR2}. The sample has been divided into single stars, resolved components, and unresolved multiples.
In the analysis that follows, all systems are presumed single unless the literature or our own observations indicate otherwise. ``Resolved components'' are DR2 entries corresponding to individual stars in binary or multi-star systems. Due to their wide separations, even the nearest of these targets are not expected to have detectable orbital motion over the time of DR2 observations, making these systems effectively single stars for the purposes of this comparison; in case any differences do become evident, however, in this work we have plotted these systems in a lighter shade of blue to distinguish them somewhat from systems with no known bound companions.

A 1:1 line has been plotted in the left panel of Figure~\ref{fig:pipi} to indicate parallaxes with perfect agreement. The agreement between RECONS and DR2 parallaxes is generally quite good, with no obvious systematic differences with distance. The systems that fall furthest from the 1:1 line tend to be unresolved multiples.
The distribution of differences for each set is plotted in the right panel of Figure~\ref{fig:pipi}, where each system's parallax difference has been scaled by its RECONS parallax and expressed as a percentage of that value. Here the difference is more clear between the distribution of unresolved multiples and that of the single stars (combined with resolved companions): both are strongly peaked at 0, indicating many systems with good parallax agreement, but the unresolved multiples distribution
is broader,
 representing more systems with discrepant parallaxes. The result is that the systems with a parallax difference of $\sim$10\% or greater tend to be unresolved multiples rather than single stars or resolved companions. On this basis, we regard the outlier ``single'' stars as potential unresolved multiples as well.

This scenario occurs because the model employed by \textit{Gaia} to fit the astrometric data for DR2 is that of a single star, thus it characterizes only proper motion and parallactic motion of the source, whereas most nearby unresolved multiples exhibit a third type of motion --- orbital motion. 
This additional motion could mimic linear proper motion enough to mislead the astrometric fit, changing the proper motion and parallax results (hence the discrepant values in Figure~\ref{fig:pipi}). 
This would occur in cases of long orbital periods that are not completely observed, orbits that are very eccentric, or orbits oriented nearly edge-on to the observer. 
Orbital motion could also create apparent scatter about the single-star model fit that would inflate the errors in that model's parameters. 
This analysis motivates our comparison of the astrometric fit of DR2 and RECONS, and our analysis from that comparison to identify criteria for selection of likely unresolved multiples in DR2.

\subsection{Comparing Astrometric Fits of Singles and Unresolved Multiples: Cutoffs for Revealing Potential Multiples in \textit{Gaia} DR2}
\label{sec:paramcomparison}

\begin{figure}\centering
\vspace{-1cm}
\includegraphics[scale=0.44]{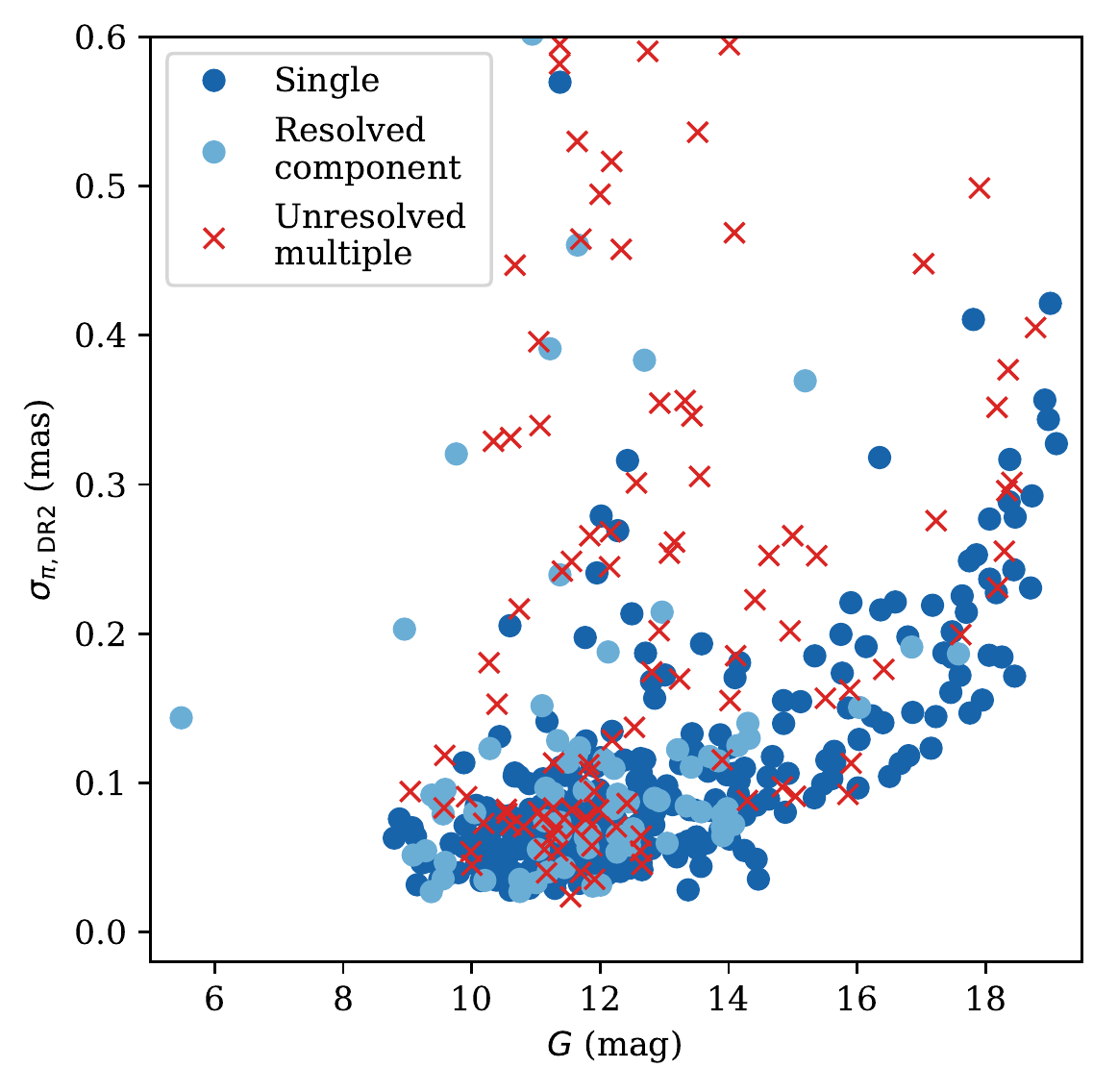}
\includegraphics[scale=0.44]{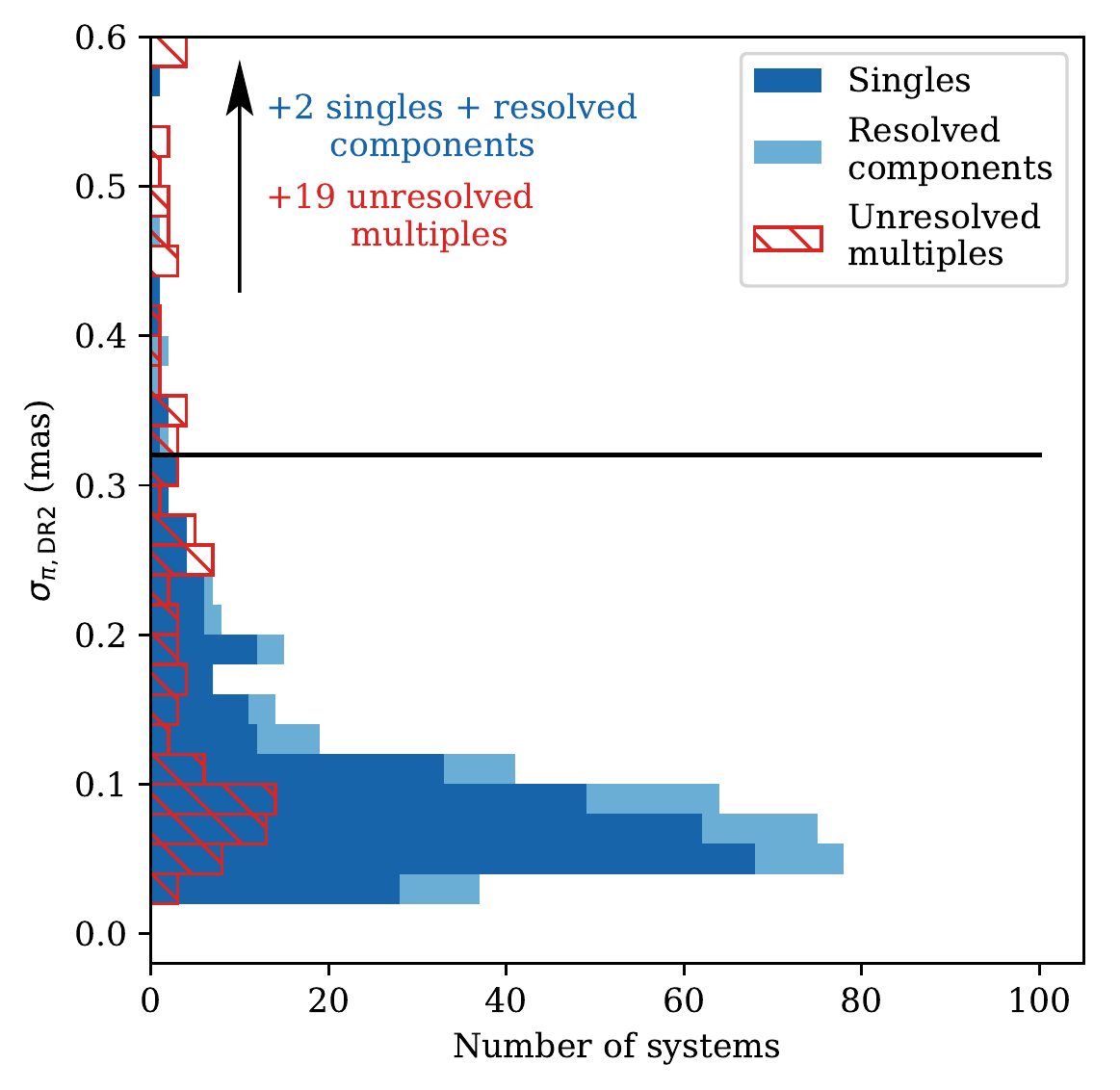}
\includegraphics[scale=0.44]{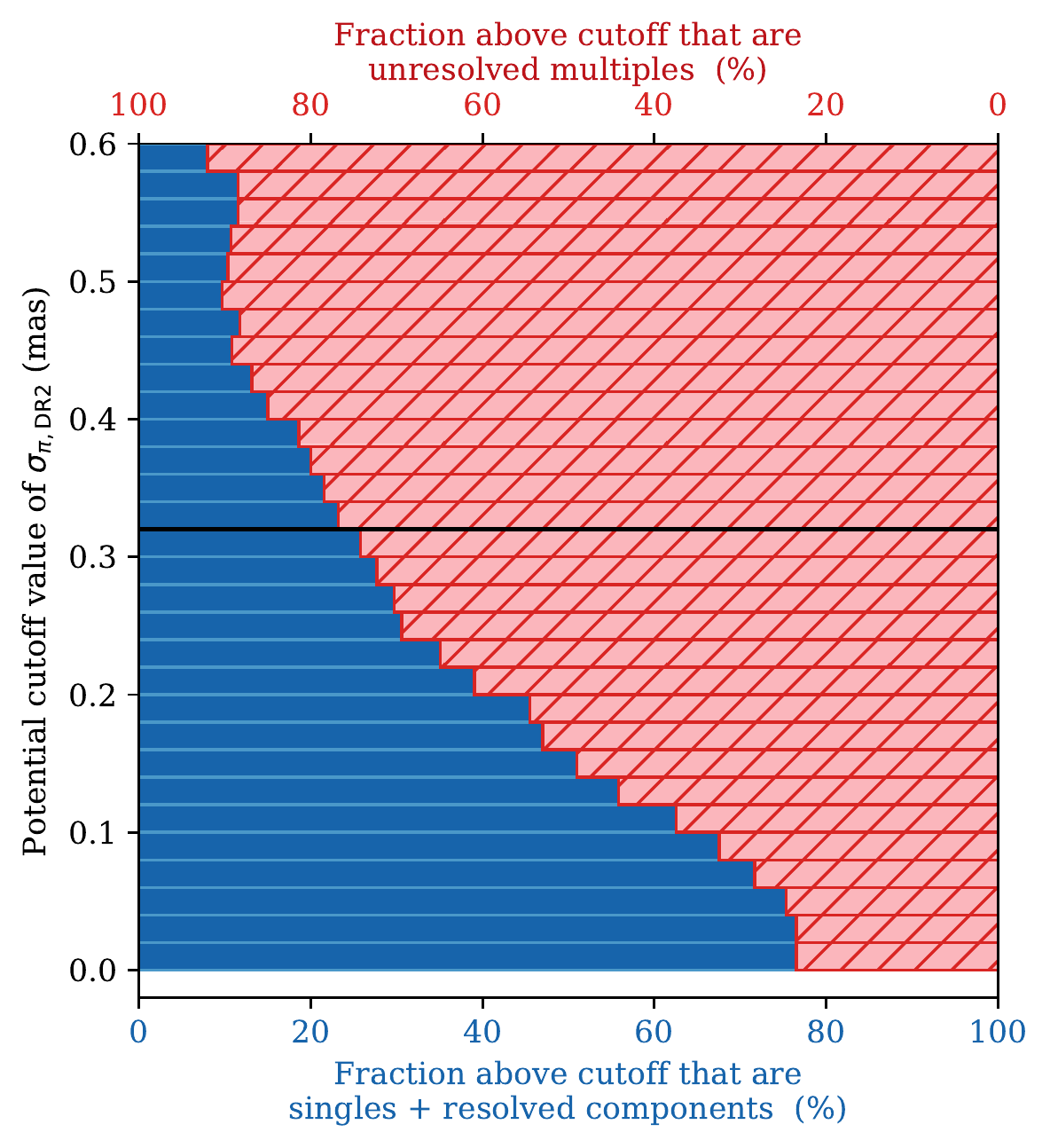}

\includegraphics[scale=0.44]{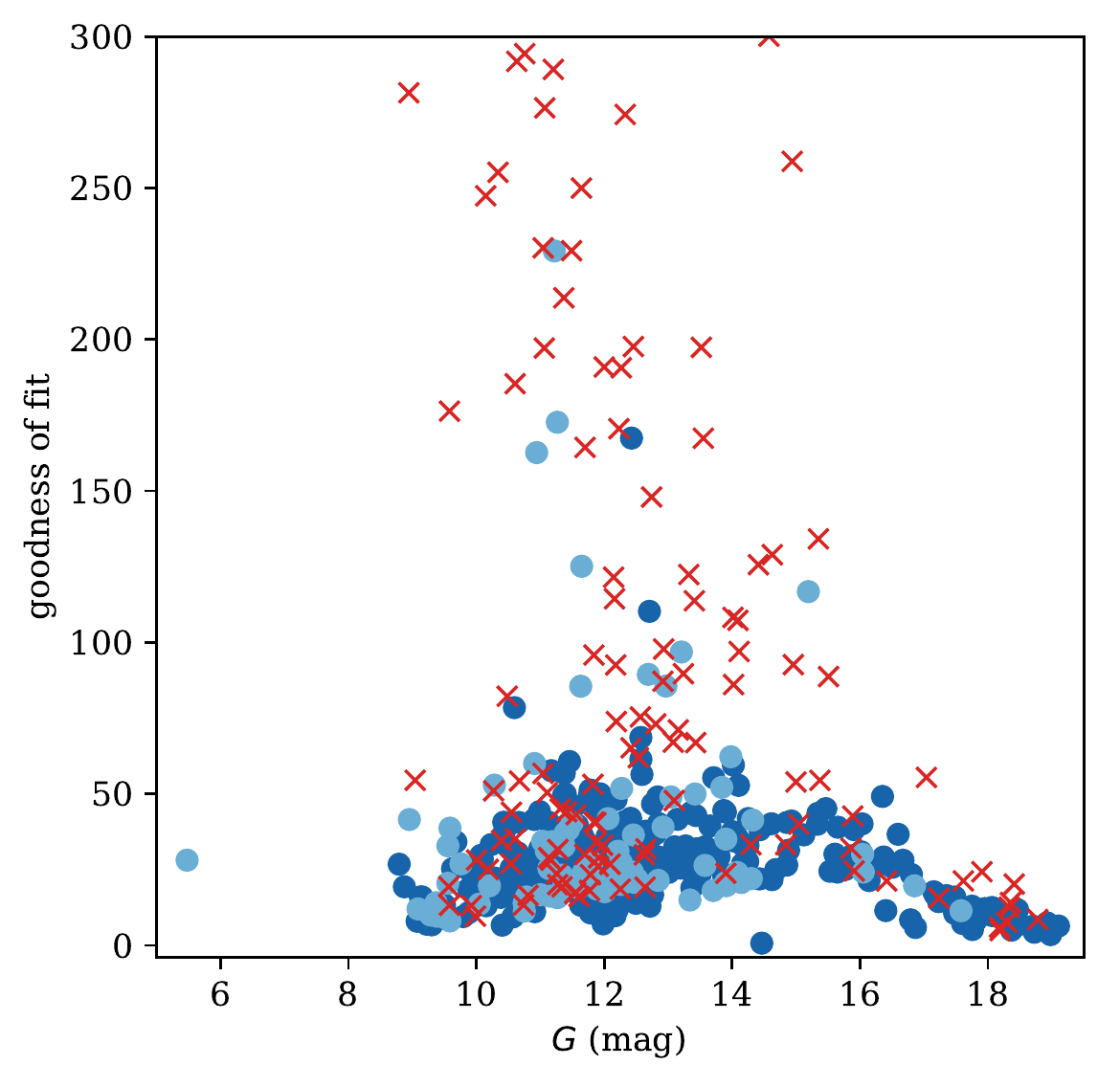}
\includegraphics[scale=0.44]{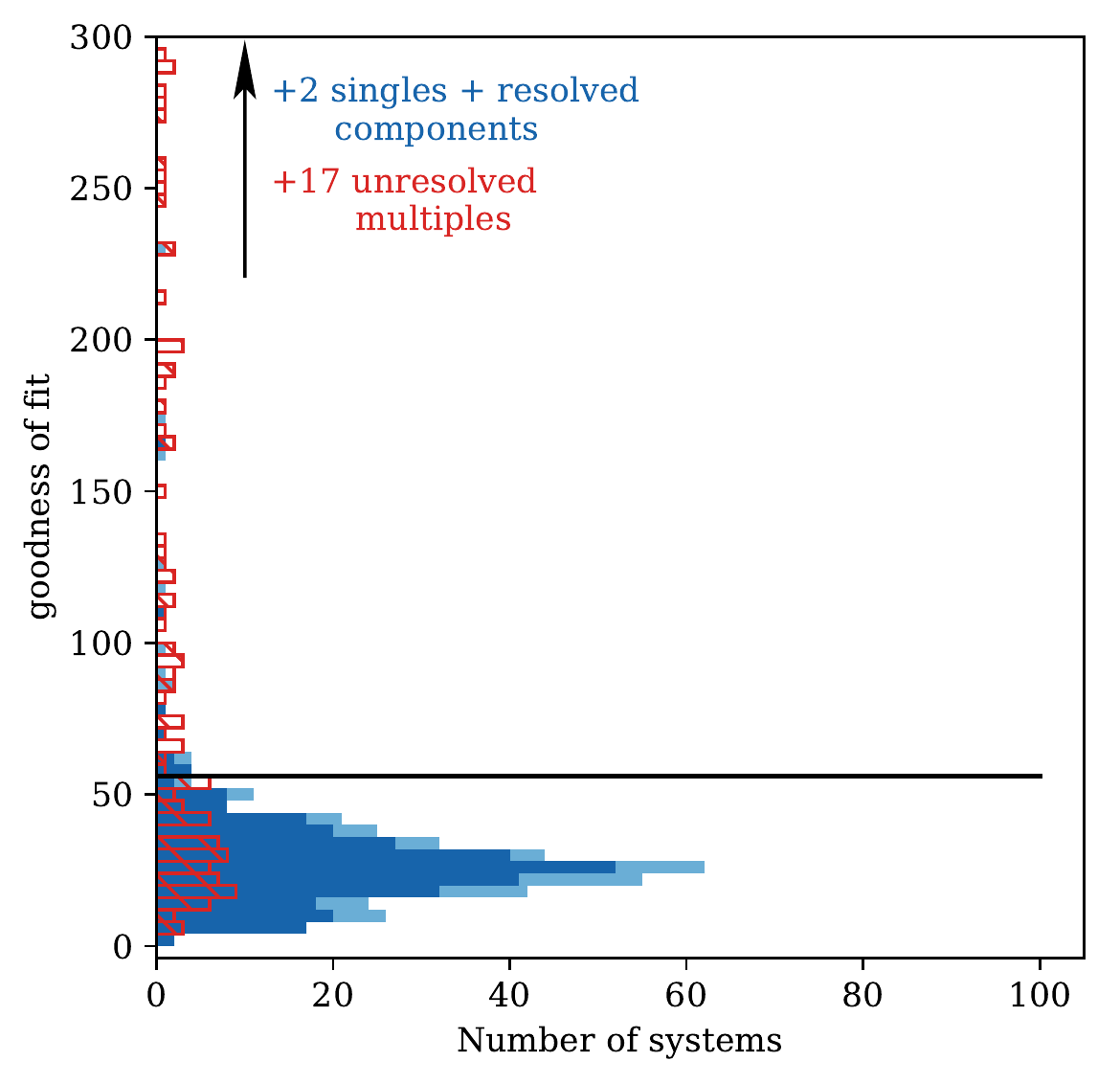}
\includegraphics[scale=0.44]{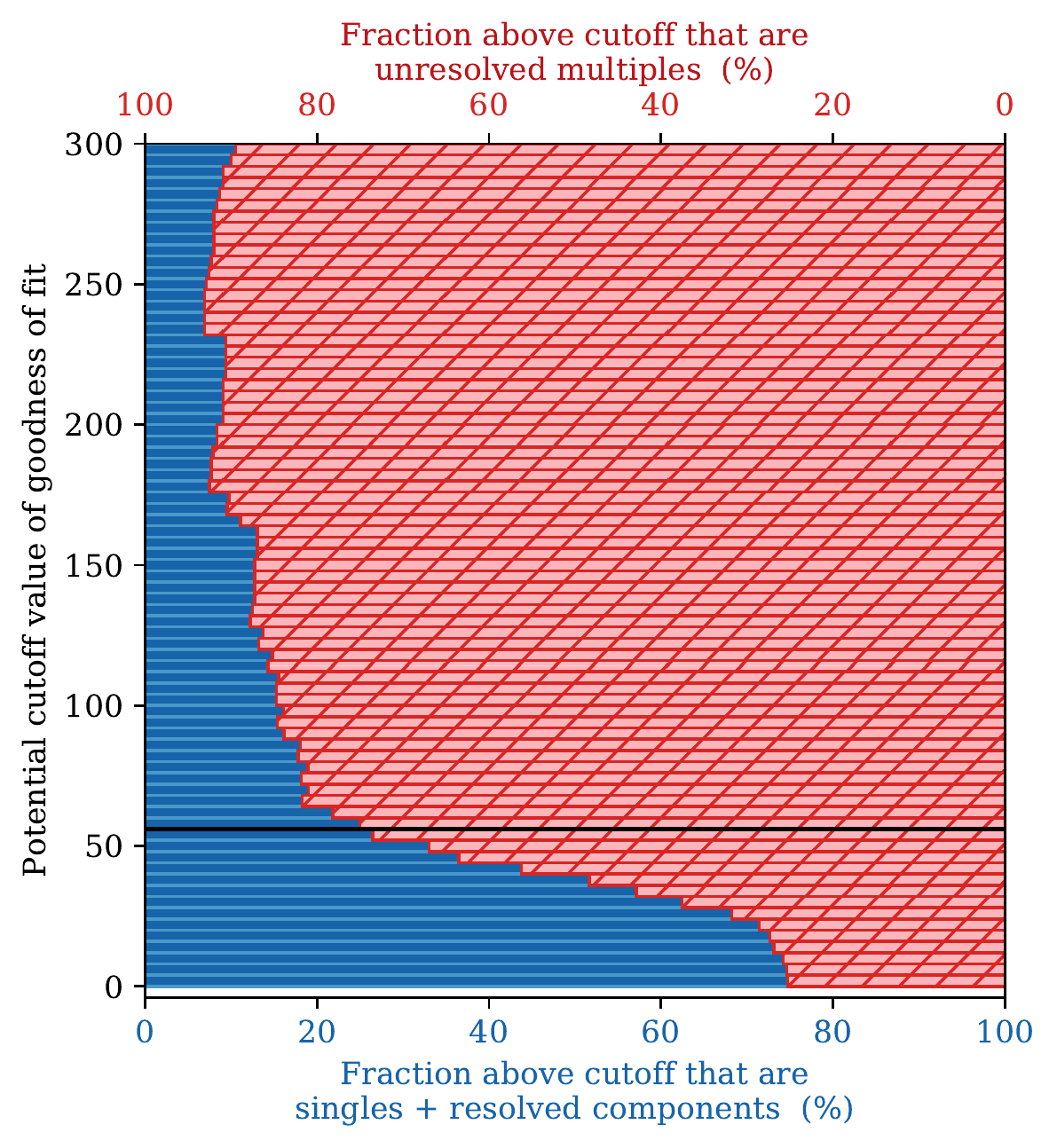}

\includegraphics[scale=0.44]{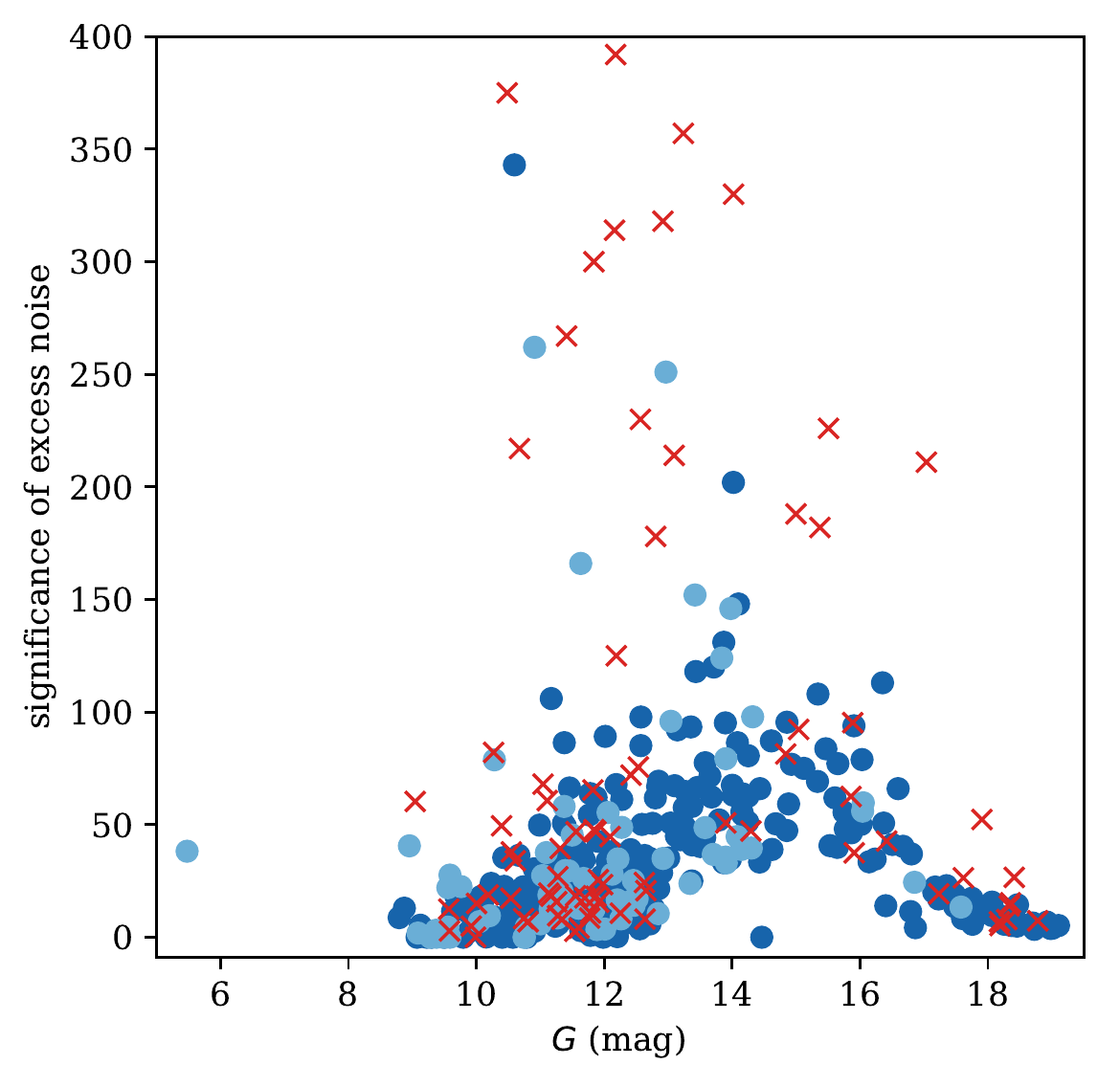}
\includegraphics[scale=0.44]{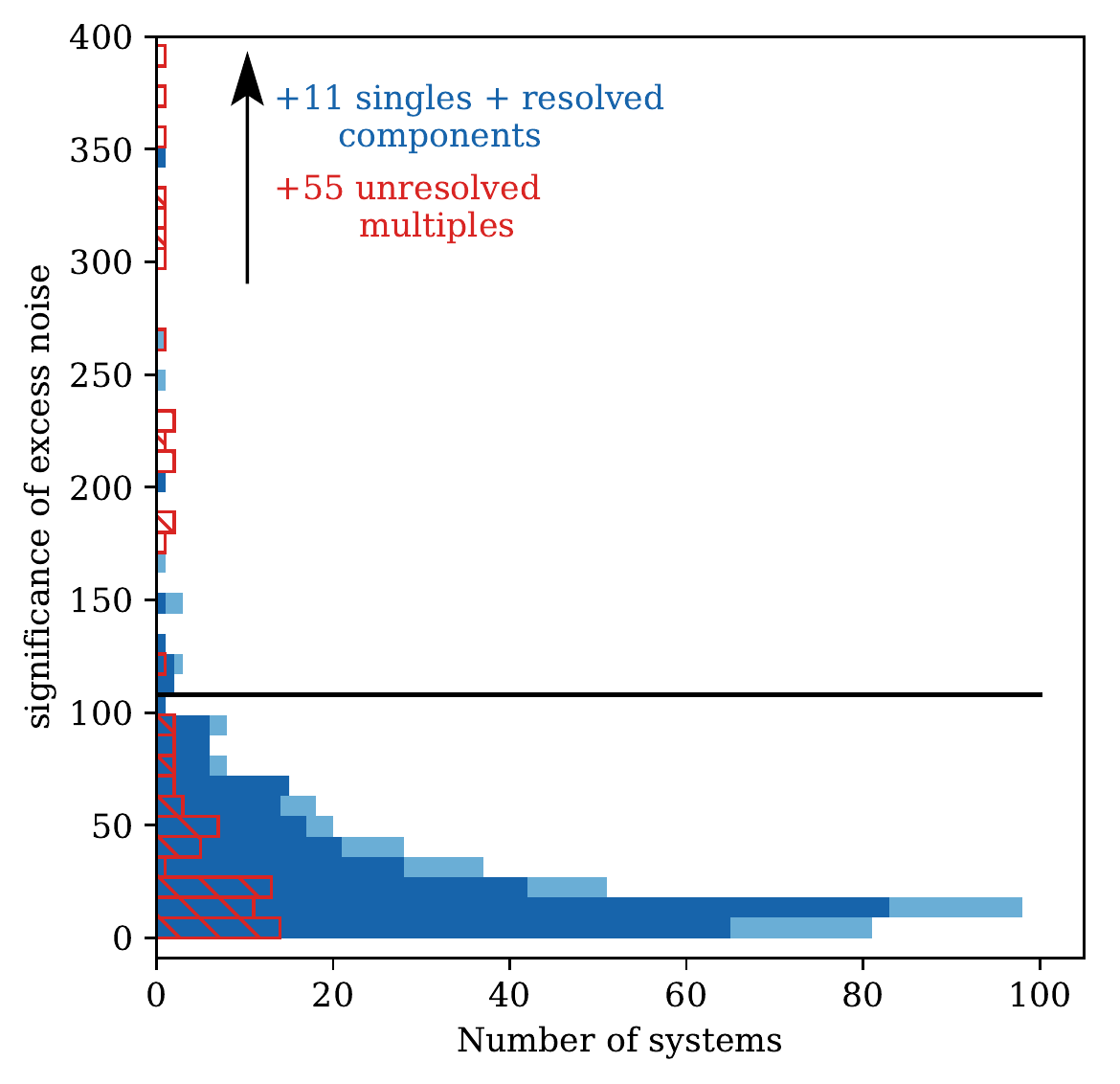}
\includegraphics[scale=0.44]{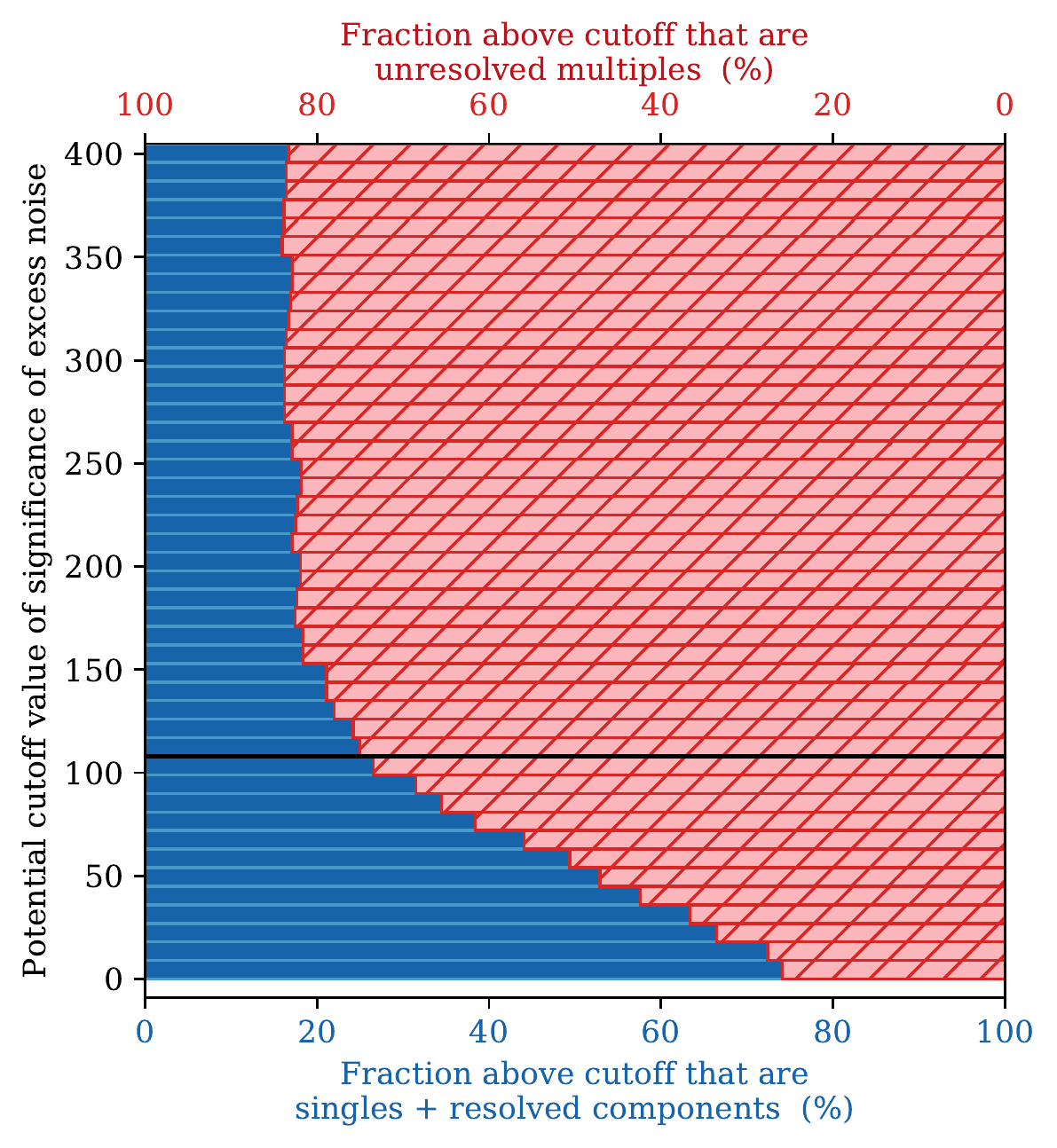}

\includegraphics[scale=0.44]{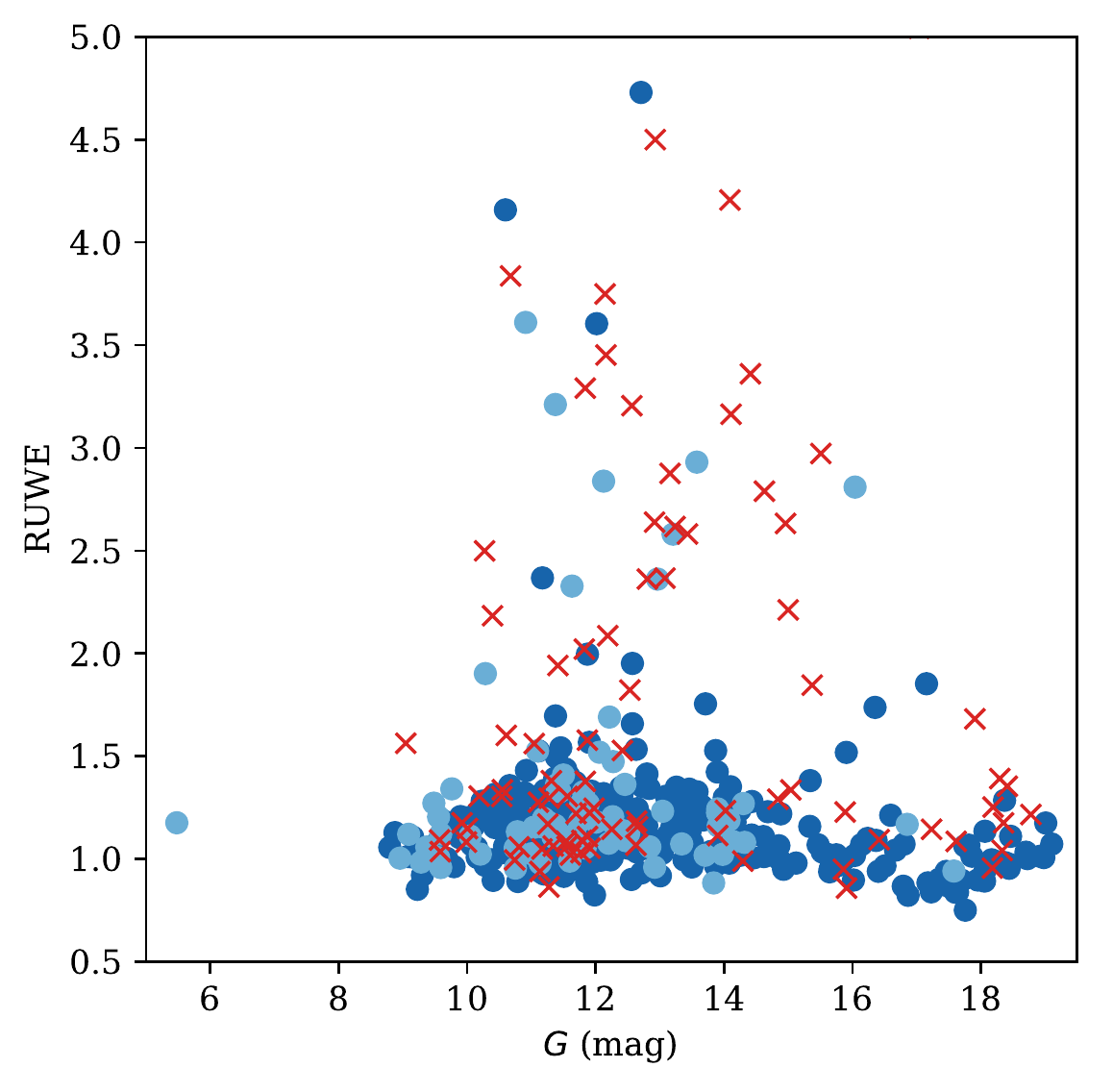}
\includegraphics[scale=0.44]{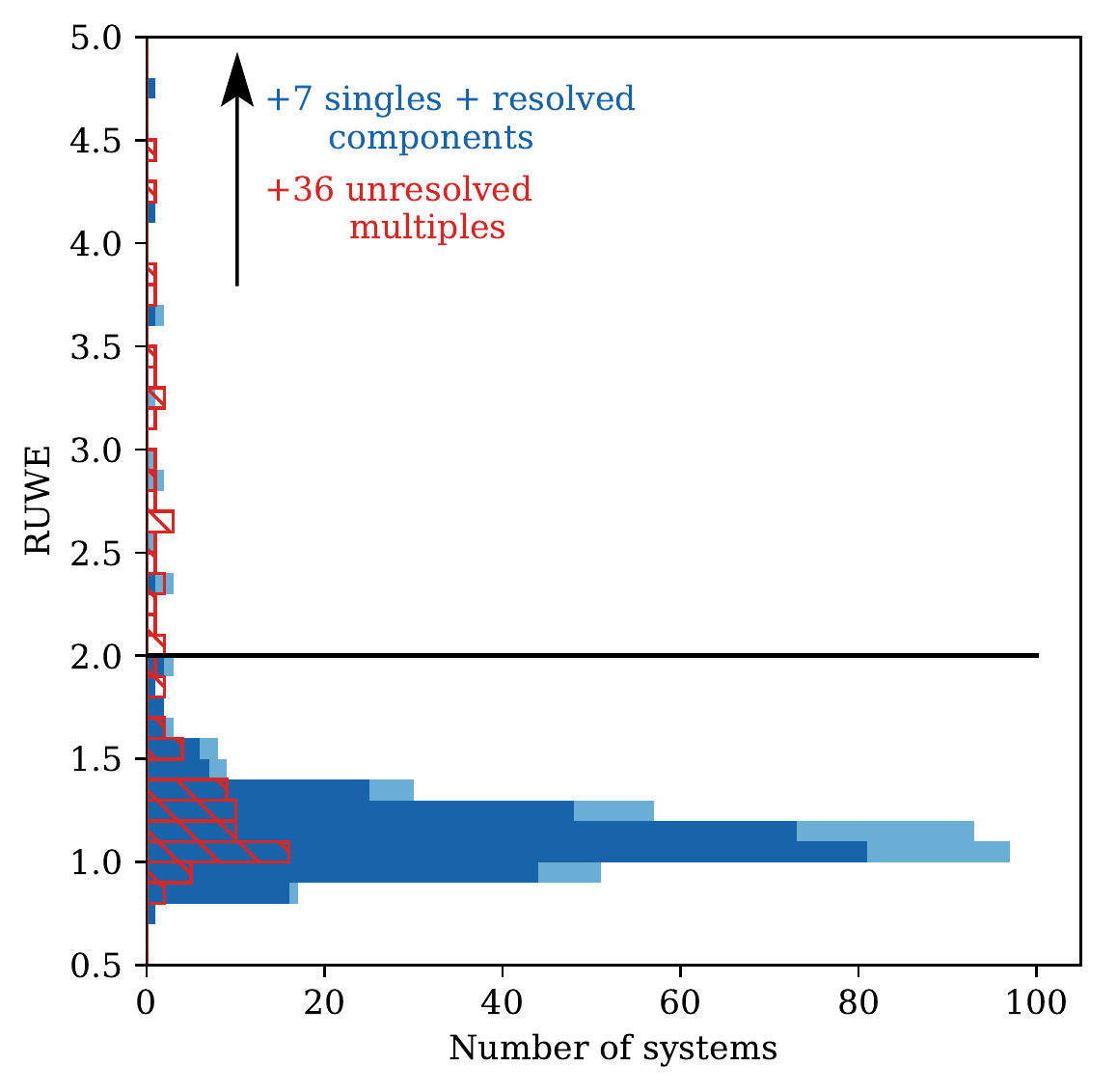}
\includegraphics[scale=0.44]{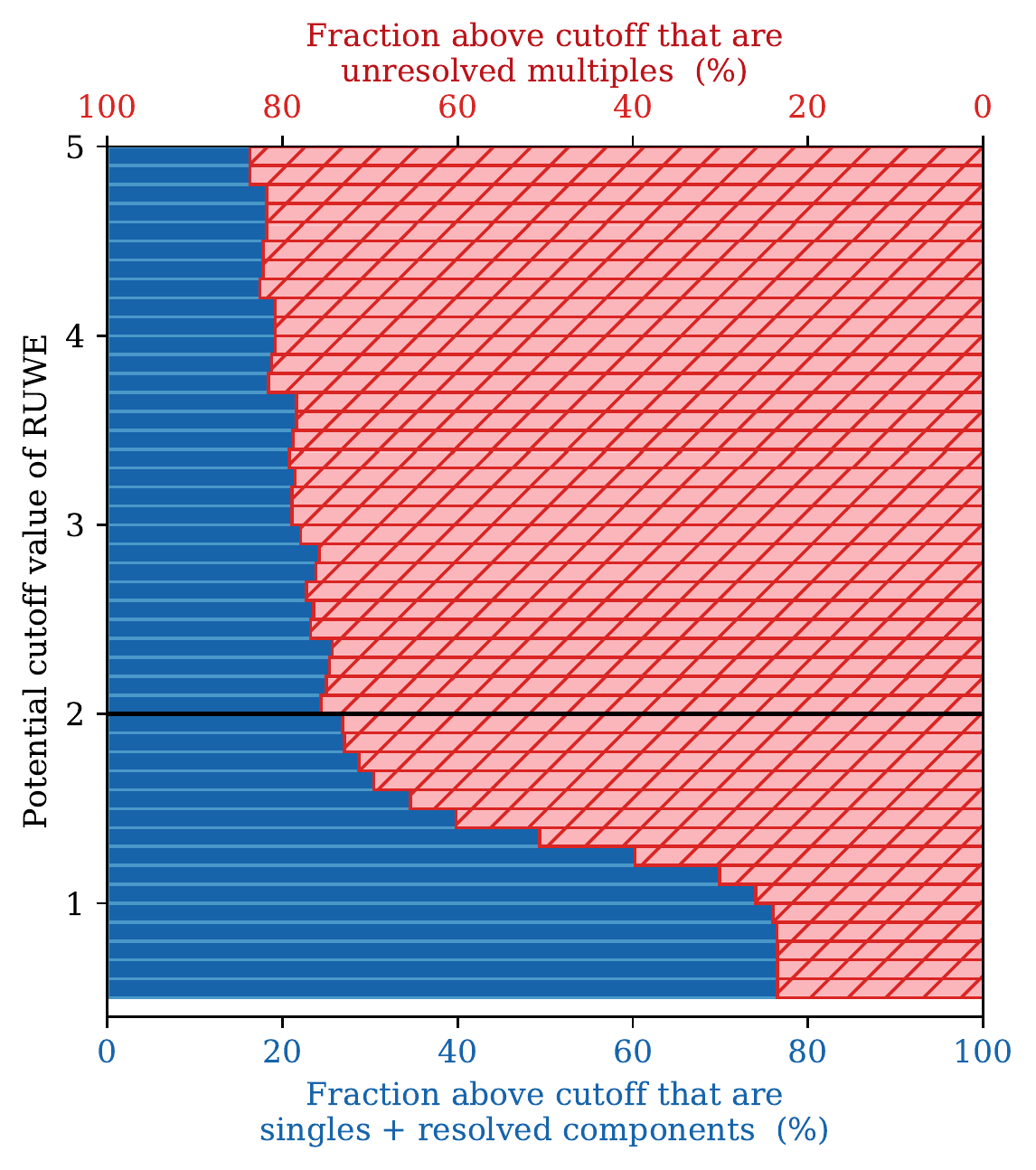}

\caption{\scriptsize Astrometric fit parameters in DR2 that are useful for selecting potential unresolved multiples. In all panels, the color scheme is the same as in Figure~\ref{fig:pipi}.
\textit{Left column:} parameters for single systems, resolved components, and unresolved multiples in \textit{Gaia} DR2, plotted against their \textit{Gaia} $G$ magnitudes. The unresolved multiples tend to have higher values of these fit parameters independent of their $G$ magnitude, indicating their poor astrometric fits in DR2.
\textit{Middle column:} Distributions of these parameters for systems in the left column, separated again by multiplicity status. Systems with values outside these plots are noted with the arrow and text on each panel. Although both distributions are peaked at low values of each parameter, the distribution of unresolved multiples extends to higher values. 
\textit{Right column:} Potential cutoff values for each parameter, showing the fraction of systems above each cutoff belonging to singles and resolved components (blue) and to unresolved multiples (red). The cutoff values above which 75\% of our systems are unresolved multiples ($\S$\ref{sec:paramcomparison}) are indicated with a horizontal black line.
\label{fig:astromparmsGOOD}}
\end{figure}

\begin{figure} \centering
\vspace{-1cm}
\includegraphics[scale=0.44]{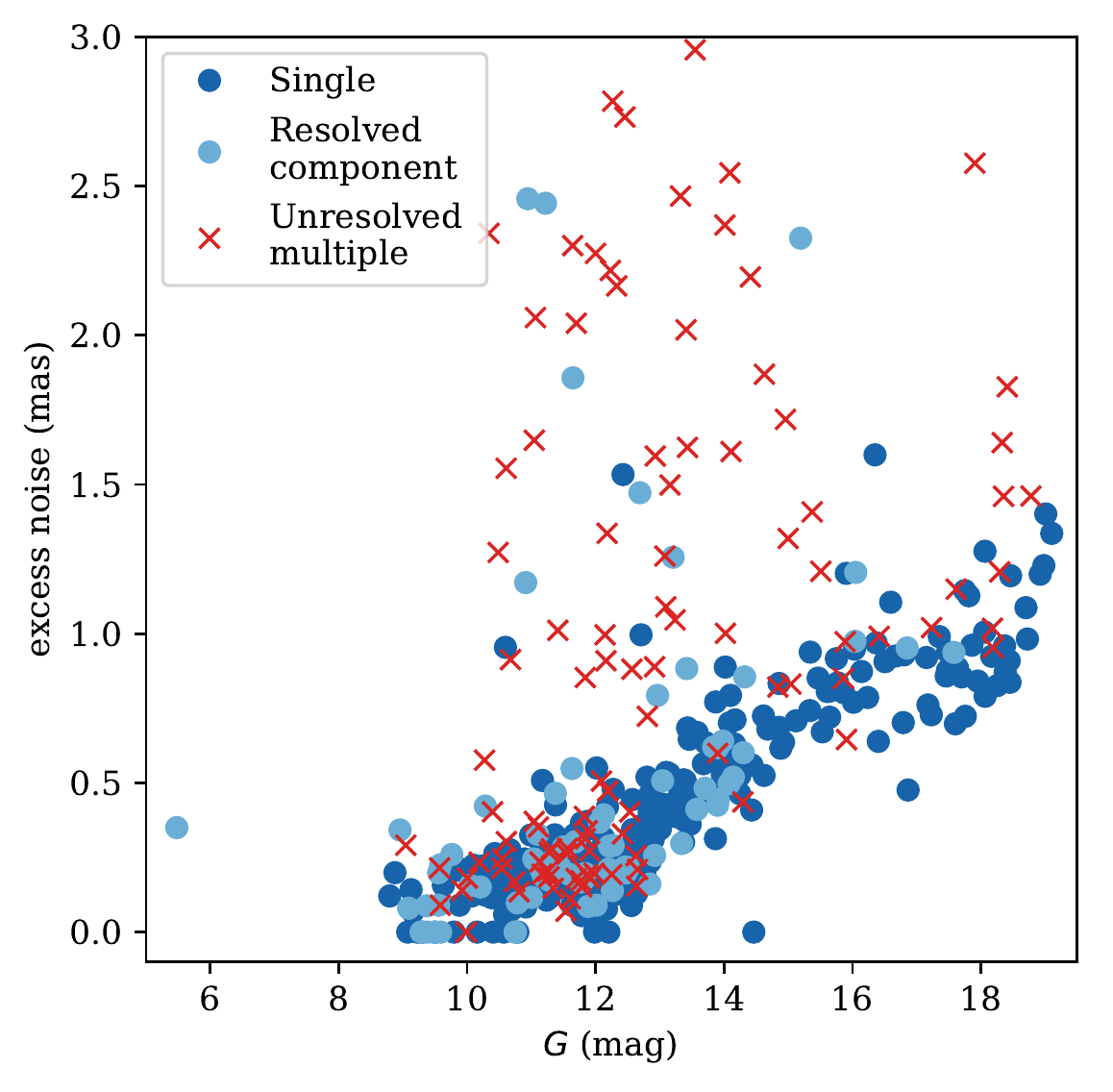}
\includegraphics[scale=0.44]{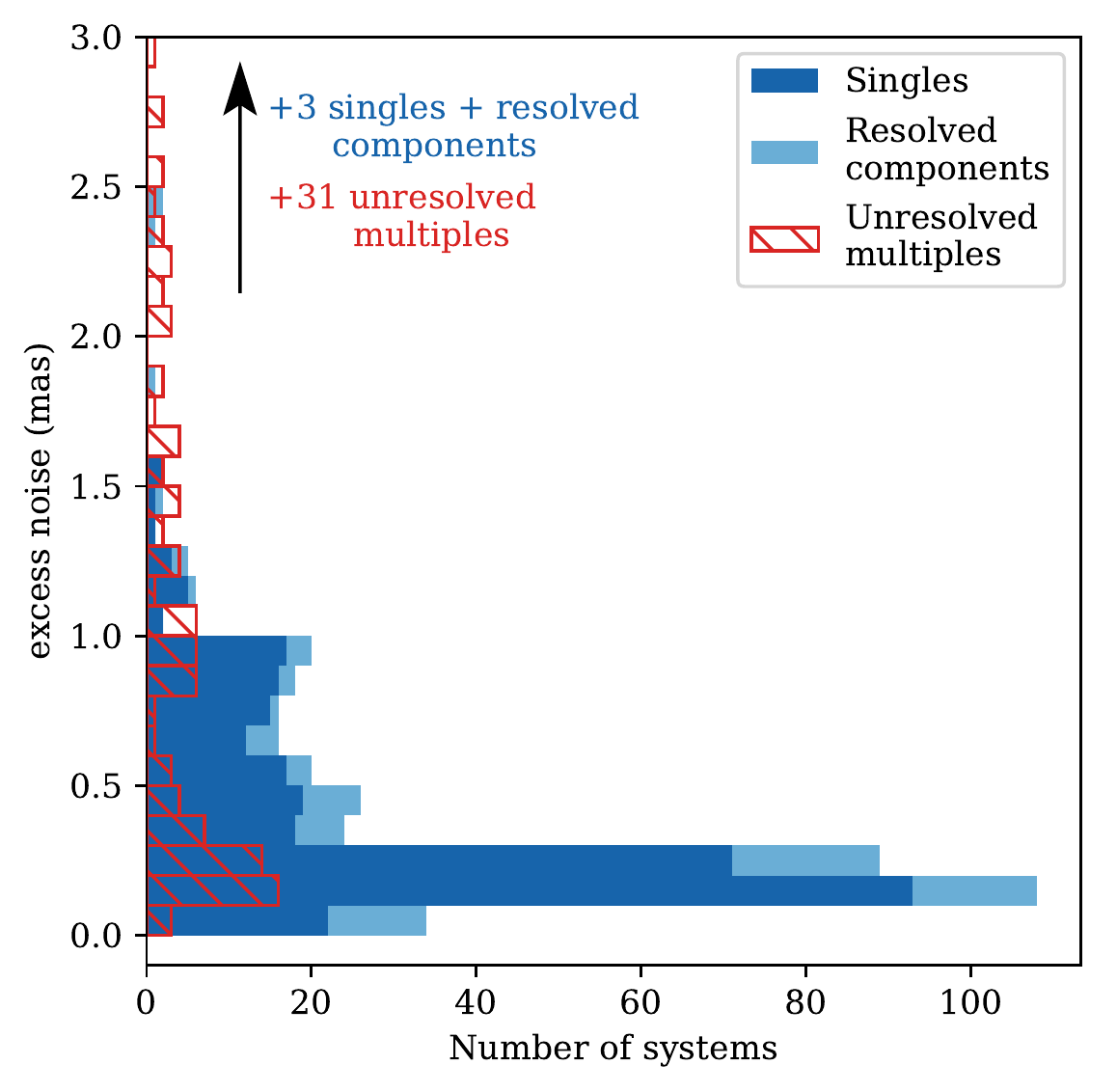}
\includegraphics[scale=0.44]{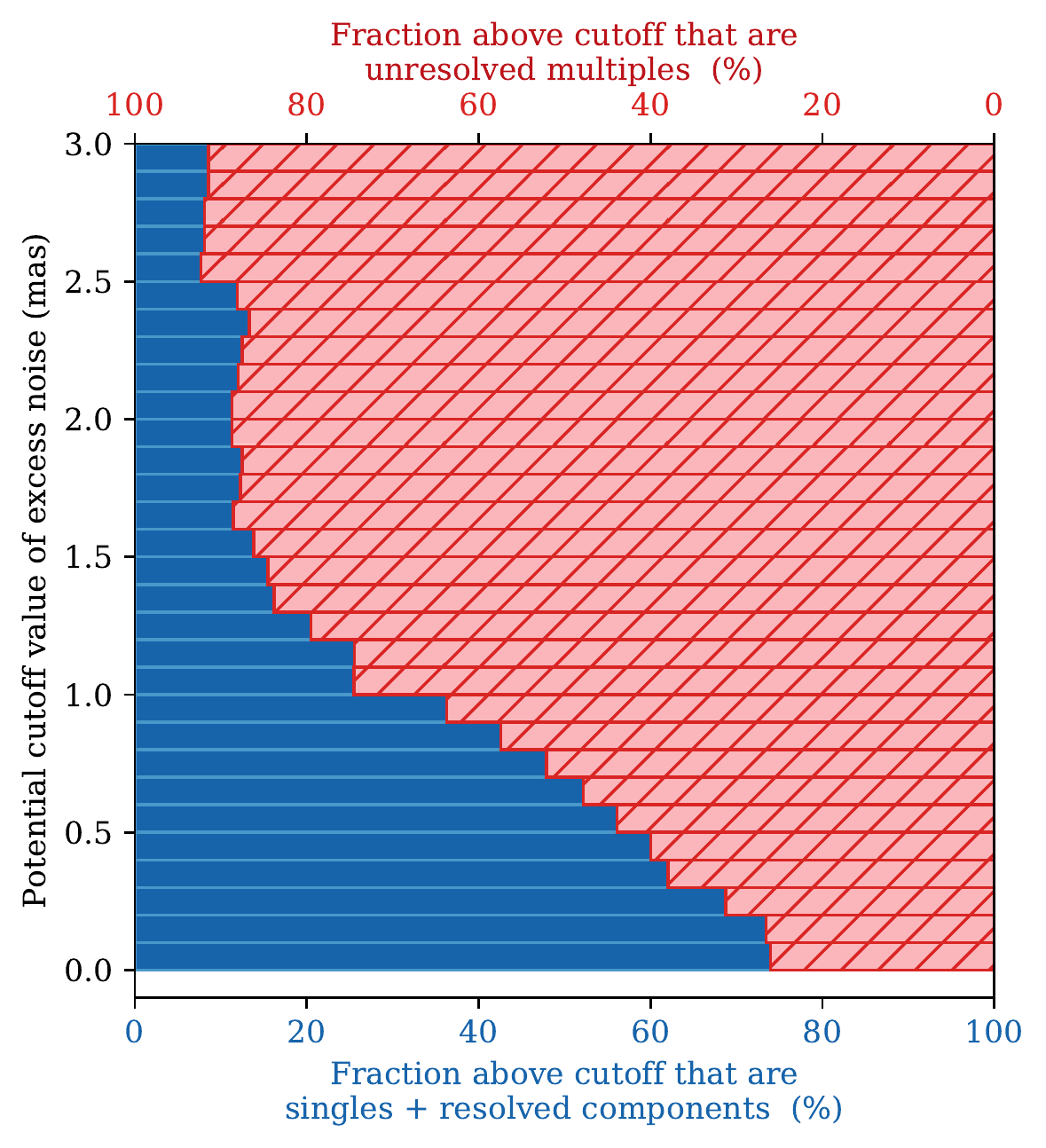}

\includegraphics[scale=0.44]{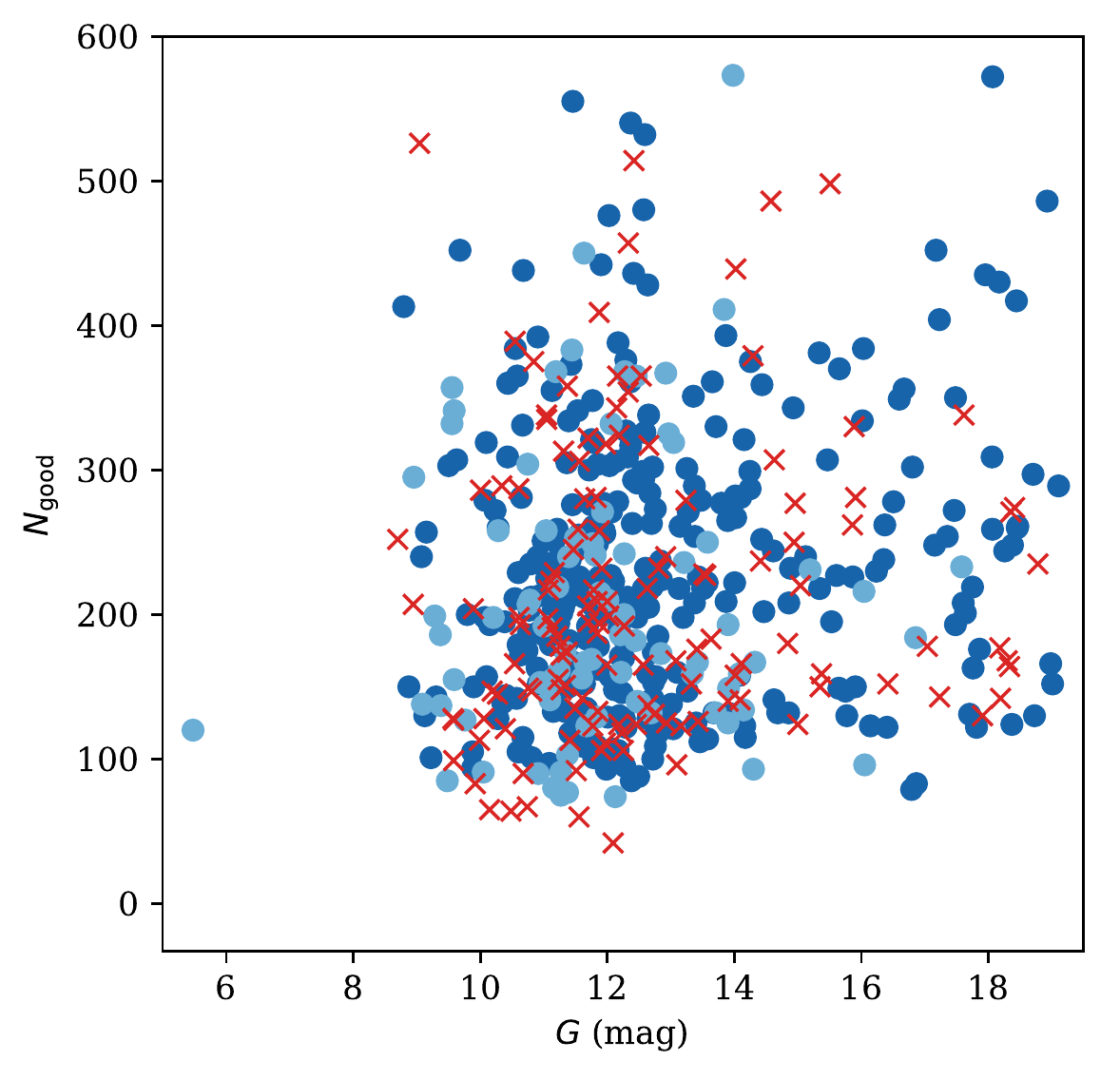}
\includegraphics[scale=0.44]{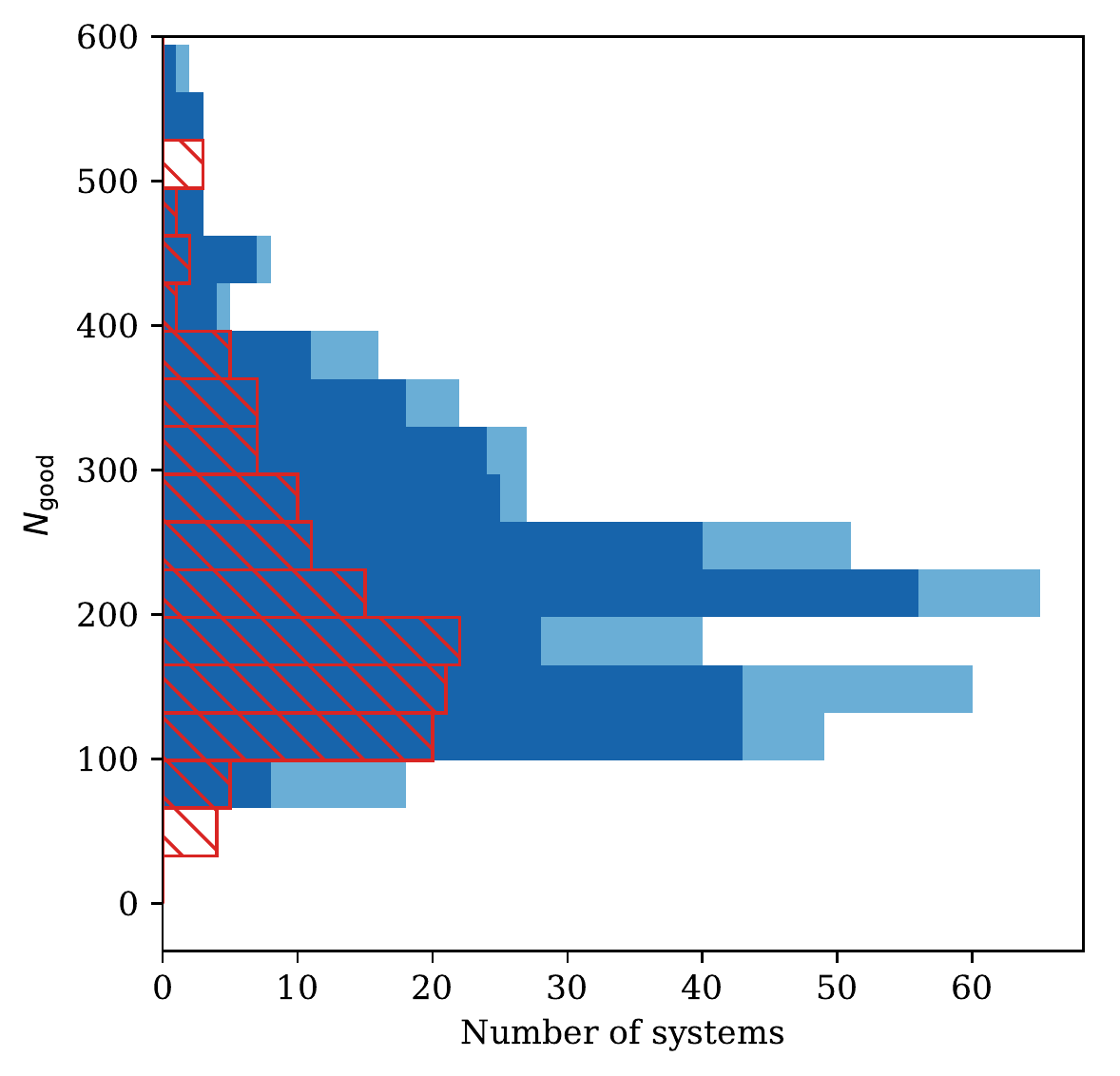}
\includegraphics[scale=0.44]{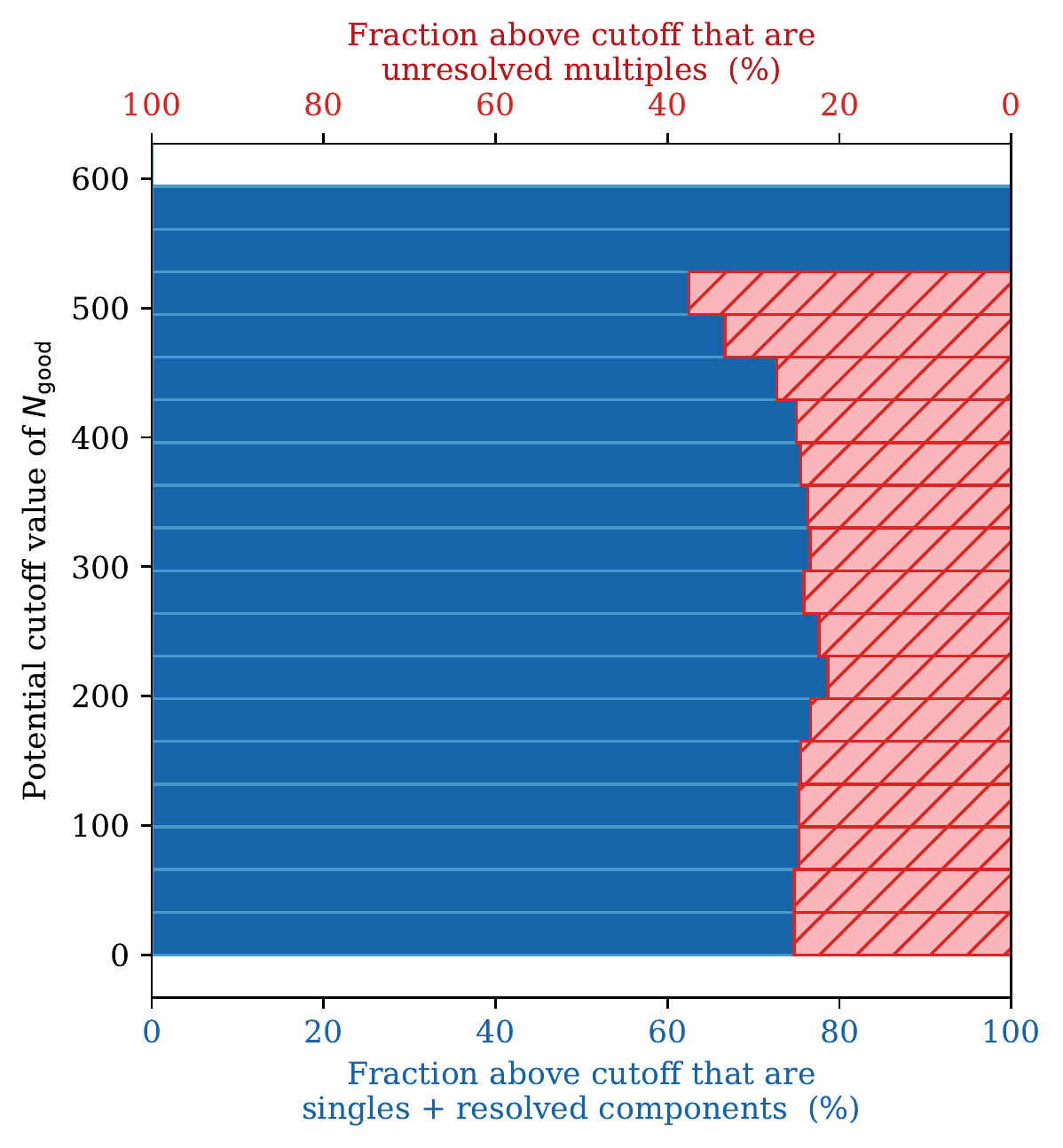}

\includegraphics[scale=0.44]{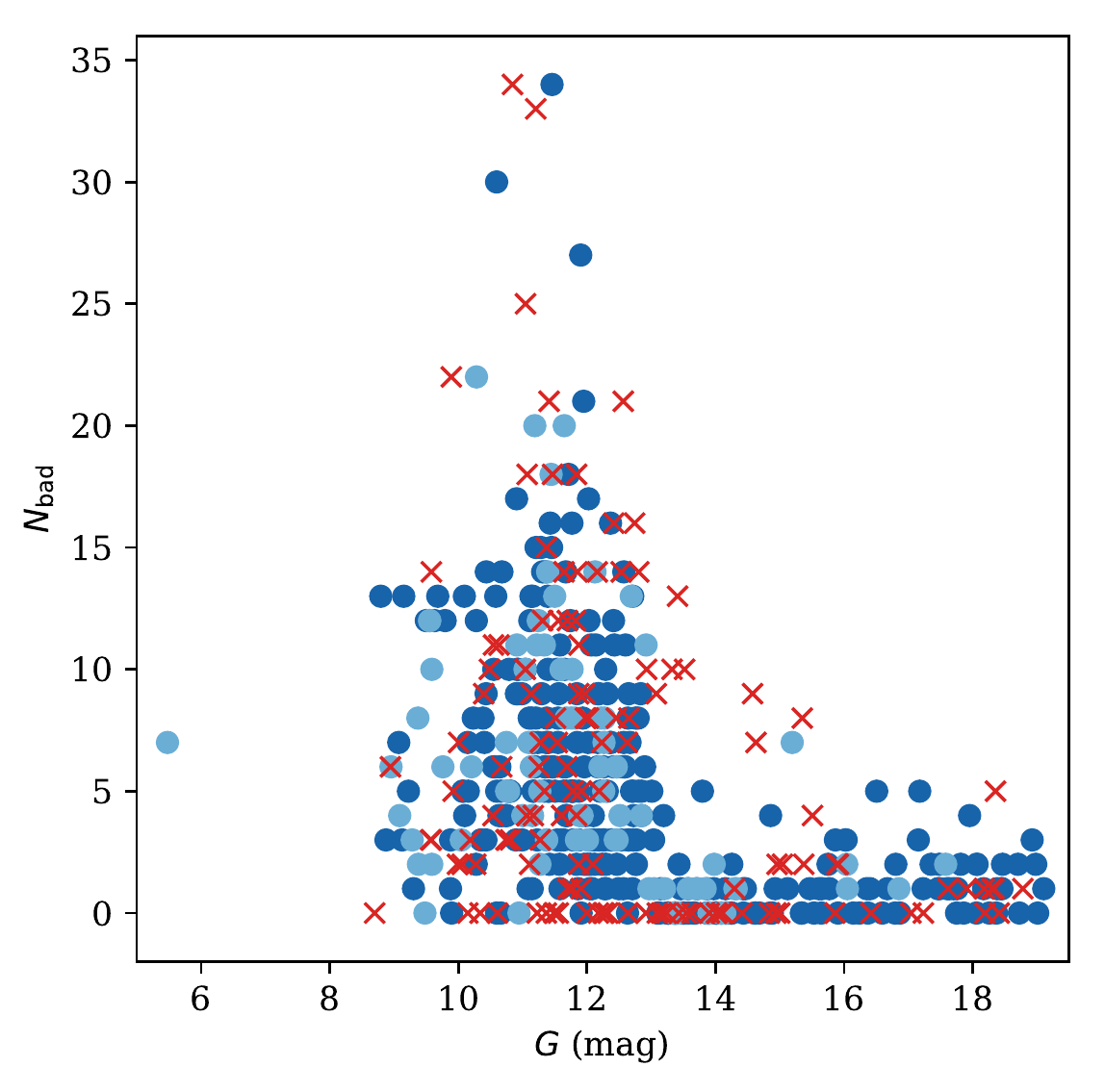}
\includegraphics[scale=0.44]{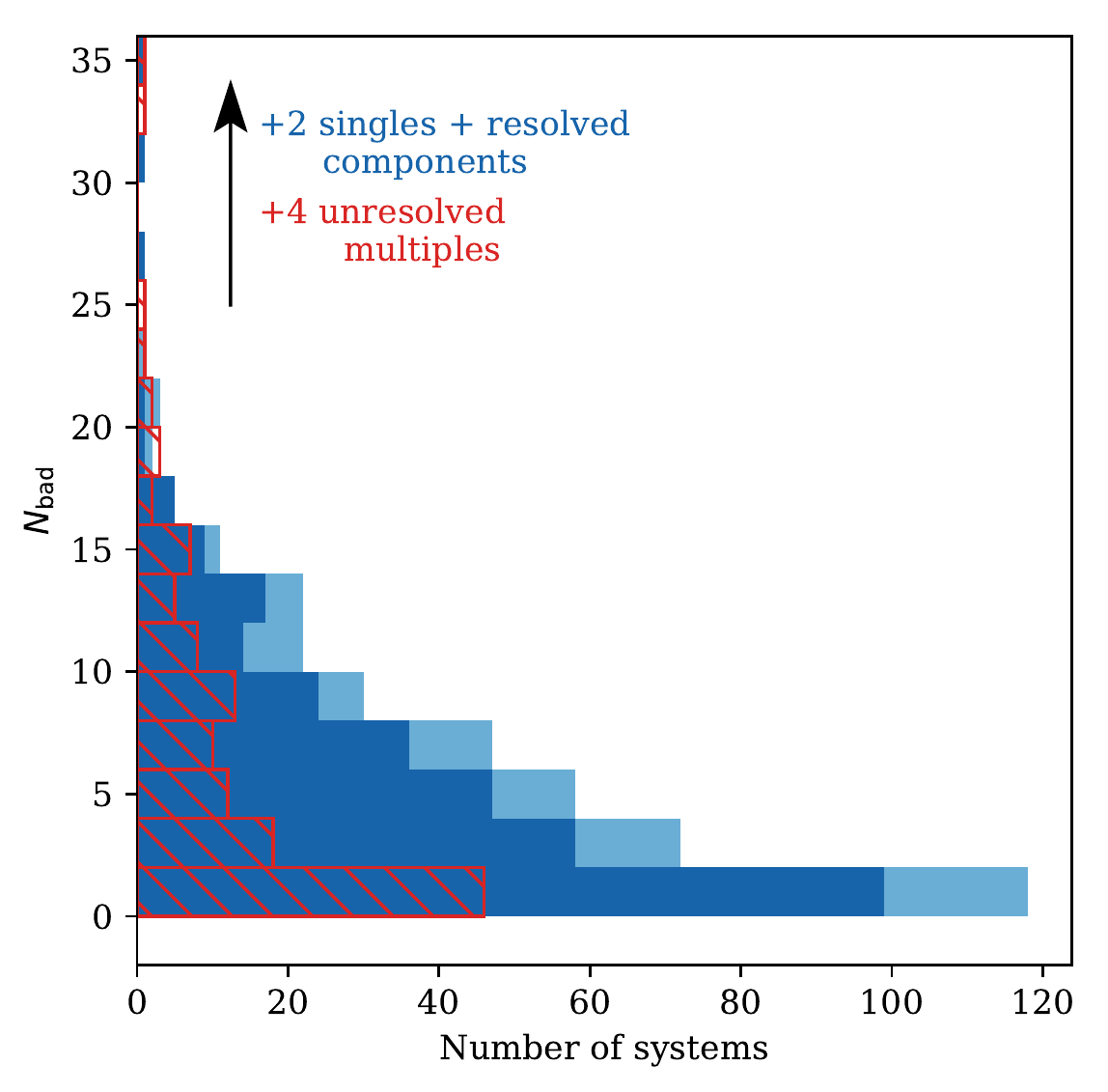}
\includegraphics[scale=0.44]{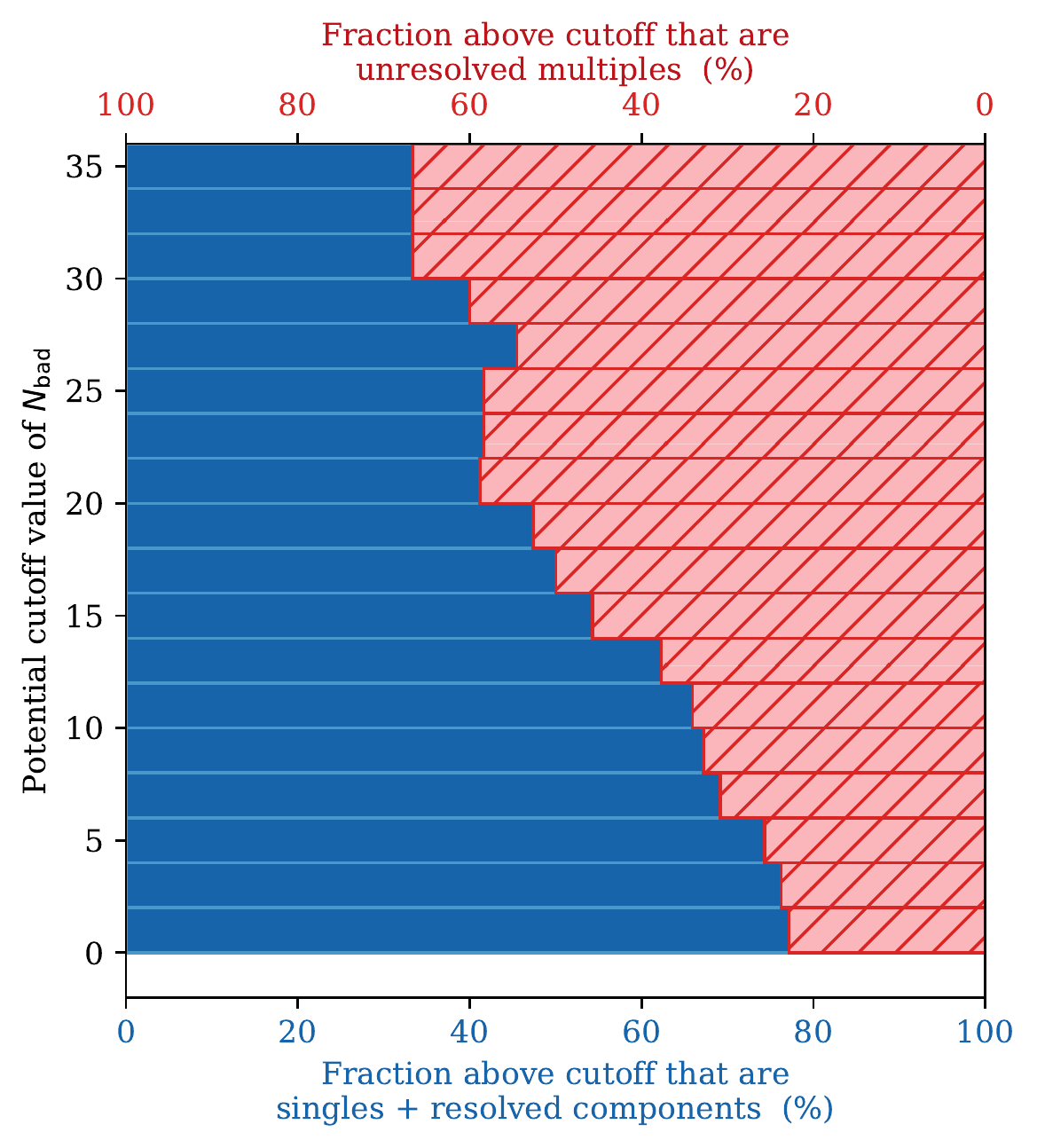}

\includegraphics[scale=0.44]{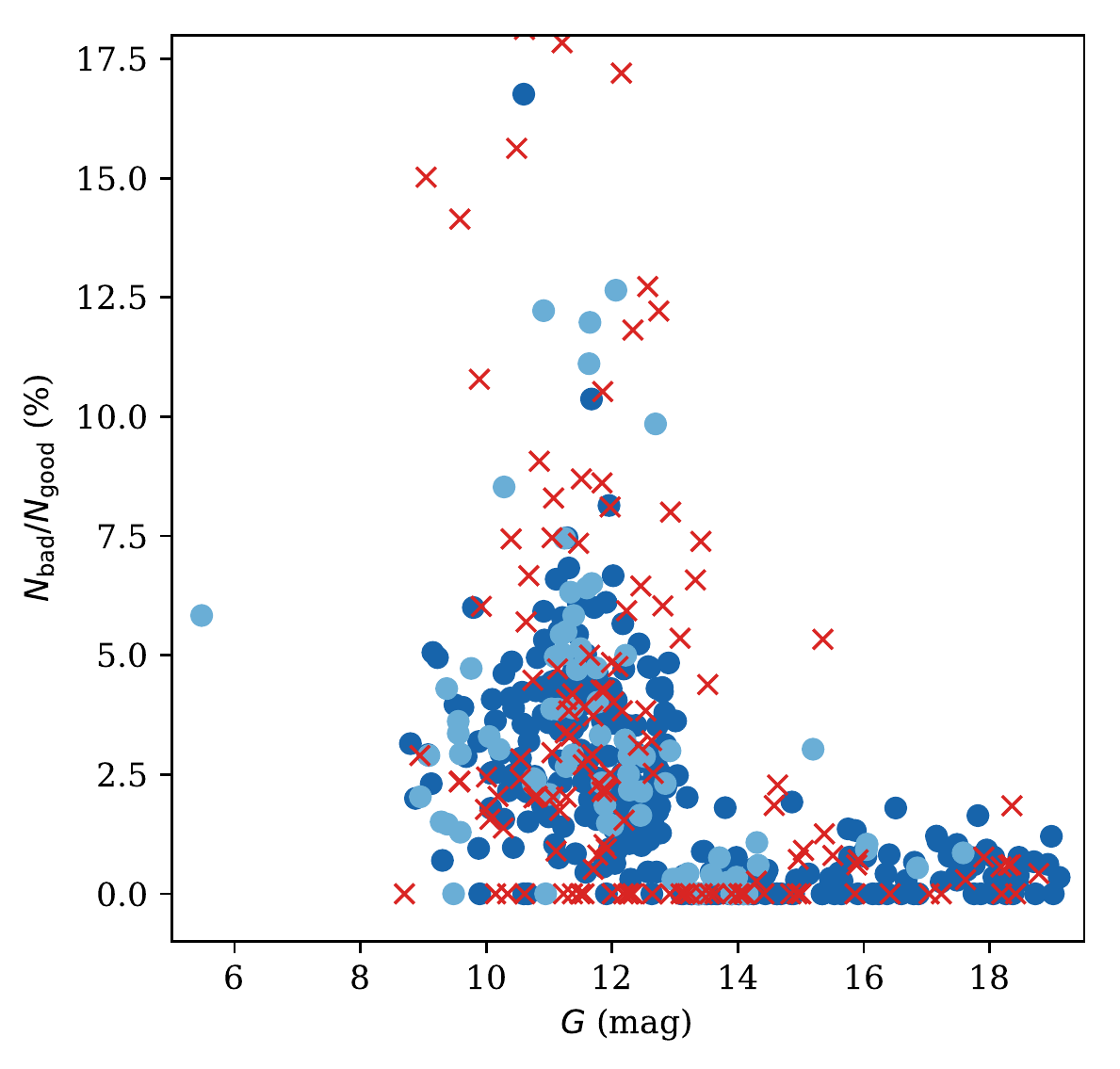}
\includegraphics[scale=0.44]{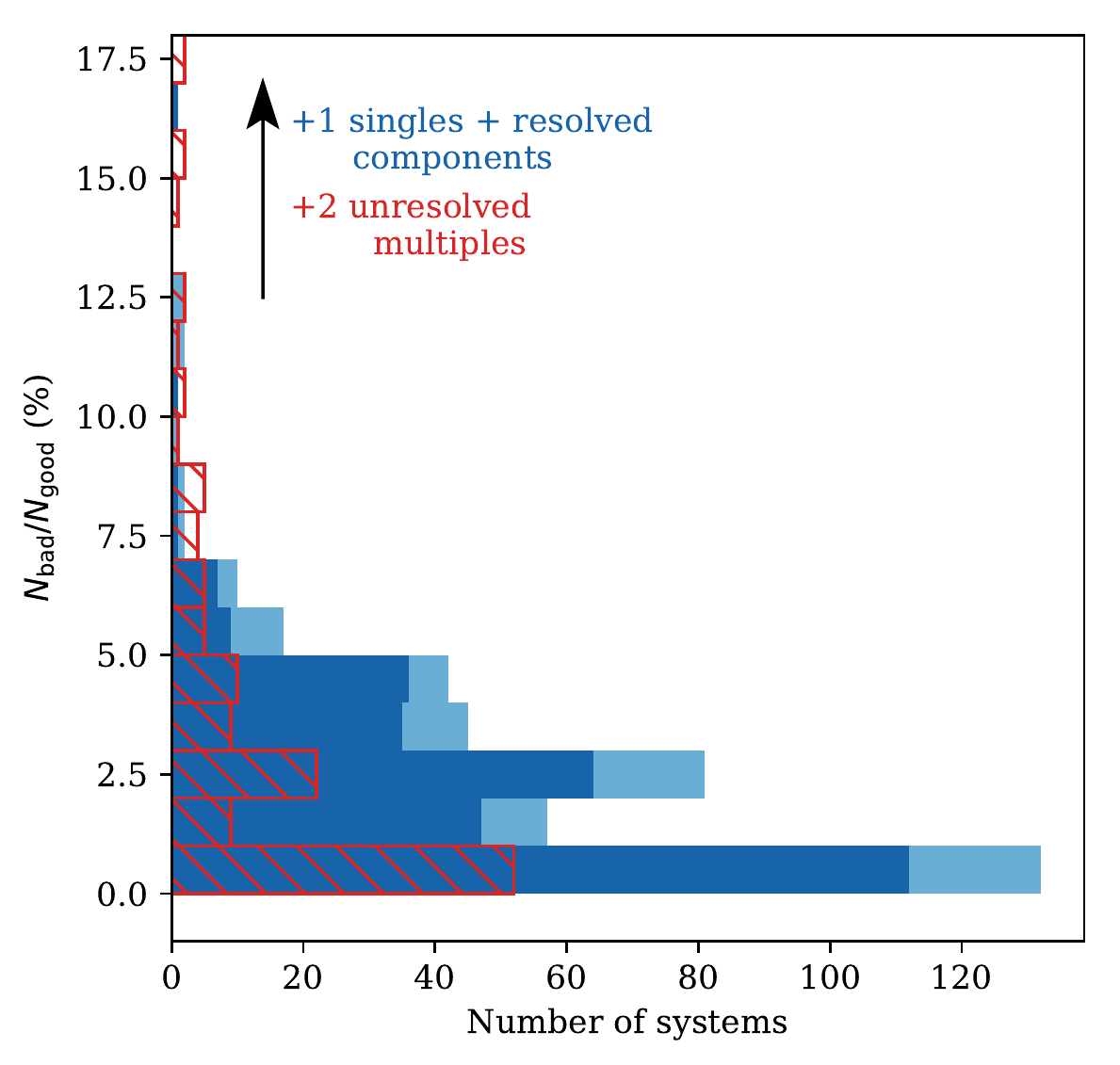}
\includegraphics[scale=0.44]{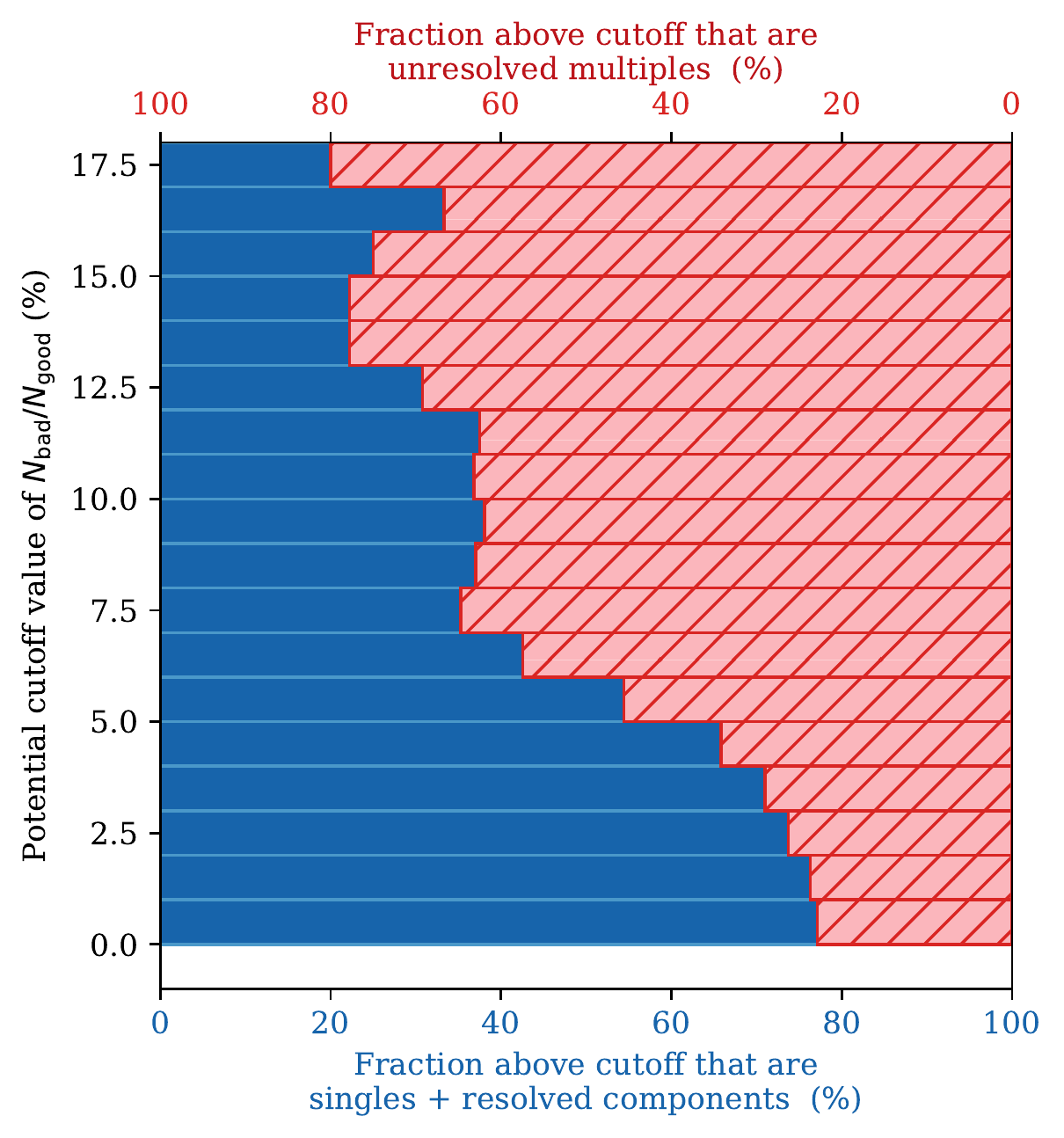}

\caption{\scriptsize Four quantities in DR2 that are less useful for choosing potential unresolved multiples. The color schemes and columns are the same as for Figure~\ref{fig:astromparmsGOOD}, and any systems with values exceeding these ranges are noted with an arrow and text on each panel.
For these quantities, the distributions of single and unresolved systems do not differ as significantly as those in Figure~\ref{fig:astromparmsGOOD}, making them less useful for identifying potential unresolved systems.
Excess noise (\texttt{astrometric\_excess\_noise}) does show a distinction between singles and multiples, but it is less useful than the very similar  \texttt{astrometric\_excess\_noise\_sig} because it strongly depends on $G$ magnitude (faint singles have values similar to brighter unresolved multiples).
\label{fig:paramsBAD}}
\end{figure}

We next compare the quality of the astrometric fit in DR2 for the single stars, resolved companions, and unresolved multiples, leveraging the same sample used in the parallax comparison in $\S$\ref{sec:picomparison}. The quality of the DR2 data and astrometric fits is described by several parameters given with each solution, which are listed in full in $\S$14 of the \textit{Gaia} DR2 documentation \citep{Ham18} and described in more detail in Appendix C of \cite{Lin18}. In the analysis that follows, we focus on a subset of eight of those parameters, of which four we find useful for selecting unresolved multiples, and four that were of interest but we find to be not useful. Described briefly (and with notation used in Figures~\ref{fig:astromparmsGOOD} and \ref{fig:paramsBAD} given in parentheses), these parameters are:

    \begin{itemize}

    \item \texttt{parallax\_error} ($\sigma_\pi$): 
    standard uncertainty in the parallax, given in milliarcseconds.

    \item \texttt{astrometric\_gof\_al} (goodness of fit): 
     describes the quality of the astrometric fit, computed as a function of the reduced $\chi^2$ of the fit --- i.e., \texttt{astrometric\_chi2\_al} and the degrees of freedom (using only \texttt{astrometric\_n\_good\_obs\_al}). This function is constructed such that these values follow a normal distribution centered around zero with a standard deviation of 1.0. 

    \item \texttt{ruwe} (RUWE): 
    renormalized unit weight error, representing the quality of the astrometric fit as a single dimensionless value. This parameter is analogous to \texttt{astrometric\_excess\_noise} (described above), but expressed in a dimensionless form intended to be easier to interpret. \texttt{ruwe} has been normalized to correct for quality issues in very bright and very faint sources and as a function of source color \citep{Lin18note}. 

    \item \texttt{astrometric\_excess\_noise\_sig}  (significance of excess noise): 
    significance of \texttt{astrometric\_excess\_noise} (described below), constructed to statistically resemble the positive half of a Gaussian distribution centered around zero with a standard deviation of 1.0. 

    \item \texttt{astrometric\_excess\_noise} (excess noise): 
    the difference between the DR2 data and the astrometric fit, given as the angle between these quantities on the sky. For each source, this value is used to indicate the noise of each observation contributing to the final solution. Values of zero indicate good astrometric fits, with positive increasing values indicating statistically higher-than-expected residuals. 

    \item \texttt{astrometric\_n\_good\_obs\_al} ($N_\mathrm{good}$): 
    number of observations that were not downweighted in the computation of the astrometric solution, analogous to the number of frames in RECONS data.

    \item \texttt{astrometric\_n\_bad\_obs\_al} ($N_\mathrm{bad}$): 
    number of observations that were downweighted in the computation of the astrometric solution, thus contributed little to the astrometric solution.

    \item In addition to the above seven parameters, we also calculate \texttt{astrometric\_n\_bad\_obs\_al} divided by \texttt{astrometric\_n\_good\_obs\_al} ($N_\mathrm{bad}/N_\mathrm{good}$) for each point, as \cite{Are18} has suggested that this quantity could be elevated for unresolved multiples.
    \end{itemize}

To establish criteria for likely multiplicity, we plot the values of these eight quantities for the single stars, resolved components, and unresolved multiples shared by the \textit{Gaia} DR2 and RECONS 25 pc samples. These visual comparisons are presented in the leftmost columns of Figures~\ref{fig:astromparmsGOOD} and \ref{fig:paramsBAD}. In each of these plots, the DR2 parameters are set against the $G$ magnitude of each source, allowing easy identification of any trend with magnitude, as faintness may degrade the astrometric fit even for single stars.

Figure~\ref{fig:astromparmsGOOD} presents four astrometric parameters that exhibit clear differences between the distributions of unresolved multiples and single stars/resolved components. These distribution differences have been judged by eye using the three types of diagnostic plots in Figure~\ref{fig:astromparmsGOOD}, as the differences are clear and the goal of this work is to identify the four most useful parameters (rather than an all-inclusive list of every useful parameter). In each panel of the far left column, the single stars and resolved components cluster primarily at low values of the given DR2 fit parameter, indicating high quality fits, whereas the unresolved multiples show a greater spread and thus much more variation in fit quality. Of the systems with poor fits indicated by these parameters, the majority are unresolved multiples. 
\textit{Note in all columns that there are many multiples off the tops of each plot, and only a few singles.} The plot limits have been set to prioritize clarity of the distributions. 
To clarify these trends, the middle column of Figure~\ref{fig:astromparmsGOOD} illustrates these distributions as histograms, with bins assigned by the Freedman-Diaconis rule applied over the range shown to accommodate the often non-Gaussian nature of these distributions and the outliers in each set. The distributions of each set of targets is not the same for each DR2 parameter, as each of those values is calculated differently and describes a different aspect of the astrometric model fit.

   To determine a useful ``cutoff'' value for each parameter, we define cutoffs that provide within a population that 75\% will be multiples.
To define these criteria, 
in the far right column of Figure~\ref{fig:astromparmsGOOD} we plot the potential cutoff values for each parameter against the ``set composition'' of the targets with values above each potential cutoff. For the systems exceeding each potential cutoff value (each value on the vertical axis), the bars at that coordinate indicate what fraction of them are unresolved multiples, and what fraction are single stars or resolved companions. A parameter that is useful for selecting potential unresolved multiples would have some point at which the balance of systems above that cutoff is dominated by the unresolved multiples, making it overwhelmingly likely that any selected system above that point is \textit{not} single. 
   
\begin{figure} \centering
\plotone{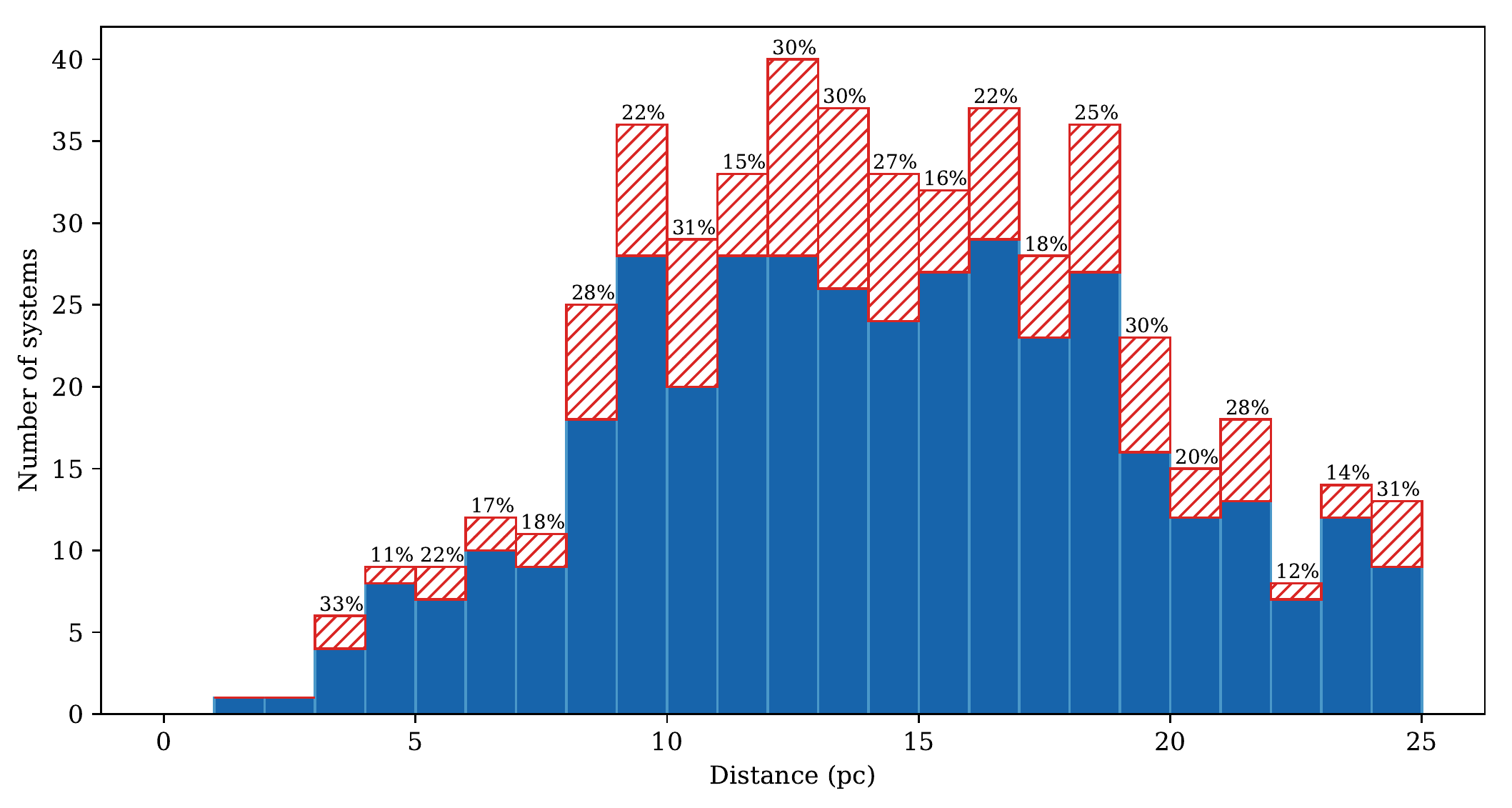}
\caption{\scriptsize Number of systems in 1 pc shells outward from the Sun. Blue bars indicate single stars, red hatched bars are the multiples unresolved in \textit{Gaia} DR2, and the black labels indicate the multiplicity in each 1 pc bin based on known unresolved multiples as a fraction of each bin's total. For bins with $\gtrsim 13$ systems, the multiplicity is similar to that observed in volume-complete surveys such as \cite{Win19}, which reports 27\% multiplicity. Although the dropoff of systems past $\sim$13 pc indicates that our survey is not volume-complete past that distance, the average multiplicity for those bins is 26\%, therefore our incompleteness is not biased toward single stars nor multiples.
\label{fig:pihist}}
\end{figure}

   This method is chosen in lieu of a Gaussian characterization and standard-deviation based criteria because most of these distributions do not follow normal distributions, and because our 25 pc sample is not volume-complete. Figure~\ref{fig:pihist} illustrates the number of systems in our sample in 1 pc shells extending outward from the Sun, with the fraction of unresolved multiples indicated by red bars and percentage labels. As the shells extend outward in equal-radius steps, their contained volume increases, thus we expect the number of systems in each shell would increase accordingly. This trend is indeed followed until $\sim$13 pc, where the number of systems in subsequent shells begins to fall, indicating volume incompleteness at these distances. Despite the paucity of systems past 13 pc, however, the fraction of multi-star systems in each bin is 20--30\% (average $26$\%), consistent with the more robust multiplicity for M dwarfs measured by \cite{Win19}. The unresolved multiples are therefore not systematically over- or underrepresented with distance. Furthermore, plotting the DR2 parameters against these distances does not reveal any distance-dependent trend, eliminating the possibility that the paucity of systems at large distances will significantly bias the positions of the cutoffs.  

Finally, we note that the unresolved multiples among the missing systems would likely have small-amplitude orbital motion or longer orbital periods, as these systems' greater distances would shrink the scales of their apparent orbits. 
Then these missing systems, if they were included in our sample, would likely have good astrometric fits in DR2, and add more to the sizes of the peaks of the distributions in Figures~\ref{fig:astromparmsGOOD} and \ref{fig:paramsBAD} rather than the outliers. The result would be minimal effect on our definitions of the cutoffs points.

   Choosing the cutoffs as the points where at least three out of four systems (75\%) above that point are unresolved multiples, we arrive at the following criteria:
   \begin{itemize}
   \item \texttt{parallax\_err} $\geq 0.32$ mas for $G \lesssim 18$ ( $\geq 0.40$ for $G \gtrsim 18$)
   \item \texttt{astrometric\_gof\_al} $\geq 56.0$
   \item \texttt{astrometric\_excess\_noise\_sig} $\geq 108.0$
   \item \texttt{ruwe} $\geq 2.0$
   \end{itemize}
Targets meeting all four of these criteria are marked with check marks in Column~16 of Table~\ref{tab:RECONS-GDR2}. 
For \texttt{parallax\_err}, this cutoff must include the caveat that systems fainter than $G \sim 18$ show elevated values regardless of duplicity, making a more appropriate cutoff in that region \texttt{parallax\_err} $\geq 0.40$ mas. Such brightness dependence is also the reason we recommend \texttt{astrometric\_excess\_noise\_sig} rather than \texttt{astrometric\_excess\_noise}; although both show clear points where the unresolved multiples distribution dominates, the dimensionless \texttt{astrometric\_excess\_noise\_sig} is more controlled with respect to $G$ (see Figure~\ref{fig:astromparmsGOOD}, third row), whereas \texttt{astrometric\_excess\_noise} (Figure~\ref{fig:paramsBAD}, fourth row) rises continuously with magnitude in the brightness range we examine here.
The values presented above are likely overestimates, as many systems currently marked ``single'' have not had sufficient observations in our astrometry to fully rule out the existence of a bound companion. Some of those systems may later be revealed as unresolved binaries.
    
    In presenting these ``cutoff'' values, we must emphasize that they do \textit{not} represent dividing lines between single and unresolved multiple systems in the DR2 results. This investigation suggests that systems with values exceeding these points are likely to be unresolved multiples, but also that systems with values below these points may be multiple \textit{or} single --- those entries are ambiguous at best. 
The astrometry of an unresolved multiple would match a single-star model in cases where the brightness ratio is large, the brightness ratio is zero (which why \cite{Are18} find insignificant difference in astrometric fit between single stars and unresolved multiples with small $\Delta$mag), or the orbital period is much longer than 22 months. 

We explored several additional parameters, and in Figure~\ref{fig:paramsBAD} present four quantities that we found to be not as useful as expected for selecting unresolved multiples.  This judgement is made by eye using the middle and third columns of Figure~\ref{fig:paramsBAD}. 
The quantity \texttt{astrometric\_excess\_noise} (first row of Figure~\ref{fig:paramsBAD}) shows a strong difference between the distributions of unresolved multiples and single stars (and resolved components), but its dependence on $G$ magnitude makes it less useful for distinguishing potential multiples than the similar parameter \texttt{astrometric\_excess\_noise\_sig}. 
The parameters \texttt{astrometric\_n\_good\_obs\_al} and \texttt{astrometric\_n\_bad\_obs\_al} show very little distinction between the distributions of unresolved multiples and single stars.
Finally, dividing the bad observations by the good observations \citep[as suggested by][]{Are18} also produces a rather weak trend (visible in the third row of Figure~\ref{fig:paramsBAD}).

\subsection{Systems Missing from \textit{Gaia} DR2}
\label{sec:missingsystems}

In addition to systems with poor astrometric fits as indicated by the fit parameters, several systems in \textit{Gaia} DR2 are missing parallaxes (i.e., have a two-parameter solution rather than a five-parameter solution) or are missing from that catalog entirely. These results indicate targets for which the five-parameter solution was an exceptionally poor fit \citep{Lin18}. For the nearby red dwarfs, 42 of these targets appear in Table~\ref{tab:RECONS-GDR2} with blank spaces in their DR2-specific fields. These systems represent 7.2\% of our 25 pc sample.

Astrometric fits poor enough to merit exclusion from DR2 could be multiples with photocentric orbital motion, and on this basis we have marked these systems as ``suspicious'' in Column~16 of Table~\ref{tab:RECONS-GDR2}. Of the 42 of these systems in the 25 pc sample, 27 (64\%) have unresolved companions or astrometric perturbations noted in Table~\ref{tab:RECONS-GDR2}, and four are resolved components of multiple systems. 
In addition to multiplicity, \cite{Are18} cite high proper motion as a primary reason for stellar point sources to be missing DR2 solutions, but only four of our non-multiple missing systems meet their benchmark of 600 mas/yr: SSS~1444-2019 (3495.1 mas/yr), 2MA~0251-0352 (2149.7 mas/yr), LHS 1918 (606.0 mas/yr), and GJ 273 (3732.1 mas/yr). 
We therefore note that the remaining seven are particularly worthy of follow-up investigation regarding their multiplicity status.


\subsection{Comparison to Similar Works}
\label{sec:otherwork}
    Other published works selecting multiples from DR2 have focused on the proper motion anomaly \citep{Bra18a, Ker19a}, or the difference in a target's proper motion measured over a few years and its motion as computed over a decades-long temporal baseline. The \textit{Gaia} DR2 catalog presents the opportunity to calculate this anomaly because of its nearly 25-year temporal separation from the \textit{Hipparcos} astrometric mission, which measured positions, proper motions, and parallaxes of more than 100{,}000 stars over 1989--1993 \citep{Per97,vanL07}. Systems with significant proper motion anomalies are likely unresolved multiples, as a system with a bound companion inducing orbital motion will have non-linear proper motion, i.e., acceleration.
    
    \cite{Bra18a} renormalizes the \textit{Hipparcos} and \textit{Gaia} DR2 errors and presents a catalog of proper motion anomalies for systems common to these two catalogs, but the faintness limits of \textit{Hipparcos} limit the overlap between this catalog and our 25 pc sample to 59 systems. The faintest system in \cite{Bra18a} has $G = 12.31$, whereas our catalog extends to $G =19.01$.
\cite{Bra18a} does not suggest a benchmark value to flag accelerating systems, and instead advocates a case-by-case approach. 
Of the 59 targets common to that catalog and our 25 pc sample, only seven have proper motions that differ by more than 20\% in their comparisons, and we confirm that four of those are unresolved multiples; we suspect the other three will turn out to be multiples as well.
Our method of DR2 parameter cutoffs ($\S$\ref{sec:paramcomparison} above) identifies eight likely unresolved multiples from the 59 common targets, with seven being confirmed multiples, but only two of those systems are flagged by the 20\% acceleration search above. 

\cite{Ker19a} compute the \textit{Hipparcos}-DR2 proper motion anomaly similarly to \cite{Bra18a}, but limit their core sample to systems within 50 pc. By taking into account radial velocities of their systems (with $\sim$70\% of these RVs not from the DR2 entries), they are able to set limits on the masses of potential companions to the anomalous systems. 
As in our comparison with \cite{Bra18a}, of the 51 systems in common between \cite{Ker19a} and our 25 pc catalog, our DR2 criteria identify eight potential multiples. Seven of these are already confirmed unresolved multiples, and seven (not the same subset) are flagged by \cite{Ker19a} as likely binaries. In addition to these, \cite{Ker19a} flag nine more systems as likely binaries (for a total of 16 of the 51) that do not meet our DR2 criteria; four of these are confirmed multiples (for a total of 11 confirmed of their 16 flagged).

These searches suggest that both methods are effective at selecting likely unresolved multiples, but they are not sensitive to the same orbits. Our approach using the DR2 fit quality seeks the systems with significant motion on DR2's 22-month observing timescale, and is poorly sensitive to systems with decades-long orbits, whereas the \cite{Bra18a} catalog is likely more sensitive to systems with significant motion on those decades-long timescales because it hinges on the \textit{Hipparcos}-DR2 25-year baseline. We also note that our approach using \textit{Gaia} DR2 parameters alone is applicable to more systems because it is not restricted by the faintness limit of \textit{Hipparcos}. For those systems highlighted by \cite{Bra18a}, however, that catalog's offered information is rich enough to constrain the dynamical masses when combined with just a single measurement of separation of the resolved components, propelling the user past the step of mapping orbits that could be decades or centuries in length \citep[demonstrated in][]{Bra19}.

    The renormalized unit weight error (\texttt{ruwe}) merits a dedicated discussion because it is intended as a straightforward, easy way to interpret of the quality of each astrometric fit to the single-star model, as has been used by some studies already to exclude (or include) multiples \citep[e.g.,][]{Sch19}. Provided as a lookup table several months after the DR2 release, \texttt{ruwe} is the reduced $\chi^2$ normalized to counter the documented statistical trends of \texttt{astrometric\_chi2\_al} with $G$ and $B_G - R_G$. The DR2 technical documentation \citep{Lin18note} suggest that solutions with \texttt{ruwe} $> 1.4$ indicate bad fits likely due to orbital motion, with enough confidence that it will be incorporated into the criteria for selecting unresolved multi-star system candidates in \textit{Gaia}'s upcoming orbital-motion pipeline \citep{Pou19}. 
Our comparison of this parameter in Figure~\ref{fig:astromparmsGOOD}
 confirms that targets with high \texttt{ruwe} values are more likely to be unresolved multiples than single stars. This result is confirmed for the FGK stars by \cite{Zie20}, who detect stellar companions with SOAR speckle interferometry for 84\% of their 135 FGK systems with \texttt{ruwe} $> 1.4$.   
In a multiplicity-focused study, \cite{Jor19} also find a correlation between duplicity and \texttt{ruwe} using a sample of bright spectroscopic multiples from SB9. Although those authors caution that the link may not be sustained for multiples fainter than their sample of $6 \leq G \leq 10$, our similar work here uses a large number of systems with $G \sim$ 10--20.
We recommend \texttt{ruwe} as an effective way to select many unresolved multiples, but note that its given format as a separate lookup table only accessible through the ESA website makes it less convenient to use than the other astrometric parameters, which are delivered alongside the DR2 solutions everywhere that catalog is accessible. 


\section{Validation of the Unresolved Multiples via SOAR Observations}
\label{sec:SOARvalidation}

Using the criteria described above in $\S$\ref{sec:gaia}, we have added 114 nearby likely multi-star red dwarfs to a sample that we are observing with speckle interferometry at the Southern Astrophysical Research (SOAR) Telescope. These targets include most of the 97 marked in Column~16 of Table~\ref{tab:RECONS-GDR2} (those within the brightness and airmass limits of SOAR), supplemented by many that meet only a subset of those criteria. The speckle interferometry is carried out using the high-resolution camera mounted on the adaptive optics module on SOAR \citep[HRCam+SAM; see][]{Tok18}. 
These observations will complement the long-term RECONS astrometry by mapping orbits shorter than $\sim$6 years through multi-epoch observations on these 25 pc M dwarfs. 
There is significant overlap in the target lists of the SOAR and CTIO/SMARTS 0.9~m programs, as these two facilities have the same latitude and sky coverage. 
Systems flagged through the \textit{Gaia} DR2 parameters described above ($\S$\ref{sec:gaia}) are well suited for this 3-year SOAR program because only systems with appreciable orbital motion in DR2's 22-month observations will have poor astrometric fits. 

Initial results have marked a promising start to this observing program, with 90\% of the DR2-selected targets already observed, of which 73\% have had companions detected. Several systems have also had orbital motion detected through these multi-epoch observations, demonstrating that this program is already capturing the fast orbits that are most needed to complement the RECONS astrometric multiples. A full description of these observations and results will be presented in a forthcoming paper.

%
\section{Conclusions} \label{sec:conclusions}
%
Efforts to complete the astronomical community's census of nearby red and brown dwarfs have expanded beyond RECONS since that program's inception in 1999, and will continue to expand in precision with future updates from \textit{Gaia} astrometry. Considering this success, 
the focus of the RECONS astrometry program has shifted toward the potential for characterization of its red dwarf systems enabled by their remarkably long temporal baseline of astrometry and relative photometry (variability).

Specific results of this paper include:

\begin{itemize}
\item The single biggest update of (and addition to) RECONS astrometry to date: 210 systems with 220 distinct proper motions and parallaxes, of which 155 are new and 65 are updates to the RECONS catalog. 

\item Nine high-quality orbits from RECONS astrometry, fit using a new technique that determines the astrometric parameters simultaneously with the orbit elements \citep[introduced in][]{Die18}.

\item These orbits represent the beginning of a project to assemble M dwarf orbits across the entire range of that extensive spectral type, with the goal of identifying any trends (or lack thereof) in the sizes and shapes of these orbits.

\item Using a set of 542 RECONS systems (with 582 distinct parallaxes), we have defined four criteria for selection of potential unresolved multiples among nearby targets in \textit{Gaia} DR2 (see $\S$\ref{sec:paramcomparison}). 
\end{itemize}

These results, in particular the DR2 unresolved multiples selection criteria, can be used to hone samples for everything from stellar astrophysics to exoplanet searches. Additional observations, such as our new speckle imaging program at SOAR, will allow us to refine these DR2 criteria by identifying new unresolved multiples. The rich set of M dwarf multiples revealed by this work will be used in the Orbital Architectures project to answer fundamental questions about the formation of multi-star systems. Ultimately, these systems can either be avoided or targeted in searches for planets orbits the nearest stars.

\acknowledgements
{
This research has been supported by RECONS members Andrew Couperus and Jennifer Winters, who provided expertise in sample definition and multiplicity, as well as several useful conversations on specific systems.  Andrei Tokovinin has played a leading role in the SOAR observations that will be key in the next phase of this work.  Colleagues at the Cerro Tololo Inter-American Observatory and the SMARTS Consortium have played integral parts in supporting the effort at the 0.9~m for over two decades, and we are indebted to all those at CTIO and SMARTS who have made this work possible.

The National Science Foundation has provided consistent support of this long-term effort under grants AST-0507711, AST-0908402, AST- 1109445, AST-141206, and AST-1715551.

This work has used data products from the Two Micron All Sky Survey, which is a joint project of the University of Massachusetts and the Infrared Processing and Analysis Center at the California Institute of Technology, funded by NASA and NSF.

This work has made use of data from the European Space Agency (ESA) mission {\it Gaia} (\url{https://www.cosmos.esa.int/gaia}), processed by the {\it Gaia} Data Processing and Analysis Consortium (DPAC, \url{https://www.cosmos.esa.int/web/gaia/dpac/consortium}). Funding for the DPAC has been provided by national institutions, in particular the institutions participating in the {\it Gaia} Multilateral Agreement.

Information was collected from several additional large database efforts: the Simbad database and the VizieR catalogue access tool, operated at CDS, Strasbourg, France; NASA's Astrophysics Data System; and the Washington Double Star Catalog maintained at the U.S. Naval Observatory.
}

\facilities{CTIO:0.9m}

\software{GaussFit~\citep{Jef87}, IRAF~\citep{Tod86,Tod93}, SExtractor~\citep{Ber96}, }


\startlongtable
\begin{longrotatetable}
\begin{deluxetable}{llrrcrrrrrrrrrrrl}
\centerwidetable
\tabletypesize{\scriptsize}
\tablecaption{\parbox{25cm}{\scriptsize New and updated results from the RECONS astrometry program. Column 16 (Notes) indicates if a result is an update to a previously-published result for this system (``update'') in the \textit{Solar Neighborhood} series, if a preliminary orbit has been fit to improve the astrometry (``orbit''), or if the astrometry shows a perturbation that does not yet permit an orbit fit (``PB'').}
\label{tab:astrRECONS}}
\tablehead{
\colhead{    } & & \colhead{R.A.   } & \colhead{Decl.  } & \colhead{      } & \colhead{                }  & \colhead{                } & \colhead{        } & \colhead{Time} & \colhead{                } & \colhead{$\pi_\mathrm{rel}$} & \colhead{$\pi_\mathrm{corr}$} & \colhead{$\pi_\mathrm{abs}$} & \colhead{$\mu$          } & \colhead{$\theta$} & \colhead{$V_\mathrm{tan}$} & \colhead{     } \\[-1em] 
\colhead{Name} & & \colhead{J2000.0} & \colhead{J2000.0} & \colhead{Filter} & \colhead{$N_\mathrm{sea}$}  & \colhead{$N_\mathrm{frm}$} & \colhead{Coverage} & \colhead{(yr)} & \colhead{$N_\mathrm{ref}$} & \colhead{(mas)             } & \colhead{(mas)              } & \colhead{(mas)             } & \colhead{(mas yr$^{-1}$)} & \colhead{(degrees)} & \colhead{(km s$^{-1}$)   } & \colhead{Notes} \\[-1em] 
\colhead{(1) } & & \colhead{(2)    } & \colhead{(3)    } & \colhead{(4)   } & \colhead{(5)             }  & \colhead{(6)             } & \colhead{(7)     } & \colhead{(8) } & \colhead{(9)             } & \colhead{(10)              } & \colhead{(11)               } & \colhead{(12)              } & \colhead{(13)           } & \colhead{(14)     } & \colhead{(15)            } & \colhead{(16) } \\[-2em] 
}
\startdata
GJ 1001         & A   & 00 04 36.46   & $-$40 44 02.7   & $R$ & 17s & 154  & 1999.64-2018.73   & 19.09 & 5  &   80.73 $\pm$    1.64 &    1.03 $\pm$    0.16 &   81.76 $\pm$    1.65 &   1643.1 $\pm$     0.30 &   156.6 $\pm$    0.0 &    95.3 & update            \\ 
LEHPM 1-0255    & \,  & 00 09 45.06   & $-$42 01 39.6   & $V$ & 5c  & 66   & 2009.75-2018.70   &  8.95 & 5  &   55.89 $\pm$    1.14 &    1.14 $\pm$    0.22 &   57.03 $\pm$    1.16 &    248.3 $\pm$     0.40 &    90.5 $\pm$    0.1 &    20.6 & update            \\ 
GJ 2005         & ABC & 00 24 44.19   & $-$27 08 24.2   & $R$ & 21s & 189  & 1999.64-2019.74   & 20.10 & 6  &  116.73 $\pm$    3.22 &    0.98 $\pm$    0.04 &  117.71 $\pm$    3.22 &    691.0 $\pm$     0.50 &   350.7 $\pm$    0.1 &    27.8 & update, PB        \\ 
GJ 1012         & \,  & 00 28 39.46   & $-$06 39 49.1   & $V$ & 8s  & 65   & 2012.96-2019.52   &  6.56 & 5  &   75.45 $\pm$    2.36 &    1.11 $\pm$    0.10 &   76.56 $\pm$    2.36 &    862.6 $\pm$     1.00 &   202.8 $\pm$    0.1 &    53.4 &                   \\ 
LP 50-078       & \,  & 00 31 04.25   & $-$72 01 06.0   & $V$ & 6s  & 61   & 2014.93-2019.95   &  5.02 & 10 &   51.58 $\pm$    1.14 &    0.91 $\pm$    0.11 &   52.49 $\pm$    1.15 &    436.0 $\pm$     0.70 &    71.8 $\pm$    0.2 &    39.4 &                   \\ 
LP 645-053      & \,  & 00 35 44.13   & $-$05 41 10.6   & $I$ & 4c  & 72   & 2015.55-2018.94   &  3.39 & 7  &   54.30 $\pm$    0.92 &    0.96 $\pm$    0.23 &   55.26 $\pm$    0.95 &    265.0 $\pm$     1.00 &   183.9 $\pm$    0.3 &    22.7 &                   \\ 
LHS 1134        & \,  & 00 43 26.01   & $-$41 17 34.0   & $V$ & 11s & 91   & 2009.78-2019.75   &  9.97 & 7  &   92.89 $\pm$    1.20 &    1.95 $\pm$    0.27 &   94.84 $\pm$    1.23 &    764.4 $\pm$     0.40 &   220.2 $\pm$    0.1 &    38.2 & update, orbit     \\ 
LHS 1140        & \,  & 00 44 59.34   & $-$15 16 17.5   & $V$ & 13s & 87   & 2003.95-2018.76   & 14.81 & 5  &   62.91 $\pm$    1.64 &    0.61 $\pm$    0.11 &   63.52 $\pm$    1.64 &    669.6 $\pm$     0.40 &   156.4 $\pm$    0.1 &    50.0 & update            \\ 
2MA 0045+1634   & \,  & 00 45 21.41   & $+$16 34 44.7   & $I$ & 7s  & 27   & 2009.63-2019.76   & 10.13 & 8  &   63.34 $\pm$    2.32 &    0.71 $\pm$    0.16 &   64.05 $\pm$    2.33 &    357.7 $\pm$     0.50 &    96.9 $\pm$    0.1 &    26.5 &                   \\ 
2MA 0050-1538   & \,  & 00 50 24.42   & $-$15 38 19.2   & $I$ & 9s  & 35   & 2010.73-2019.77   &  9.03 & 9  &   42.08 $\pm$    1.66 &    0.43 $\pm$    0.08 &   42.51 $\pm$    1.66 &    519.2 $\pm$     0.40 &   202.8 $\pm$    0.1 &    57.9 &                   \\ 
GJ 1025         & \,  & 01 00 56.37   & $-$04 26 56.5   & $V$ & 12s & 88   & 2000.57-2018.67   & 18.09 & 6  &   84.13 $\pm$    1.44 &    0.57 $\pm$    0.05 &   84.70 $\pm$    1.44 &   1322.3 $\pm$     0.20 &    70.4 $\pm$    0.0 &    74.0 & update, PB        \\ 
LTT  573        & \,  & 01 01 24.65   & $-$01 05 58.6   & $I$ & 6s  & 70   & 2013.94-2018.94   &  5.00 & 6  &   32.30 $\pm$    1.18 &    2.88 $\pm$    0.13 &   35.18 $\pm$    1.19 &    278.5 $\pm$     0.90 &    86.4 $\pm$    0.3 &    37.5 &                   \\ 
GJ 1028         & \,  & 01 04 53.81   & $-$18 07 28.7   & $R$ & 8s  & 65   & 2012.95-2019.94   &  6.99 & 7  &  101.66 $\pm$    1.27 &    1.21 $\pm$    0.14 &  102.87 $\pm$    1.28 &   1376.7 $\pm$     0.60 &    69.8 $\pm$    0.1 &    63.4 &                   \\ 
SSS 0109-5100   & \,  & 01 09 01.51   & $-$51 00 49.5   & $I$ & 11s & 70   & 2009.75-2019.63   &  9.89 & 9  &   62.27 $\pm$    0.72 &    1.04 $\pm$    0.04 &   63.31 $\pm$    0.72 &    222.8 $\pm$     0.20 &    68.0 $\pm$    0.1 &    16.7 &                   \\ 
GJ 54.1         & \,  & 01 12 30.65   & $-$16 59 56.1   & $V$ & 7s  & 59   & 2003.85-2018.94   & 15.10 & 5  &  268.92 $\pm$    4.30 &    1.67 $\pm$    0.44 &  270.59 $\pm$    4.32 &   1349.6 $\pm$     1.00 &    62.7 $\pm$    0.1 &    23.6 &                   \\ 
DEN 0113-5429   & \,  & 01 13 16.41   & $-$54 29 13.8   & $R$ & 15s & 123  & 1999.91-2019.77   & 19.86 & 7  &   56.35 $\pm$    0.79 &    1.62 $\pm$    0.17 &   57.97 $\pm$    0.81 &    390.0 $\pm$     0.10 &    72.2 $\pm$    0.0 &    31.9 & update, PB        \\ 
G 34-023        & \,  & 01 22 10.30   & $+$22 09 02.7   & $V$ & 4c  & 54   & 2015.56-2018.93   &  3.37 & 7  &   84.29 $\pm$    1.70 &    1.39 $\pm$    0.16 &   85.68 $\pm$    1.71 &    274.7 $\pm$     1.50 &   123.1 $\pm$    0.6 &    15.2 &                   \\ 
LP 768-113      & \,  & 01 33 58.01   & $-$17 38 23.8   & $R$ & 9s  & 73   & 2008.70-2019.76   & 11.06 & 5  &   68.22 $\pm$    1.40 &    3.63 $\pm$    0.19 &   69.72 $\pm$    1.49 &    156.3 $\pm$     0.60 &   167.8 $\pm$    0.4 &    10.3 &                   \\ 
2MA 0138-7320   & \,  & 01 38 21.52   & $-$73 20 58.3   & $I$ & 8s  & 63   & 2009.74-2018.94   &  9.21 & 5  &   32.75 $\pm$    1.32 &    0.14 $\pm$    0.02 &   32.89 $\pm$    1.32 &    123.9 $\pm$     0.50 &   347.9 $\pm$    0.4 &    17.9 &                   \\ 
2MA 0141+1804   & \,  & 01 41 03.25   & $+$18 04 50.1   & $I$ & 6s  & 26   & 2009.56-2015.93   &  6.37 & 6  &   40.16 $\pm$    1.79 &    0.83 $\pm$    0.10 &   40.99 $\pm$    1.79 &    407.8 $\pm$     1.00 &    97.4 $\pm$    0.2 &    47.2 &                   \\ 
L 870-044       & AB  & 01 46 36.84   & $-$08 38 58.1   & $V$ & 6s  & 64   & 2013.67-2018.97   &  5.30 & 8  &   38.97 $\pm$    1.75 &    0.81 $\pm$    0.15 &   39.78 $\pm$    1.76 &    448.0 $\pm$     1.00 &   112.2 $\pm$    0.2 &    53.4 & PB                \\ 
L 88-043        & \,  & 01 53 37.08   & $-$66 53 34.1   & $R$ & 8s  & 86   & 2005.71-2018.93   & 13.22 & 6  &   78.85 $\pm$    3.20 &    2.97 $\pm$    0.24 &   81.82 $\pm$    3.21 &    420.0 $\pm$     0.80 &    65.8 $\pm$    0.2 &    24.3 & update            \\ 
GJ 83.1         & \,  & 02 00 12.96   & $+$13 03 07.1   & $V$ & 10s & 89   & 2010.74-2019.63   &  8.88 & 5  &  223.18 $\pm$    2.00 &    1.57 $\pm$    0.32 &  224.75 $\pm$    2.03 &   2068.1 $\pm$     0.60 &   148.3 $\pm$    0.0 &    43.6 &                   \\ 
LHS 1326        & \,  & 02 02 16.24   & $+$10 20 13.9   & $V$ & 13s & 59   & 2006.78-2019.93   & 13.15 & 5  &  108.10 $\pm$    1.88 &    0.28 $\pm$    0.03 &  108.38 $\pm$    1.88 &    736.9 $\pm$     0.50 &   248.3 $\pm$    0.1 &    32.2 &                   \\ 
LHS 1339        & \,  & 02 05 48.55   & $-$30 10 36.0   & $V$ & 15s & 98   & 2003.94-2019.75   & 15.81 & 5  &  105.55 $\pm$    1.38 &    0.59 $\pm$    0.07 &  106.14 $\pm$    1.38 &    553.1 $\pm$     0.30 &   280.9 $\pm$    0.1 &    24.7 &                   \\ 
LHS 1367        & \,  & 02 15 08.05   & $-$30 40 01.3   & $I$ & 8s  & 58   & 2012.88-2019.93   &  7.05 & 5  &   71.16 $\pm$    1.53 &    0.32 $\pm$    0.03 &   71.48 $\pm$    1.53 &    838.2 $\pm$     0.60 &   115.0 $\pm$    0.1 &    55.6 &                   \\ 
LHS 1375        & \,  & 02 16 29.86   & $+$13 35 12.7   & $V$ & 11s & 78   & 2009.75-2019.94   & 10.20 & 8  &  105.65 $\pm$    2.34 &    1.04 $\pm$    0.26 &  106.69 $\pm$    2.35 &    652.2 $\pm$     1.00 &   130.9 $\pm$    0.2 &    29.0 &                   \\ 
2MA 0228+1639   & \,  & 02 28 42.44   & $+$16 39 32.9   & $I$ & 6s  & 21   & 2010.75-2015.93   &  5.18 & 8  &   46.82 $\pm$    3.14 &    0.65 $\pm$    0.09 &   47.47 $\pm$    3.14 &    582.9 $\pm$     1.30 &   137.6 $\pm$    0.2 &    58.2 &                   \\ 
GJ 102          & \,  & 02 33 37.18   & $+$24 55 37.8   & $R$ & 10s & 87   & 2010.74-2019.94   &  9.20 & 8  &  100.42 $\pm$    2.76 &    1.49 $\pm$    0.18 &  101.91 $\pm$    2.77 &    671.6 $\pm$     0.70 &   176.0 $\pm$    0.1 &    31.2 &                   \\ 
GJ 105          & B   & 02 36 04.91   & $+$06 53 12.6   & $V$ & 5s  & 36   & 2010.97-2014.92   &  3.95 & 5  &  128.96 $\pm$    4.93 &    1.05 $\pm$    0.33 &  130.01 $\pm$    4.94 &   2317.6 $\pm$     3.40 &    51.2 $\pm$    0.2 &    84.5 &                   \\ 
\enddata

\tablecomments{The first 30 lines of this Table are shown to illustrate its form and content.}
\end{deluxetable}
\end{longrotatetable}

\startlongtable
\begin{longrotatetable}
\begin{deluxetable}{llrrrcrrrrrrrrrcc}
\centerwidetable
\tabletypesize{\scriptsize}
\tablecaption{\scriptsize Astrometric solutions from RECONS and \textit{Gaia} DR2 for red dwarfs within 25 pc common to both these catalogs. Columns 7--14 reproduce the parameters characterizing the DR2 astrometric fits. Column 15 (``class.'') indicates the classification given to each system in the plots of Figures~\ref{fig:pipi}, \ref{fig:astromparmsGOOD}, and \ref{fig:paramsBAD}: res = resolved companion, unr = unresolved multiple, PB = perturbation in astrometric residuals (but companion not yet confirmed), no label = presumed single. \\
Check marks in Column 16 (``sus.'') indicate the system that meet all four criteria given in $\S$\ref{sec:paramcomparison} for suspicion of being unresolved multiples. Systems that are missing parallaxes in DR2 have been included in that set. \\
Reference codes for RECONS parallaxes: * = This work, Bar17 = \cite{Bar17}, Ben16 = \cite{Ben16}, Dav15 = \cite{Dav15}, Hen18 = \cite{Hen18}, Jao05 = \cite{Jao05}, Jao11 = \cite{Jao11}, Jao17 = \cite{Jao17}, Lur14 = \cite{Lur14}, Rie10 = \cite{Rie10}, Rie14 = \cite{Rie14}, Rie18 = \cite{Rie18}, Sub09 = \cite{Sub09}, Win17 = \cite{Win17}
\label{tab:RECONS-GDR2}}
\tablehead{
\colhead{    } & \colhead{} & \colhead{       } & \colhead{       } & \colhead{RECONS            } & \colhead{    } & \colhead{\textit{Gaia} DR2 } & \colhead{         } & \colhead{      } & \colhead{          } & \colhead{                 } & \colhead{                } & \colhead{$N_\mathrm{bad}$  } & \colhead{    } & \colhead{     } & \colhead{      } & \colhead{    } \\[-1em] 
\colhead{    } & \colhead{} & \colhead{R.A.   } & \colhead{Decl.  } & \colhead{$\pi_\mathrm{abs}$} & \colhead{    } & \colhead{$\pi_\mathrm{abs}$} & \colhead{Goodness-} & \colhead{Excess} & \colhead{Excess    } & \colhead{                 } & \colhead{                } & \colhead{$/N_\mathrm{good}$} & \colhead{    } & \colhead{$G$  } & \colhead{      } & \colhead{    } \\[-1em] 
\colhead{Name} & \colhead{} & \colhead{J2000.0} & \colhead{J2000.0} & \colhead{(mas)             } & \colhead{Ref.} & \colhead{(mas)             } & \colhead{of-fit   } & \colhead{noise } & \colhead{noise sig.} & \colhead{$N_\mathrm{good}$} & \colhead{$N_\mathrm{bad}$} & \colhead{(\%)              } & \colhead{RUWE} & \colhead{(mag)} & \colhead{class.} & \colhead{sus.} \\[-1em] 
\colhead{(1) } & \colhead{} & \colhead{(2)    } & \colhead{(3)    } & \colhead{(4)               } & \colhead{(5) } & \colhead{(6)               } & \colhead{(7)      } & \colhead{(8)   } & \colhead{(9)       } & \colhead{(10)             } & \colhead{(11)            } & \colhead{(12)              } & \colhead{(13)} & \colhead{(14) } & \colhead{(15)  } & \colhead{(16)} \\[-2em] 
}
\startdata
GJ 1001          & BC  & 00 04 36.46  & $-$40 44 02.7   & $  77.02 \pm    2.07$ & Die14 & $   82.095 \pm    0.377$ &    14.161 &   1.460 &    15.3 & 271 & 5   &   1.8 &  1.176 &  18.35 & unr &   \\ 
GJ 1001          & A   & 00 04 36.46  & $-$40 44 02.7   & $  81.76 \pm    1.65$ & *     & $   81.228 \pm    0.114$ &    40.520 &   0.285 &    45.4 & 253 & 13  &   5.1 &  1.409 &  11.50 & res &   \\ 
GJ 1002          & \,  & 00 06 43.19  & $-$07 32 17.0   & $ 207.18 \pm    3.09$ & Dav15 & $  206.213 \pm    0.128$ &    51.302 &   0.366 &    63.8 & 261 & 5   &   1.9 &  1.280 &  11.78 &     &   \\ 
LTT 57           & \,  & 00 08 17.37  & $-$57 05 52.9   & $  75.17 \pm    2.11$ & Win17 & $   78.115 \pm    0.061$ &    24.222 &   0.172 &    14.0 & 213 & 9   &   4.2 &  1.123 &  10.94 &     &   \\ 
G 131-026        & AB  & 00 08 53.92  & $+$20 50 25.4   & $  54.13 \pm    1.35$ & Rie14 & $   55.255 \pm    0.761$ &   333.638 &   4.059 &  6740.0 & 165 & 8   &   4.8 & 21.268 &  11.99 & unr & $\checkmark$ \\ 
LEHPM 1-0255     & AB  & 00 09 45.06  & $-$42 01 39.6   & $  57.03 \pm    1.16$ & *     & $   57.200 \pm    0.245$ &   121.537 &   0.996 &   516.0 & 343 & 59  &  17.2 &  3.749 &  12.15 & unr &   \\ 
GJ 1005          & AB  & 00 15 28.07  & $-$16 08 01.8   & $ 169.52 \pm    0.97$ & *     & $                      $ &   247.417 &   8.177 & 16100.0 & 65  & 0   &   0.0 &        &  10.15 & unr & $\checkmark$  \\ 
LHS 1050         & \,  & 00 15 49.25  & $+$13 33 22.3   & $  85.85 \pm    2.57$ & Rie10 & $   81.871 \pm    0.087$ &    24.271 &   0.187 &    15.0 & 150 & 7   &   4.7 &  1.220 &  11.40 &     &   \\ 
NLTT 1261        & \,  & 00 24 24.63  & $-$01 58 20.0   & $  82.43 \pm    2.22$ & Rie18 & $   79.965 \pm    0.221$ &    36.677 &   1.105 &    66.0 & 349 & 0   &   0.0 &  1.213 &  16.60 &     &   \\ 
GJ 2005          & ABC & 00 24 44.19  & $-$27 08 24.2   & $ 117.71 \pm    3.22$ & *     & $                      $ &    47.837 &   1.090 &   214.0 & 96  & 0   &   0.0 &        &  13.09 & unr & $\checkmark$  \\ 
LP 349-025       & AB  & 00 27 55.99  & $+$22 19 32.3   & $  66.31 \pm    1.48$ & *     & $                      $ &   258.788 &   6.254 &  6230.0 & 250 & 0   &   0.0 &        &  14.94 & unr & $\checkmark$  \\ 
GJ 1012          & \,  & 00 28 39.46  & $-$06 39 49.1   & $  76.56 \pm    2.36$ & *     & $   74.775 \pm    0.100$ &    27.903 &   0.245 &    21.4 & 163 & 3   &   1.8 &  1.322 &  10.89 &     &   \\ 
LP 50-078        & \,  & 00 31 04.25  & $-$72 01 06.0   & $  52.49 \pm    1.15$ & *     & $   50.679 \pm    0.041$ &    22.201 &   0.163 &    11.3 & 212 & 5   &   2.4 &  1.124 &  12.32 &     &   \\ 
GIC 50           & ABC & 00 32 53.14  & $-$04 34 07.0   & $  52.61 \pm    1.05$ & Rie18 & $   52.445 \pm    0.137$ &    61.981 &   0.403 &    75.6 & 365 & 14  &   3.8 &  1.822 &  12.53 & unr &   \\ 
LP 645-053       & \,  & 00 35 44.13  & $-$05 41 10.6   & $  55.26 \pm    0.95$ & *     & $   55.851 \pm    0.092$ &    33.298 &   0.502 &    55.7 & 321 & 1   &   0.3 &  1.096 &  14.15 &     &   \\ 
LHS 1134         & \,  & 00 43 26.01  & $-$41 17 34.0   & $  94.84 \pm    1.23$ & *     & $   97.728 \pm    0.082$ &    43.135 &   0.280 &    46.9 & 306 & 12  &   3.9 &  1.305 &  11.56 & PB  &   \\ 
LHS 1140         & \,  & 00 44 59.34  & $-$15 16 17.5   & $  63.52 \pm    1.64$ & *     & $   66.700 \pm    0.067$ &    28.682 &   0.173 &    14.9 & 284 & 7   &   2.5 &  1.117 &  12.67 &     &   \\ 
2MA 0045+1634    & \,  & 00 45 21.41  & $+$16 34 44.7   & $  64.05 \pm    2.33$ & *     & $   65.015 \pm    0.227$ &    10.612 &   0.924 &     9.8 & 430 & 1   &   0.2 &  0.997 &  18.17 &     &   \\ 
LHS 124          & \,  & 00 49 29.05  & $-$61 02 32.7   & $  48.62 \pm    1.39$ & Jao11 & $   51.258 \pm    0.041$ &    23.933 &   0.146 &     9.6 & 251 & 7   &   2.8 &  1.140 &  11.16 &     &   \\ 
2MA 0050-1538    & \,  & 00 50 24.42  & $-$15 38 19.2   & $  42.51 \pm    1.66$ & *     & $   40.319 \pm    0.243$ &     8.325 &   0.909 &     7.5 & 417 & 1   &   0.2 &  0.955 &  18.44 &     &   \\ 
L 87-002         & \,  & 00 57 12.48  & $-$64 15 24.0   & $  56.25 \pm    1.38$ & Win17 & $   56.958 \pm    0.037$ &    22.986 &   0.150 &    10.5 & 257 & 6   &   2.3 &  0.928 &  11.19 &     &   \\ 
GJ 1025          & \,  & 01 00 56.37  & $-$04 26 56.5   & $  84.70 \pm    1.44$ & *     & $   80.924 \pm    0.082$ &    33.028 &   0.197 &    23.4 & 318 & 8   &   2.5 &  1.248 &  11.98 & PB  &   \\ 
L 87-010         & \,  & 01 04 06.95  & $-$65 22 27.3   & $  85.51 \pm    1.47$ & Win17 & $   80.193 \pm    0.043$ &    22.852 &   0.150 &     9.7 & 198 & 2   &   1.0 &  1.062 &  12.47 &     &   \\ 
GJ 1028          & \,  & 01 04 53.81  & $-$18 07 28.7   & $ 102.87 \pm    1.28$ & *     & $  102.322 \pm    0.075$ &    35.187 &   0.263 &    35.0 & 302 & 13  &   4.3 &  1.058 &  12.71 &     &   \\ 
SSS 0109-5100    & \,  & 01 09 01.51  & $-$51 00 49.5   & $  63.31 \pm    0.72$ & *     & $   62.853 \pm    0.113$ &    28.113 &   0.925 &    40.4 & 356 & 1   &   0.3 &  1.044 &  16.67 &     &   \\ 
LP 647-013       & \,  & 01 09 51.20  & $-$03 43 26.4   & $  94.11 \pm    0.85$ & Hen18 & $   94.398 \pm    0.318$ &    49.172 &   1.599 &   113.0 & 238 & 1   &   0.4 &  1.738 &  16.35 &     &   \\ 
LP 707-016       & \,  & 01 10 17.53  & $-$11 51 17.6   & $  53.94 \pm    1.47$ & Win17 & $   54.428 \pm    0.111$ &    50.023 &   0.295 &    49.4 & 334 & 13  &   3.9 &  1.496 &  11.39 &     &   \\ 
GJ 54            & AB  & 01 10 22.90  & $-$67 26 41.9   & $ 124.94 \pm    2.08$ & Hen18 & $  121.449 \pm    1.193$ &   558.636 &   8.605 & 31600.0 & 252 & 0   &   0.0 & 43.292 &   8.70 & unr & $\checkmark$  \\ 
LP 467-016       & AB  & 01 11 25.41  & $+$15 26 21.6   & $  45.79 \pm    1.78$ & Rie14 & $                      $ &           &         &         &     &     &       &        &        & unr & $\checkmark$  \\ 
SCR 0111-4908    & \,  & 01 11 47.52  & $-$49 08 08.9   & $  54.26 \pm    1.39$ & Win17 & $   53.733 \pm    0.090$ &    39.993 &   0.742 &    69.2 & 381 & 0   &   0.0 &  1.159 &  15.34 &     &   \\ 
\enddata

\tablecomments{The first 30 lines of this Table are shown to illustrate its form and content.}
\end{deluxetable}
\end{longrotatetable}

\startlongtable
\begin{longrotatetable}
\movetabledown=25mm
\begin{deluxetable}{lclrrrrrrrrrrr}
\tablewidth{25cm}
\tabletypesize{\scriptsize}
\tablecaption{\parbox{23cm}{\scriptsize Elements for the nine photocentric orbits presented in Figures~\ref{fig:orbits_cal}, \ref{fig:orbits1}, and \ref{fig:orbits2}. Orbits were fit simultaneously with proper motion and parallax to RECONS astrometry data using the routine of \cite{Die18}. \\
Three systems are calibration systems, fit using the same routine as the science systems above. For these cases, the reference for the values appears to the right of the system name: * = work, Ben16 = \cite{Ben16}. }
\label{tab:orbelements}}
\tablehead{
\colhead{    } & \colhead{}& \colhead{R.A.   } & \colhead{Decl.  } & \colhead{$\pi$} & \colhead{$\mu_\mathrm{RA}$} & \colhead{$\mu_\mathrm{DEC}$}& \colhead{$P$    } & \colhead{$a$  } & \colhead{$e$} & \colhead{$i$  } & \colhead{$\Omega$} & \colhead{$\omega$} & \colhead{$T_0$ }   \\[-1em]
\colhead{Name} & \colhead{}& \colhead{J2000.0} & \colhead{J2000.0} & \colhead{(mas)} & \colhead{(mas yr$^{-1}$)  } & \colhead{(mas yr$^{-1}$)   }& \colhead{(years)} & \colhead{(mas)} & \colhead{   } & \colhead{(deg)} & \colhead{(deg)   } & \colhead{(deg)   } & \colhead{(year)}   \\[-1em]
\colhead{(1) } & \colhead{}& \colhead{(2)    } & \colhead{(3)    } & \colhead{(4)  } & \colhead{(5)              } & \colhead{(6)               }& \colhead{(7)    } & \colhead{(8)  } & \colhead{(9)} & \colhead{(10) } & \colhead{(11)    } & \colhead{(12)    } & \colhead{(13)  }   \\[-2em]
}
\startdata
\cutinhead{ New Systems}
LP 349-025 AB   & *  & 00 27 55.99  & $+$22 19 32.3  & $66.31\pm 1.48$ & $397.6\pm 0.2$&$-161.9\pm 0.2$ & $7.37 \pm 0.19$  &  $14.63 \pm 1.34$ & $0.38 \pm 0.16$ & $ 97.47 \pm 5.20$ & $ 49.77 \pm  4.29$        & $145.06 \pm 27.20$      & $1996.05 \pm 0.50$ \\
LHS 1582 AB     & *  & 03 43 22.08  & $-$09 33 50.9  & $47.62\pm 0.95$ & $404.0\pm 0.1$& $308.1\pm 0.1$ & $5.23 \pm 0.02$  &  $23.70 \pm 0.68$ & $0.27 \pm 0.05$ & $143.95 \pm 4.39$ & $111.99 \pm  6.57$        & $ 14.90 \pm 12.53$      & $1995.82 \pm 0.13$ \\
LTT 6288 AB     & *  & 15 45 41.62  & $-$43 30 29.0  & $48.22\pm 0.72$ &$-272.2\pm 0.1$&$-366.1\pm 0.1$ & $9.87 \pm 0.06$  &  $36.20 \pm 1.14$ & $0.51 \pm 0.06$ & $ 89.95 \pm 0.80$ & $154.16 \pm  1.00$        & $191.57 \pm  4.37$      & $1994.01 \pm 0.18$ \\
USN 2101+0307 AB& *  & 21 01 04.80  & $+$03 07 04.7  & $52.91\pm 0.91$ &$1009.2\pm 0.1$& $-29.6\pm 0.2$ & $7.53 \pm 0.05$  &  $33.69 \pm 1.64$ & $0.55 \pm 0.05$ & $ 36.71 \pm 4.63$ & $172.22 \pm  8.85$        & $ 57.17 \pm  8.64$      & $2004.14 \pm 0.12$ \\
LEHPM 1-4771 AB & *  & 22 30 09.41  & $-$53 44 55.5  & $64.35\pm 0.99$ & $-64.4\pm 0.2$&$-739.7\pm 0.1$ & $5.79 \pm 0.06$  &  $22.39 \pm 0.98$ & $0.30 \pm 0.07$ & $122.49 \pm 2.31$ & $ 87.94 \pm  2.49$        & $ 32.28 \pm 15.86$      & $2005.29 \pm 0.28$ \\
LTT 9828 AB     & *  & 23 59 44.77  & $-$44 05 00.3  & $58.87\pm 1.14$ & $-33.4\pm 0.1$& $256.6\pm 0.1$ &$11.17 \pm 0.07$  &  $34.65 \pm 0.90$ & $0.38 \pm 0.04$ & $ 33.95 \pm 4.40$ & $123.78 \pm  6.68$        & $141.01 \pm 10.11$      & $1991.74 \pm 0.24$ \\
\cutinhead{ Calibration Systems}
GJ 1005 AB & *       & 00 15 28.07  & $-$16 08 01.8  &$169.52\pm 0.97$ & $597.4\pm 0.2$&$-605.6\pm 0.1$ & $4.56 \pm 0.01$  &  $60.04 \pm 1.06$ & $0.32 \pm 0.03$ & $143.51 \pm 2.14$ & $ 60.78 \pm  3.50$        & $166.06 \pm  6.28$      & $1999.93 \pm 0.06$ \\
           & Ben16   &              &                &                 &               &                & $4.56 \pm 0.01$  &                   & $0.36 \pm 0.01$ & $146.1  \pm 0.2 $ & $ 62.8  \pm  0.4$\tablenotemark{$\dagger$} & $166.6 \pm 0.5$\tablenotemark{$\dagger$} & $1995.36 \pm 0.01$ \\
           &         &              &                &                 &               &                &                  &                   &                 &                   &                           &                         &                    \\
GJ 234 AB  & *       & 06 29 23.39  & $-$02 48 48.8  &$241.82\pm 1.21$ & $699.7\pm 0.7$&$-693.0\pm 0.4$ &$16.63 \pm 0.14$  & $294.94 \pm 2.49$ & $0.35 \pm 0.01$ &  $54.26 \pm 0.38$ &  $31.58 \pm  0.61$        &  $43.55 \pm  2.13$      & $1999.38 \pm 0.19$ \\
           & Ben16   &              &                &                 &               &                &$16.62 \pm 0.03$  &                   & $0.38 \pm 0.01$ &  $53.2  \pm 0.1 $ & $30.60  \pm  0.1$\tablenotemark{$\dagger$} & $40.4 \pm  0.1$\tablenotemark{$\dagger$} & $1999.27 \pm 0.01$ \\
           &         &              &                &                 &               &                &                  &                   &                 &                   &                           &                         &                    \\
GJ 748 AB  & *       & 19 12 14.60  & $+$02 53 11.0  & $99.89\pm 1.08$ &$1787.7\pm 0.1$&$-502.2\pm 0.1$ & $2.49 \pm 0.01$  &  $29.01 \pm 0.96$ & $0.43 \pm 0.05$ & $131.00 \pm 3.51$ & $161.69 \pm  3.99$        & $142.07 \pm  7.81$      & $2002.97 \pm 0.01$ \\
           & Ben16   &              &                &                 &               &                & $2.47 \pm 0.01$  &                   & $0.45 \pm 0.01$ & $131.6  \pm 0.3 $ & $179.6  \pm  0.2$\tablenotemark{$\dagger$} & $207.3 \pm 0.4$\tablenotemark{$\dagger$} & $1995.86 \pm 0.01$ \\
\enddata

\tablenotetext{\dagger}{angle rotated $180^\circ$ to match the corresponding orientation for a photocentric orbit, as \cite{Ben16} reports the relative orbits.}
\tablecomments{Semi-major axis ($a$) is not included for the Ben16 orbits because that work fit the systems' relative orbits, whereas ours fits the photocentric orbits, hence that parameter is not comparable.}
\end{deluxetable}
\end{longrotatetable}

\end{document}